\documentclass[a4paper,11pt]{article}
\usepackage{jheppub} 
\usepackage{lineno}
\usepackage{amssymb}
\usepackage{graphicx}
\usepackage{bm}

\newcommand*\xbar[1]{%
  \hbox{%
    \vbox{%
      \hrule height 0.5pt 
      \kern0.5ex
      \hbox{%
        \kern-0.1em
        \ensuremath{#1}%
        \kern-0.1em
      }%
    }%
  }%
}

\arxivnumber{2604.23329} 

\title{\boldmath Some approximate renormalization group invariants for supersymmetric extensions of the Standard Model and the Yukawa unification}

\allowdisplaybreaks

\author{K.D. Krylov, D.M. Rystsov, and K.V. Stepanyantz}
\affiliation{Department  of Theoretical Physics, Moscow State University,\\
Leninskie Gory, 119991 Moscow, Russia}

\emailAdd{kirdkr@mail.ru, scall2017@mail.ru, stepan@m9com.ru}

\abstract{For supersymmetric extensions of the Standard Model we construct some expressions that include Yukawa couplings for the third and second generations and receive relatively small quantum corrections. This implies that they slightly depend on scale and are therefore approximate renormalization group invariants. Using these invariants we try to analyse possible relations between the Yukawa couplings at the unification scale $M_X$ for some values of $\mbox{tg}\,\beta$ for the MSSM and its extensions containing certain additional superfields forming complete $SU(5)$ multiplets.}

\begin{document}
\maketitle
\flushbottom

\section{Introduction}
\hspace*{\parindent}

The quantum numbers of various fields with respect to the $SU(3)\times SU(2)\times U(1)$ group imply that the Standard Model may be a low energy remnant of a Grand Unified Theory with a wider gauge symmetry. The simplest such theory \cite{Georgi:1974sy} is based on the group $SU(5)$. In this case the fermions of a single generation (including the right-handed neutrinos) can be placed into three irreducible representations, $1+5+\xbar{10}$ if we deal with the right-handed fermions performing the charge conjugation for the left-handed ones. These representations appear in the branching rule for the representation $\xbar{16}$ of $SO(10)$ \cite{Slansky:1981yr}. That is why in Grand Unified Theories based on the group $SO(10)$ \cite{Fritzsch:1974nn,Georgi:1974my} the fermions of one generation can be placed into a single irreducible representation. It is also possible to consider Grand Unified Theories based on the group $E_6$ \cite{Gursey:1975ki} (see \cite{King:2020ldn} for a recent review), which contains the maximal subgroup $SO(10)\times U(1)$. (The use of large gauge groups $E_7$ and $E_8$ faces the difficulty that all their representations are vectorlike \cite{Slansky:1981yr}, and in this case one has to construct a mechanism for parity breaking.)

One of the most famous predictions of Grand Unified Theories is the unification of the gauge couplings $\alpha_1=\alpha_2=\alpha_3$, where

\begin{equation}\label{Gauge_Couplings}
\alpha_1 = \frac{5}{3}\cdot\frac{e_1^2}{4\pi};\qquad \alpha_2 = \frac{e_2^2}{4\pi};\qquad \alpha_3 = \frac{e_3^2}{4\pi}.
\end{equation}

\noindent
The factor $5/3$ in the definition of the coupling $\alpha_1$ encodes the value of the Weinberg angle at the unification scale, for which the prediction is $\sin^2\theta_W=3/8$ \cite{Georgi:1974sy}. The renormalization group running of couplings in the Standard Model does not agree with this prediction (see, e.g., Ref. \cite{Arason:1991ic}), but in its supersymmetric extensions the running couplings really meet at a single point with much better accuracy \cite{Ellis:1990wk,Amaldi:1991cn,Langacker:1991an}. Certainly, this is a very convincing evidence in favor of both supersymmetry \cite{Gates:1983nr,West:1990tg,Buchbinder:1998qv} and Grand Unification \cite{Mohapatra:1986uf,Raby:2017ucc}, so that in what follows we will deal only with the (softly broken) supersymmetric theories.

Another interesting prediction of Grand Unification is the existence of certain relations between Yukawa couplings leading in turn to some relations between elementary particle masses. For instance, the simplest model based on the group $SU(5)$ predicts that at the unification scale the Yukawa matrices satisfy the relation $Y_D = (Y_E)^T$, so that the masses of down quarks and charged leptons should coincide \cite{Georgi:1974sy,Chanowitz:1977ye,Buras:1977yy}.\footnote{The detailed analysis of the mass relations in the supersymmetric case at the quantum level can be found in \cite{Einhorn:1981sx,Inoue:1982pi,Ibanez:1983di,Ibanez:1983wi}.} However, this prediction cannot be valid simultaneously for all generations. Really, the expression

\begin{equation}\label{Standard_RGI}
\frac{m_e\, m_s}{m_\mu\, m_d}
\end{equation}

\noindent
receives rather small quantum corrections (because of the hierarchical structure of the Yukawa couplings) and is therefore an approximate renormalization group invariant (RGI) \cite{Raby:2017ucc}. Its value is close to $1/9$, so the prediction $m_e=m_d$, $m_\mu = m_s$ cannot be satisfied even at the unification scale. A beautiful solution to this contradiction was proposed by Georgi and Jarlskog \cite{Georgi:1979df}. They suggested that the Yukawa matrices $Y_D$ and $Y_E$ have the structure

\begin{equation}\label{Georgi_Jarlskog}
Y_D =\left(
\begin{array}{ccc}
0 & A & 0\\
A & B & 0\\
0 & 0 & C
\end{array}
\right);\qquad
Y_E =\left(
\begin{array}{ccc}
0 & A & 0\\
A & -3B & 0\\
0 & 0 & C
\end{array}
\right),
\end{equation}

\noindent
where $C\gg B\gg A$ are dimensionless constants. Then, the seesaw-like mechanism gives the mass relations

\begin{equation}
m_b = m_\tau,\qquad m_s\approx \frac{1}{3}m_\mu,\qquad m_d \approx 3m_e,
\end{equation}

\noindent
which are in much better agreement with the experimental value of the RGI (\ref{Standard_RGI}). The factor $(-3)$ for the second generation can be obtained under the assumptions that the corresponding Higgs superfield comes from the representation 45 of the group $SU(5)$ and the corresponding Yukawa term has the structure $5\times\xbar{10}\times 45$ \cite{Nanopoulos:1978kz,Frampton:1979wf}. In theories based on the group $SO(10)$ this factor may appear from the interaction $\,\xbar{16}\,\times\,\xbar{16}\,\times 126$ \cite{Georgi:1979ga}. This implies that realistic supersymmetric Grand Unified Theories are (much) more complicated than the simplest models, see, e.g., \cite{Raby:2017ucc} and references therein.

Theories with larger gauge symmetry predict more restrictive relations between Yukawa couplings. For instance, in the models based on the group $SO(10)$ the simplest Yukawa interaction $\xbar{16}\times\xbar{16}\times 10$ leads to the relation $Y_U = Y_D = Y_E = -Y_\nu$ (see, e.g., Eq. (\ref{16_Yukawa_Relations}) in what follows), which contradicts the experimental data supplemented by the renormalization group running of couplings. Due to the relation

\begin{equation}
\xbar{16} \times \xbar{16} = 10_s + 120_a + \xbar{126}_s,
\end{equation}

\noindent
it is also reasonable to consider the invariants $\,\xbar{16}\times \xbar{16}\times 120$ and $\,\xbar{16}\times\xbar{16}\times 126$, which are antisymmetric and symmetric under the exchange of two representations $\,\xbar{16}\,$, respectively. The corresponding predictions were obtained in \cite{Mohapatra:1979nn,Nath:2001uw}. Comparing them (or other similar predictions) with experimental data, one can use different values of unknown parameters of the Minimal Supersymmetric Standard Model (MSSM), such as $\mbox{tg}\,\beta$, or even add some new superfields in addition to those present in the MSSM. Therefore, it is highly desirable to understand if it is possible to achieve agreement by a special choice of free parameters and/or field content of the theory or such agreement can never be reached. For this purpose, one may use expressions similar to (\ref{Standard_RGI}) which do not receive (large) quantum corrections and can be calculated at low energies. If they allowed for predicting some Yukawa relations, then it would be much easier to choose the gauge group and to construct the Yukawa interaction in Grand Unified Theories.

In principle, there are two independent all-loop exact RGIs in the MSSM \cite{Rystsov:2024soq} constructed with the help of the NSVZ equations \cite{Novikov:1983uc,Jones:1983ip,Novikov:1985rd,Shifman:1986zi} for this theory \cite{Shifman:1996iy,Korneev:2021zdz} and the nonrenormalization theorem for the superpotential \cite{Grisaru:1979wc}. However, they involve the Yukawa couplings for all generations and are not convenient for obtaining simple mass relations. That is why it is better to construct approximate RGIs which involve fewer Yukawa couplings. To do this, we will use the fact that the Yukawa couplings have the hierarchical structure. This implies that in radiative corrections one can neglect the small values of the Yukawa couplings for the first and second generations. Of course, doing this, it is necessary to use a certain textures for the Yukawa matrices. The structure of these matrices compatible with the experimental data was analysed in \cite{Hall:1993ni}. The options 4) for the Yukawa couplings in Ref. \cite{Hall:1993ni} correspond to small non-diagonal terms $(Y_U)_{3I}$, $(Y_U)_{I3}$, $(Y_D)_{3I}$, and $(Y_D)_{I3}$. This will always be assumed in this paper. In fact, we will investigate only the renormalization group behaviour of the components $(Y_i)_{33}$ and $(Y_i)_{22}$, where $i=U,\,D,\,E$, assuming that the other components of the Yukawa matrices are much smaller. For this purpose, we will construct approximate RGIs from these couplings and analyse them. In particular, we would like to investigate whether it is possible to construct phenomenologically satisfactory relations between Yukawa couplings, which may possibly be obtained from the first principles on the base certain group invariants.

The paper is organized as follows. In Sect.~\ref{Section_MSSM_RG} we recall some basic information about the MSSM. After that, in Sect.~\ref{Section_RGI_Yukawa} we construct certain approximate RGIs which slightly depend on scale in the MSSM and its extensions containing some exotic superfields (under the assumption that their Yukawa interaction is rather small in comparison with the Yukawa interaction for the third generation). In the next Sect.~\ref{Section_Yukawa_MSSM} we verify these results by numerical analysis for two different values of $\mbox{tg}\,\beta$. In particular, it is demonstrated that the constructed approximate RGIs indeed slightly depend on scale. However, the values of the Yukawa couplings for the third and second generations in the MSSM with a single supersymmetric threshold $M_{\mbox{\scriptsize SUSY}}$ do not seem to satisfy simple relations at the unification scale. Somewhat better ratios of the Yukawa couplings at the unification scale can be obtained in this case for extensions of the MSSM containing some additional superfields forming complete $SU(5)$ multiplets. They are investigated in Sect.~\ref{Section_Yukawa_With_Extra_Matter}. Nevertheless, it seems very difficult to derive even these ratios from group theory, at least, from cubic invariants. As an example, in Sect.~\ref{Section_Yukawa_Relation} we consider the $E_6$ invariant $\,\xbar{27}\,\times\,\xbar{27}\,\times \,\xbar{351}\,'$ and try to derive Yukawa relations which mostly resemble those which follow from the renormalization group analysis. However, although the resulting equations capture certain features of the Yukawa relations, they lead to different ratios of the Yukawa couplings at the unification scale. The results are briefly summarized and discussed in Conclusion. Some auxiliary equations and technical details of calculations are presented in Appendices.

\section{Some aspects of the MSSM and its renormalization}
\hspace*{\parindent}\label{Section_MSSM_RG}

The MSSM is the simplest (softly broken) supersymmetric theory extending the Standard Model. Like the Standard Model, it is a gauge theory based on the group $SU(3)\times SU(2)\times U(1)$. This in particular implies that there are three gauge coupling constants which are defined by Eq. (\ref{Gauge_Couplings}). Quarks, leptons, and Higgs scalars in the supersymmetric case are components of chiral superfields. (Certainly, the chiral superfields corresponding to left-handed fermions include the charge-conjugate left-handed quark and lepton fields.) Unlike the Standard Model, the MSSM contains two Higgs doublets inside the chiral superfields $H_u$ and $H_d$. Then, the MSSM superpotential is given by the expression\footnote{For simplicity, here we consider the theory without the right-handed neutrinos.}

\begin{eqnarray}\label{Superpotential_For_MSSM}
&&\hspace*{-9mm} W = \left(Y_U\right)_{IJ}
\left(\widetilde U\ \widetilde D \right)^{a}_I
\left(
\begin{array}{cc}
0 & 1\\
-1 & 0
\end{array}
\right)
\left(
\begin{array}{c}
H_{u1}\\ H_{u2}
\end{array}
\right) U_{aJ}
+ \left(Y_D\right)_{IJ}
\left(\widetilde U\ \widetilde D \right)^{a}_I
\left(
\begin{array}{cc}
0 & 1\\
-1 & 0
\end{array}
\right)
\left(
\begin{array}{c}
H_{d1}\\ H_{d2}
\end{array}
\right)
\nonumber\\
&&\hspace*{-9mm} \times D_{aJ} + \left(Y_E\right)_{IJ} \left(\widetilde N\ \widetilde E \right)_{I}
\left(
\begin{array}{cc}
0 & 1\\
-1 & 0
\end{array}
\right)
\left(
\begin{array}{c}
H_{d1}\\ H_{d2}
\end{array}
\right) E_J
+ \bm{\mu} \left(H_{u1}\ H_{u2} \right)
\left(
\begin{array}{cc}
0 & 1\\
-1 & 0
\end{array}
\right)
\left(
\begin{array}{c}
H_{d1}\\ H_{d2}
\end{array}
\right).
\end{eqnarray}

\noindent
This superpotential contains the dimensionless $3\times 3$ Yukawa matrices $Y_U$, $Y_D$, $Y_E$ and the parameter $\bm{\mu}$ with the dimension of mass. The chiral superfields

\begin{equation}
Q^a_I = \left(\begin{array}{c}\widetilde U^a\\ \widetilde D^a\end{array}\right)_I\qquad\mbox{and}\qquad L_I = \left(\begin{array}{c}\widetilde N\\ \widetilde E\end{array}\right)_I
\end{equation}

\noindent
include the charge-conjugate left-handed quarks and leptons, respectively. The superfields $D_{aI}$, $U_{aI}$, and $E_I$ contain right-handed down quarks, right-handed upper quarks, and right-handed charged leptons, respectively. The capital letters numerate generations and range from 1 to 3. The lower and upper color indices $a$ correspond to the fundamental and antifundamental representations of $SU(3)$, respectively. The part of the action corresponding to the superpotential in our notation  is written in the form

\begin{equation}
\Delta S = \frac{1}{2} \int d^4x\, d^2\theta\, W + \mbox{c.c.}
\end{equation}

The lowest scalar components of the Higgs superfields $H_u$ and $H_d$ (which are denoted by $h_u$ and $h_d$, respectively) acquire the vacuum expectation values

\begin{equation}\label{Hoggs_VeVs}
(h_u)_0 = \left(
\begin{array}{c}
0\\ v_u
\end{array}
\right);\qquad
(h_d)_0 = \left(
\begin{array}{c}
v_d\\ 0
\end{array}
\right),
\end{equation}

\noindent
which break the original gauge symmetry down to $SU(3)\times U(1)_{em}$. After that, the superpotential (\ref{Superpotential_For_MSSM}) produces the mass matrices for fermions, which are given by the expressions $M_u = v_u Y_U$, $M_d = -v_d Y_D$, and $M_e = -v_d Y_E$. The usual masses are equal to the absolute values of the eigenvalues of these matrices and have the hierarchical structure. They are the largest for the third generation and the smallest for the first one.

Instead of $v_u$ and $v_d$ it is more convenient to use the parameters

\begin{equation}\label{Tg_Beta_Definition}
v \equiv \sqrt{v_u^2+v_d^2} = \sqrt{\frac{2m_z^2}{e_1^2 + e_2^2}}\approx 174.\,\mbox{GeV};\qquad \mbox{tg}\,\beta\equiv \frac{v_u}{v_d}.
\end{equation}

\noindent
At present, the value of $\mbox{tg}\,\beta$ is unknown.

The renormalization group running of the gauge couplings and the renormalization of chiral matter superfields are encoded in the $\beta$-functions and anomalous dimensions

\begin{eqnarray}
&& \beta_i(\alpha,Y) \equiv \frac{d\alpha_i}{d\ln\mu}\bigg|_{\alpha_0,Y_0 =\mbox{\scriptsize const}};\qquad \gamma_{i}(\alpha,Y) \equiv 2\,Z_i^{-1/2}\, \frac{d\, Z_i^{1/2}}{d\ln\mu}\bigg|_{\alpha_0,Y_0 =\mbox{\scriptsize const}},
\end{eqnarray}

\noindent
where the derivatives should be calculated at fixed values of the bare couplings (which are marked by the subscript 0); $i=1,2,3$ for the $\beta$-functions and $i=Q,L,U,D,E, H_u,H_d$ for the anomalous dimensions. Note that, in our notation, the renormalization constants for the chiral matter superfields are defined by the equations $\phi_i \equiv \sqrt{Z_i} \phi_{i,R}$, where $\phi_{i,R}$ is the corresponding renormalized superfield.

As in any supersymmetric theory, due to the absence of divergent quantum corrections to the superpotential \cite{Grisaru:1979wc}, the renormalization of the Yukawa couplings and of the parameter $\bm{\mu}$ is determined by the renormalization of chiral matter superfields,

\begin{eqnarray}\label{Yukawa_Beta_Functions}
&& \frac{d Y_U}{d\ln\mu} = \frac{1}{2} \Big(\gamma_{H_u} Y_U + (\gamma_{Q})^T Y_U + Y_U \gamma_U\Big);\qquad \frac{d Y_D}{d\ln\mu} = \frac{1}{2} \Big(\gamma_{H_d} Y_D + (\gamma_{Q})^T Y_D + Y_D \gamma_D\Big);\nonumber\\
&& \frac{d Y_E}{d\ln\mu} = \frac{1}{2} \Big(\gamma_{H_d} Y_E + (\gamma_{L})^T Y_E + Y_E \gamma_E\Big);\qquad \frac{d\bm{\mu}}{d\ln\mu} = \frac{1}{2}\Big(\gamma_{H_u} + \gamma_{H_d}\Big)\bm{\mu}.
\end{eqnarray}

In the $\overline{\mbox{DR}}$ scheme (when a theory is regularized by dimensional reduction \cite{Siegel:1979wq} supplemented by modified minimal subtraction \cite{Bardeen:1978yd}) the MSSM renormalization group functions are known up to the three-loop approximation \cite{Jack:2004ch}. Note that starting from the three-loop approximation the $\beta$-functions become scheme-dependent. The anomalous dimensions depend on the renormalization prescription starting from the two-loop approximation. For an arbitrary renormalization prescription (supplementing the higher covariant derivative regularization \cite{Slavnov:1971aw,Slavnov:1972sq,Slavnov:1977zf} in the superfield formulation \cite{Krivoshchekov:1978xg,West:1985jx}) the three-loop $\beta$-function and the two-loop anomalous dimensions for the MSSM can be found in \cite{Haneychuk:2022qvu}. However, the higher-order quantum corrections are small, and in this paper we will need only the two-loop $\beta$-functions and the one-loop anomalous dimensions of the chiral matter superfields. For completeness, the expressions for them in the notation adopted in this paper are collected in Appendix \ref{Appendix_Lowest_RGFs}.

The initial conditions for the renormalization group equations in the MSSM are obtained from the values of the gauge and Yukawa couplings for the Standard Model. Namely, for simplicity, we assume that all superpartners have the same mass $M_{\mbox{\scriptsize SUSY}}$. If the threshold corrections \cite{Hall:1993gn,Blazek:2001sb,Tobe:2003bc,Auto:2003ys} (see also \cite{Antusch:2025fpm} and references therein) are ignored, the relations between couplings at the scale $M_{\mbox{\scriptsize SUSY}}$ take the form

\begin{equation}\label{Boundary}
\alpha_i = \alpha_i^{\mbox{\scriptsize (SM)}};\qquad
(Y_U)_{IJ} = \frac{(Y_U)_{IJ}^{\mbox{\scriptsize (SM)}}}{\sin\beta};\qquad
(Y_D)_{IJ} = \frac{(Y_D)_{IJ}^{\mbox{\scriptsize (SM)}}}{\cos\beta};\qquad
(Y_E)_{IJ} = \frac{(Y_E)_{IJ}^{\mbox{\scriptsize (SM)}}}{\cos\beta},
\end{equation}

\noindent
where the superscript $\mbox{(SM)}$ implies that the corresponding quantities pertain to the Standard Model.

\section{Approximate RGIs}
\hspace*{\parindent}\label{Section_RGI_Yukawa}

An interesting question is whether the MSSM is able to describe the physics beyond the Standard Model or it is necessary to modify it essentially. Of course, there are various aspects of this problem which go far beyond the scope of this paper. Here we will only focus on the analysis of the renormalization group behaviour of the Yukawa couplings for the third and second generations and possibility to achieve their unification. For this purpose, we will try to construct certain approximate RGIs and analyse their consequences. Certainly, it is desirable that these RGIs do not essentially depend on the presence of possible exotic superfields which can be included in addition to the MSSM field content. The renormalization group running of the gauge couplings is very sensitive to the presence of the exotic (super)fields in the nontrivial representations of the gauge group, see, e.g., Eqs. (\ref{Delta_Beta1}) --- (\ref{Delta_Beta3}) in Appendix \ref{Appendix_Lowest_RGFs}. That is why it is reasonable to construct possible RGIs from the Yukawa couplings only, under the assumption that the Yukawa interaction of the exotic superfields is much smaller than that for the third generation of the usual MSSM superfields.

In what follows we will consider a modification of the MSSM containing some exotic superfields which can be combined into the complete $SU(5)$ multiplets. (In this case the unification of the gauge couplings remains unbroken, at least, in the one-loop approximation \cite{Ghilencea:1997mu,Amelino-Camelia:1998xgn}.) For simplicity, here we assume that the masses of all exotic superfields are the same and their Yukawa interactions are negligibly small.

An interesting example of an approximate RGI constructed in Appendix \ref{Appendix_RGIs_Derivation}, which is almost scale independent for any such theory, can be written in the form

\begin{equation}\label{Yukawa_RGI1}
I_1 = \bigg|\Big(\frac{(Y_U)_{33}}{(Y_D)_{33}}\Big)^3 \Big(\frac{(Y_D)_{22}}{(Y_U)_{22}}\Big)^5\bigg|.
\end{equation}

\noindent
The derivative of the logarithm of this expression with respect to $\ln\mu$ calculated with the help of Eqs. (\ref{Yukawa_Beta_Functions}) and (\ref{Gamma_Q_1Loop}) --- (\ref{Gamma_Hd_1Loop}) does not contain the (large) couplings $\alpha_3$, $\left[(Y_U)_{33}\right]^2$, and $\left[(Y_D)_{33}\right]^2$, and is given by

\begin{equation}\label{I1_Derivative}
\frac{d\ln I_1}{d\ln\mu} = \frac{\alpha_1}{5\pi} + \frac{1}{8\pi^2} \left[(Y_E)_{33}\right]^2 + \ldots
\end{equation}

\noindent
Here dots denote terms proportional to the remaining (relatively small) Yukawa couplings for the second and first generations, the higher order terms, and the contributions coming from possible exotic superfields. (In the lowest order the contributions of the exotics are proportional to the corresponding Yukawa couplings, which we assume to be small.)

The renormalization group running of the expression (\ref{Yukawa_RGI1}) divided by $\mbox{tg}^2\beta$ is presented in what follows at the top left in Figs. \ref{Figure_Yukawa_Running_MSSM_Option1} and \ref{Figure_Yukawa_Running_MSSM_Option2} for the MSSM with two different values of $\mbox{tg}\,\beta$ and at the top left in Figs. \ref{Figure_Yukawa_Running_Exotics_Option1} and \ref{Figure_Yukawa_Running_Exotics_Option2} for certain theories which differs from the MSSM by the presence of some new chiral matter superfields, again, for two different values of $\mbox{tg}\,\beta$. We see that in all cases the expression $I_1$ remains almost constant throughout the whole evolution from the supersymmetric threshold up to the unification scale. Note that in Fig. \ref{Figure_Yukawa_Running_MSSM_Option2}, for comparison, by dots we also present a plot of the function $|((Y_U)_{33}/(Y_D)_{33})^5 ((Y_D)_{22}/(Y_U)_{22})^3|/\mbox{tg}^2\beta$ which differs from $I_1/\mbox{tg}^2\beta$ by the formal replacement of powers $3\leftrightarrow 5$. We see that this expression essentially larger depends on scale than $I_1$, so that $I_1$ is indeed an approximate RGI.

The value of $I_1$ at low energies is given by the expression

\begin{equation}\label{I1_Estimate}
I_1(M_{\mbox{\scriptsize SUSY}}) = \mbox{tg}^2\beta\, \Bigg|\bigg(\frac{(Y_U^{\mbox{\scriptsize (SM)}})_{33}}{(Y_D^{\mbox{\scriptsize (SM)}})_{33}}\bigg)^3 \bigg(\frac{(Y_D^{\mbox{\scriptsize (SM)}})_{22}}{(Y_U^{\mbox{\scriptsize (SM)}})_{22}}\bigg)^5\Bigg|_{M_{\mbox{\scriptsize SUSY}}}.
\end{equation}

\noindent
However, at present the value of $\mbox{tg}\,\beta$ is not known. Therefore, it is also desirable to construct such an RGI that can be estimated at low energies without involving this parameter. This can be achieved under a certain additional assumption. Namely, let us suppose that at the unification scale $M_X$ the Yukawa couplings of the third generation satisfy the condition

\begin{equation}\label{X_Definition}
\left|(Y_U)_{33}\right| = x \left|(Y_D)_{33}\right|,
\end{equation}

\noindent
where $x$ is a certain real number, and the absolute value of $\left|(Y_U)_{33}\right| - x \left|(Y_D)_{33}\right|$ remains small throughout the whole renormalization group evolution. (Certainly, the fulfilment of this condition should be verified in each particular case.) Under this assumption it is possible to obtain (see Appendix \ref{Appendix_RGIs_Derivation} for details) that the expression

\begin{equation}\label{Yukawa_RGI2}
I_2 = \bigg|\frac{(Y_U)_{33}}{(Y_D)_{33}}\cdot \frac{(Y_D)_{22}}{(Y_U)_{22}}\cdot \bigg(\frac{(Y_E)_{33}}{(Y_D)_{33}}\cdot \frac{(Y_D)_{22}}{(Y_E)_{22}}\bigg)^{\frac{2(x^2-1)}{x^2+3}}\bigg|
\end{equation}

\noindent
slightly depends on scale and is therefore an approximate RGI.

The renormalization group evolution of the expression $I_2$ for the MSSM with two different values of $\mbox{tg}\,\beta$ is plotted at the top right in Figs. \ref{Figure_Yukawa_Running_MSSM_Option1} and \ref{Figure_Yukawa_Running_MSSM_Option2}. The analogous plots for two supersymmetric theories obtained from the MSSM by adding some new superfields are presented in Figs. \ref{Figure_Yukawa_Running_Exotics_Option1} and \ref{Figure_Yukawa_Running_Exotics_Option2}. These figures demonstrate that the expression $I_2$ only slightly depends on scale above the supersymmetric threshold. Let us also pay attention that the renormalization group running of the Yukawa couplings for the third generation presented at the bottom left in the above mentioned figures really occurs in such a way that the deviations from the equality (\ref{X_Definition}) are relatively small. Therefore, the expression $I_2$ is really an approximate RGI.

If the threshold corrections are neglected, then at low energies this expression can be approximately estimated as

\begin{equation}\label{I2_Estimate}
I_2(M_{\mbox{\scriptsize SUSY}}) = \Bigg|\frac{(Y_U^{\mbox{\scriptsize (SM)}})_{33}}{(Y_D^{\mbox{\scriptsize (SM)}})_{33}}\cdot \frac{(Y_D^{\mbox{\scriptsize (SM)}})_{22}}{(Y_U^{\mbox{\scriptsize (SM)}})_{22}}\cdot \bigg(\frac{(Y_E^{\mbox{\scriptsize (SM)}})_{33}}{(Y_D^{\mbox{\scriptsize (SM)}})_{33}}\cdot \frac{(Y_D^{\mbox{\scriptsize (SM)}})_{22}}{(Y_E^{\mbox{\scriptsize (SM)}})_{22}}\bigg)^{\frac{2(x^2-1)}{x^2+3}}\Bigg|_{M_{\mbox{\scriptsize SUSY}}}.
\end{equation}

\noindent
For analyzing phenomenological predictions following from the approximate scale independence of $I_1$ and $I_2$, it is convenient to introduce the notations

\begin{eqnarray}\label{Y_ED_Relations}
&& \Delta_1\equiv \frac{I_1(M_X)}{I_1(M_{\mbox{\scriptsize SUSY}})}\approx 1;\qquad \Delta_2\equiv \frac{I_2(M_X)}{I_2(M_{\mbox{\scriptsize SUSY}})}\approx 1;\qquad y \equiv \Bigg|\frac{(Y_E)_{33}}{(Y_D)_{33}}\cdot \frac{(Y_D)_{22}}{(Y_E)_{22}} \Bigg|_{M_X};\qquad \nonumber\\
&& y_0 \equiv \Bigg|\frac{(Y_E)_{33}}{(Y_D)_{33}}\cdot \frac{(Y_D)_{22}}{(Y_E)_{22}} \Bigg|_{M_{\mbox{\scriptsize SUSY}}} = \Bigg|\frac{(Y_E^{\mbox{\scriptsize (SM)}})_{33}}{(Y_D^{\mbox{\scriptsize (SM)}})_{33}}\cdot \frac{(Y_D^{\mbox{\scriptsize (SM)}})_{22}}{(Y_E^{\mbox{\scriptsize (SM)}})_{22}} \Bigg|_{M_{\mbox{\scriptsize SUSY}}}.
\end{eqnarray}

\noindent
(Note that the Georgi--Jarlskog textures (\ref{Georgi_Jarlskog}) correspond to $y=1/3$.) The equalities $\Delta_{1,2}\approx 1$ follow from the approximate renormalization group invariance of the expressions $I_1$ and $I_2$. Therefore, to make approximate estimates of some values, the deviations of $\Delta_{1,2}$ from 1 can be neglected.

Then, comparing the values of the expression $I_2$ at the scales $M_X$ and $M_{\mbox{\scriptsize SUSY}}$, under the assumption (\ref{X_Definition}), we obtain

\begin{equation}\label{Y2_Estimate_From_I2}
\bigg|\frac{(Y_D)_{22}}{(Y_U)_{22}}\bigg|_{M_X} = \frac{1}{x}\cdot\Bigg|\frac{(Y_U^{\mbox{\scriptsize (SM)}})_{33}}{(Y_D^{\mbox{\scriptsize (SM)}})_{33}}\cdot \frac{(Y_D^{\mbox{\scriptsize (SM)}})_{22}}{(Y_U^{\mbox{\scriptsize (SM)}})_{22}}\Bigg|_{M_{\mbox{\scriptsize SUSY}}} \cdot \bigg(\frac{y_0}{y}\bigg)^{\frac{2(x^2-1)}{x^2+3}} \cdot \Delta_2.
\end{equation}

\noindent
An analogous estimate can be derived from the approximate RGI $I_1$,

\begin{equation}\label{Y2_Estimate_From_I1}
\bigg|\frac{(Y_D)_{22}}{(Y_U)_{22}}\bigg|_{M_X} =
x^{-3/5}\, \mbox{tg}^{2/5}\beta\cdot \Bigg| \bigg(\frac{(Y_U^{\mbox{\scriptsize (SM)}})_{33}}{(Y_D^{\mbox{\scriptsize (SM)}})_{33}}\bigg)^{3/5}\cdot
\frac{(Y_D^{\mbox{\scriptsize (SM)}})_{22}}{(Y_U^{\mbox{\scriptsize (SM)}})_{22}}\Bigg|_{M_{\mbox{\scriptsize SUSY}}} \cdot \Delta_1^{1/5}.
\end{equation}

\noindent
Comparing Eqs. (\ref{Y2_Estimate_From_I2}) and (\ref{Y2_Estimate_From_I1}), it is possible to relate the value of $\mbox{tg}\,\beta$ to the value of $x$ by the equation

\begin{equation}\label{Tg_Beta}
\mbox{tg}\,\beta = \frac{1}{x}\cdot \Bigg|\frac{(Y_U^{\mbox{\scriptsize (SM)}})_{33}}{(Y_D^{\mbox{\scriptsize (SM)}})_{33}}\Bigg|_{M_{\mbox{\scriptsize SUSY}}} \cdot \bigg(\frac{y_0}{y}\bigg)^{\frac{5(x^2-1)}{x^2+3}}
\cdot \Delta_1^{-1/2} \Delta_2^{5/2}.
\end{equation}

Using the method based on RGIs, it is also possible to estimate the ratio $|(Y_E)_{22}/(Y_D)_{22}|$ at the unification scale. For this purpose, let us consider the expression

\begin{equation}\label{I3_RGI}
I_3 \equiv \bigg|\frac{(Y_E)_{22}}{(Y_D)_{22}}\bigg|\cdot \Big(\frac{\mu}{M_{\mbox{\scriptsize SUSY}}}\Big)^{-4\alpha_3(\mu)/3\pi}\cdot \Big(\frac{M_X}{M_{\mbox{\scriptsize SUSY}}}\Big)^{4\alpha(M_X)/3\pi},
\end{equation}

\noindent
where $\mu$ is a scale. Differentiating this expression with respect to $\ln\mu$, we obtain

\begin{equation}\label{I3_Derivative}
\frac{d\ln I_3}{d\ln\mu} = -\frac{\alpha_1}{3\pi} + \frac{2(3-n)}{3\pi^2} (\alpha_3)^2 \ln\frac{\mu}{M_{\mbox{\scriptsize SUSY}}} + \ldots,
\end{equation}

\noindent
where dots denote one-loop terms proportional to the Yukawa couplings of the second and first generations and all higher-order contributions except for one two-loop term written explicitly. (This term is proportional to $(\alpha_3)^2$ and contains $\ln(\mu/M_{\mbox{\scriptsize SUSY}})$, which can be sufficiently large.)  We also use the notation

\begin{equation}
n\equiv \frac{1}{2} n_5 + \frac{3}{2}n_{10} + \frac{7}{2}n_{15} + 5n_{24} +\ldots,
\end{equation}

\noindent
where $n_5$ is the total number of the representations $5$ and $\,\xbar{5}\,$ formed by chiral matter superfields with masses lower than $\mu$, etc. Nevertheless, this term is still suppressed by the second power of the coupling $\alpha_3$. Therefore, it is possible to assume that the second term in the right hand side of Eq. (\ref{I3_Derivative}) is not too large for all reasonable values of $n$. Then the expression (\ref{I3_RGI}) is an approximate RGI for all appropriate theories even in the case when the action contains a certain number of exotic superfields.

Comparing the values of $I_3$ at the unification scale $M_X$ and at the scale of supersymmetry breaking $M_{\mbox{\scriptsize SUSY}}$, we see that

\begin{equation}\label{Y22_Ration_From_I3}
\bigg|\frac{(Y_D)_{22}}{(Y_E)_{22}}\bigg|_{M_X} = \bigg|\frac{(Y_D)_{22}^{\mbox{\scriptsize (SM)}}}{(Y_E)_{22}^{\mbox{\scriptsize (SM)}}}\bigg|_{M_{\mbox{\scriptsize SUSY}}}\cdot \Big(\frac{M_X}{M_{\mbox{\scriptsize SUSY}}}\Big)^{4\alpha(M_X)/3\pi} \cdot \Delta_3,
\end{equation}

\noindent
where the change of the approximate RGI $I_3$ is encoded in the expression

\begin{equation}
\Delta_3 \equiv \frac{I_3(M_X)}{I_3(M_{\mbox{\scriptsize SUSY}})} \approx 1.
\end{equation}

The renormalization group running of the expression $I_3$ for the MSSM with $\mbox{tg}\,\beta=28.$ is presented in Fig. \ref{Figure_I3_Scale_Dependence}. From this plot we see that $I_3$ does not depend strongly on the scale and can be considered as an approximate RGI. However, its scale dependence is nevertheless essential. It can be reduced by a certain modification in the case when the field content of the exotic superfields is known. For instance, for the MSSM we can consider the expression

\begin{equation}\label{I3_Tilde_RGI}
\widetilde I_3 \equiv \bigg|\frac{(Y_E)_{22}}{(Y_D)_{22}}\bigg|\cdot \Big(\frac{\alpha_3}{\alpha(M_X)}\Big)^{8/9} \Big(\frac{\alpha_1}{\alpha(M_X)}\Big)^{10/99},
\end{equation}

\noindent
which in this case depends significantly less on the scale. (Its derivative contains only one-loop contributions proportional to the Yukawa couplings for the first two generations and the higher order terms.) The plot of this function for the MSSM with $\mbox{tg}\,\beta=28$ presented in Fig. \ref{Figure_I3_Scale_Dependence} demonstrates that this expression is really almost scheme independent in this case. Then, for the MSSM a more precise estimate of the ratio under consideration reads as

\begin{equation}\label{I3_Tilde_Estimate}
\bigg|\frac{(Y_E)_{22}}{(Y_D)_{22}}\bigg|_{M_X} = \bigg|\frac{(Y_E)_{22}^{\mbox{\scriptsize (SM)}}}{(Y_D)_{22}^{\mbox{\scriptsize (SM)}}}\bigg|_{M_{\mbox{\scriptsize SUSY}}}\cdot \Big(\frac{\alpha_3(M_{\mbox{\scriptsize SUSY}})}{\alpha(M_X)}\Big)^{8/9} \Big(\frac{\alpha_1(M_{\mbox{\scriptsize SUSY}})}{\alpha(M_X)}\Big)^{10/99} \cdot \widetilde\Delta_3,
\end{equation}

\noindent
where

\begin{equation}
\widetilde\Delta_3 \equiv \frac{\widetilde I_3(M_X)}{\widetilde I_3(M_{\mbox{\scriptsize SUSY}})} \approx 1.
\end{equation}

\section{Renormalization group running of the Yukawa couplings in the MSSM}
\hspace*{\parindent}\label{Section_Yukawa_MSSM}

After constructing the analytical expressions for approximate RGIs, it is necessary to conduct a numerical analysis of their phenomenological applications. To perform this analysis, for simplicity, we take the unique supersymmetric threshold $M_{\mbox{\scriptsize SUSY}} = 4\, \mbox{TeV}$.\footnote{In general, the threshold effects caused by different masses of various superpartners are very important \cite{Hall:1993gn,Blazek:2001sb,Tobe:2003bc,Auto:2003ys}, but here we consider only the simplest situation.} Below this threshold we use the two-loop renormalization group equations for the gauge couplings and the one-loop renormalization group equations for the Yukawa couplings taken from Ref. \cite{Arason:1991ic}.\footnote{More precise results can be found in \cite{Mihaila:2013wma}.} (For completeness, the corresponding renormalization group functions are also presented in Appendix \ref{Appendix_Lowest_RGFs}.) Above the threshold we use the one-loop expressions for the anomalous dimensions obtained with the help of Eqs. (\ref{Yukawa_Beta_Functions}) and (\ref{Gamma_Q_1Loop}) --- (\ref{Gamma_Hd_1Loop}). For the gauge $\beta$-functions we use the NSVZ equations (in the form in which they are presented in Ref. \cite{Korneev:2021zdz}). In these NSVZ equations the anomalous dimensions of the chiral matter superfields were taken in the one-loop approximation. (This allows not only obtaining all two-loop contributions, but also taking into account certain higher order contributions, which are, however, very small.)

As initial conditions, we take the values of gauge and Yukawa couplings at the mass of the $Z$ boson presented in \cite{Antusch:2025fpm} for the $\overline{\mbox{MS}}$ scheme. (They were calculated starting from the experimental data collected in \cite{ParticleDataGroup:2024cfk}.) In fact, we use only the values of $(Y_i)_{33}$ and $(Y_i)_{22}$ neglecting the mixing and (relatively small) Yukawa couplings for the first generation. Namely, we use the following values of couplings at $m_Z$ scale:

\begin{eqnarray}\label{Gauge_Couplings_Values}
&& \sqrt{\frac{5}{3}} e_1 = 0.461228 \pm 0.000026;\qquad e_2 = 0.65096\pm 0.00004;\qquad e_3 = 1.2123\pm 0.0046;\qquad\nonumber\\
&& \vphantom{1}\\
\label{Yukawa_Couplings_Values}
&& (Y_U^{\mbox{\scriptsize (SM)}})_{33} = 0.967\pm 0.004;\qquad\qquad\qquad\quad\ \, (Y_U^{\mbox{\scriptsize (SM)}})_{22} = (3.56 \pm 0.06)\cdot 10^{-3};\qquad\nonumber\\
&& (Y_D^{\mbox{\scriptsize (SM)}})_{33} = (1.630\pm 0.009)\cdot 10^{-2};\qquad\qquad (Y_D^{\mbox{\scriptsize (SM)}})_{22} = (3.06 \pm 0.04)\cdot 10^{-4};\qquad\nonumber\\
&& (Y_E^{\mbox{\scriptsize (SM)}})_{33} = (0.99378\pm 0.00014)\cdot 10^{-2};\qquad (Y_E^{\mbox{\scriptsize (SM)}})_{22} = (5.85042 \pm 0.00075)\cdot 10^{-4}.\qquad\nonumber\\
\end{eqnarray}

\noindent
Solving the renormalization group equations for the Standard Model (corresponding to the renormalization group functions (\ref{SM_Beta1_2Loops}) --- (\ref{SM_Gamma_Ye})) with these initial conditions, for reasonable $M_{\mbox{\scriptsize SUSY}}$ values (within the range $1$ --- $10$ TeV) we approximately obtain

\begin{eqnarray}
&& y_0 \approx 0.328 + 0.0021\cdot\ln\Big(\frac{M_{\mbox{\scriptsize SUSY}}}{4\,\mbox{TeV}}\Big);\nonumber\\
&& \bigg|\frac{(Y_U^{\mbox{\scriptsize (SM)}})_{33}}{(Y_D^{\mbox{\scriptsize (SM)}})_{33}}\cdot  \frac{(Y_D^{\mbox{\scriptsize (SM)}})_{22}}{(Y_U^{\mbox{\scriptsize (SM)}})_{22}}\bigg|_{M_{\mbox{\scriptsize SUSY}}} \approx 5.394+0.069\cdot\ln\Big(\frac{M_{\mbox{\scriptsize SUSY}}}{4\,\mbox{TeV}}\Big);\nonumber\\
&& \bigg|\frac{(Y_U^{\mbox{\scriptsize (SM)}})_{33}}{(Y_D^{\mbox{\scriptsize (SM)}})_{33}} \bigg|_{M_{\mbox{\scriptsize SUSY}}} \approx 62.57+0.74\cdot\ln\Big(\frac{M_{\mbox{\scriptsize SUSY}}}{4\,\mbox{TeV}}\Big).
\end{eqnarray}

\begin{figure}[h]
\begin{picture}(0,7)
\put(1.8,0.4){\includegraphics[scale=0.31]{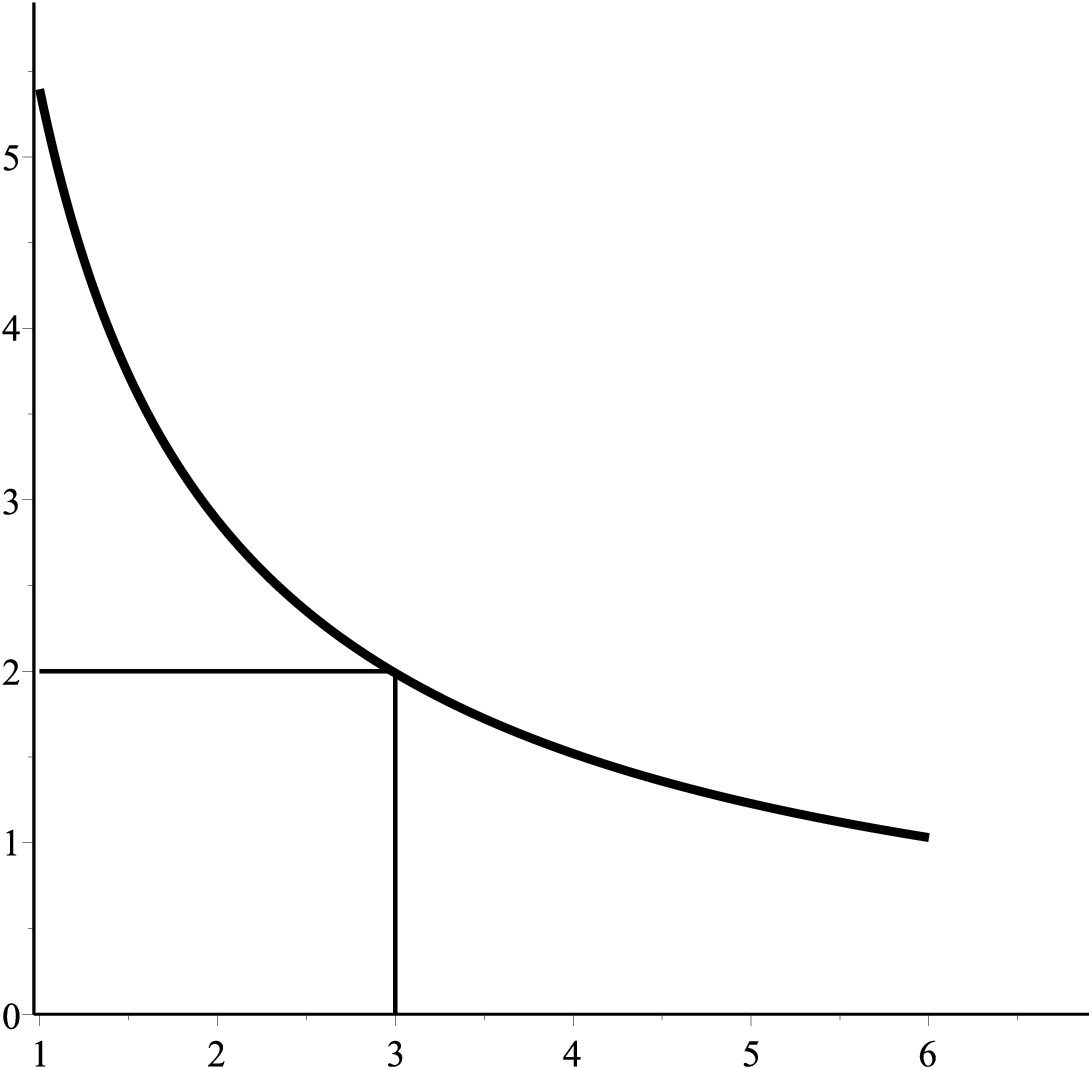}}
\put(0,5.9){$\Big|\frac{(Y_D)_{22}}{(Y_U)_{22}}\Big|_{M_X}$}
\put(7.4,0.2){$x$}

\put(9.1,0.4){\includegraphics[scale=0.31]{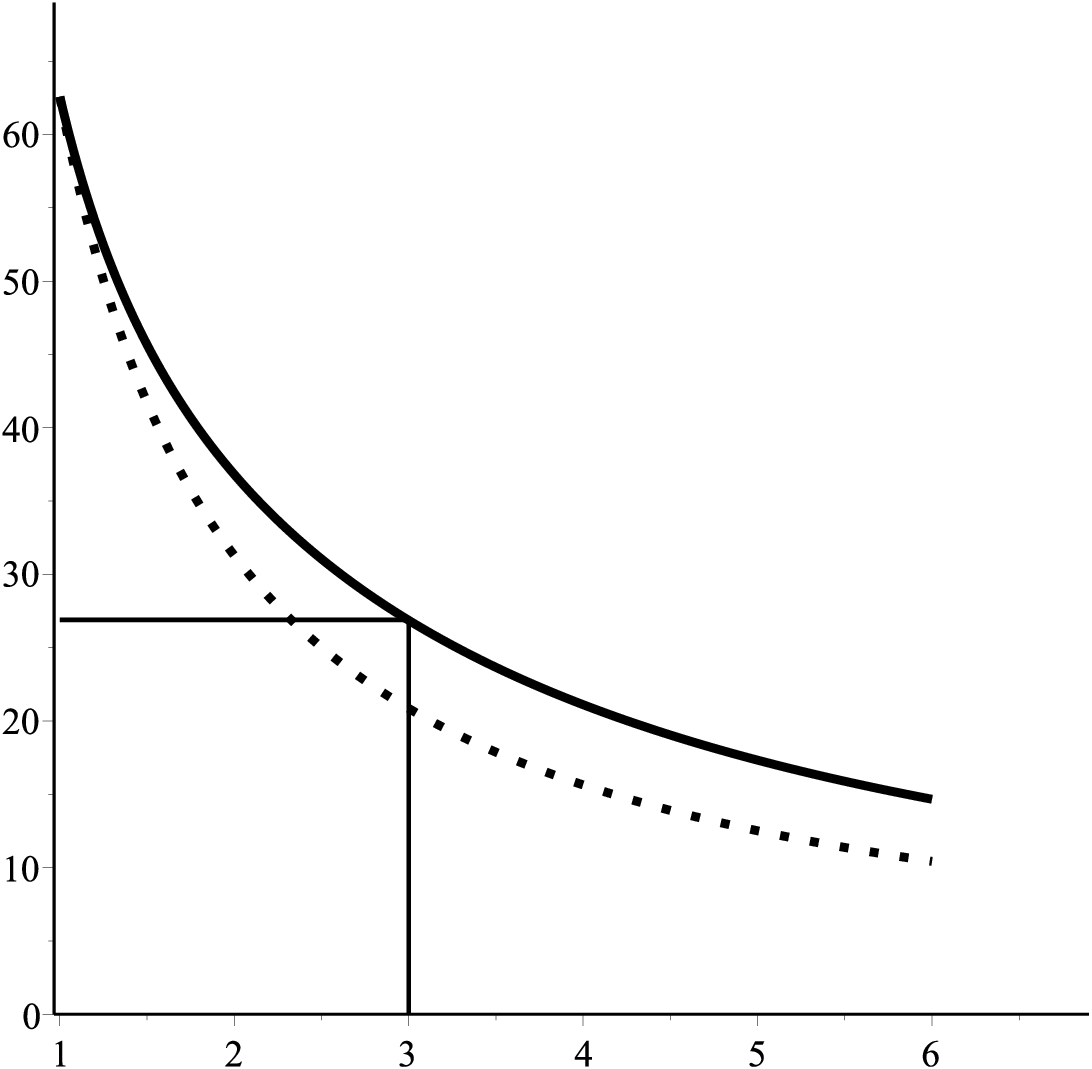}}
\put(8.5,6.0){$\mbox{tg}\,\beta$}
\put(8.5,2.7){\scriptsize $26.9$}
\put(14.7,0.2){$x$}
\end{picture}
\caption{Approximate dependences of the ratio $|(Y_D)_{22}/(Y_U)_{22}|_{M_X}$ (given by Eq. (\ref{Y2_Estimate_From_I2})) and of $\mbox{tg}\,\beta$ (given by Eq. (\ref{Tg_Beta})) as functions of the parameter $x$ defined by Eq. (\ref{X_Definition}) for the MSSM with $M_{\mbox{\scriptsize SUSY}}=4\,\mbox{TeV}$ (solid line). For comparison, in the right plot we also provide the tree curve $1/x\cdot |(Y_U^{\mbox{\scriptsize (SM)}})_{33}/(Y_D^{\mbox{\scriptsize (SM)}})_{33}|_{M_{\mbox{\scriptsize SUSY}}}$ (dotted line). Drawing these plots we took $y_0=0.328$, $y=0.305$ and set $\Delta_{1,2}\to 1$.}\label{Figure_X_Dependence}
\end{figure}

As we have already mentioned, the Georgi--Jarlskog textures correspond to $y=1/3$. However, for the MSSM its value is slightly different. This can be seen, e.g., from the tables presented in \cite{Antusch:2025fpm}. Namely, for a rather large range of $\mbox{tg}\,\beta$ values $y\approx 0.305\approx (3.28)^{-1}$. Substituting this value and $y_0=0.328$ into Eqs. (\ref{Y2_Estimate_From_I2}) and (\ref{Tg_Beta}) we can find the dependence of these expressions on the parameter $x$. For $M_{\mbox{\scriptsize SUSY}}=4\,\mbox{TeV}$ the corresponding plots are presented in Fig. \ref{Figure_X_Dependence} on the left and on the right, respectively. Together with the plot of $\mbox{tg}\,\beta$, on the right we also present the naive tree curve $1/x\cdot |(Y_U^{\mbox{\scriptsize (SM)}})_{33}/(Y_D^{\mbox{\scriptsize (SM)}})_{33}|_{M_{\mbox{\scriptsize SUSY}}}$ plotted by dots.
In particular, we see that the results obtained with the help of Eq. (\ref{Tg_Beta}) are in general essentially different for $x\ne 1$.

However, the most difficult task is to determine the value of $x$. The relations like (\ref{X_Definition}) are usually obtained on the basis of certain group theory arguments. For instance, the $SO(10)$ invariant $\,\xbar{16}\,\times\,\xbar{16}\,\times 10$ leads to the equality $\left|Y_U\right| = \left|Y_D\right|$ (see, e.g., Appendix \ref{Appendix_Invariants} for details), while the relations $\left|Y_E\right| = \left|Y_D\right|$ and
$\left|Y_E\right| = 3 \left|Y_D\right|$ follow from the invariants $5\times\,\xbar{10}\,\times 5$ and $5\times\,\xbar{10}\,\times 45$ of the group $SU(5)$, respectively. However, from the left plot in Fig. \ref{Figure_X_Dependence} we see that the relations $\left|(Y_U)_{33}\right| = \left|(Y_D)_{33}\right|$ (corresponding to $x=1$) and $\left|(Y_U)_{22}\right|= \left|(Y_D)_{22}\right|$ cannot be satisfied simultaneously. In fact, this plot essentially restricts possible MSSM relations between the Yukawa couplings of the third and second generations at the unification scale. Certainly, the group theory arguments are unable to produce arbitrary values of $x$ and $|(Y_D)_{22}/(Y_U)_{22}|$. The simplest group invariants lead to the ratios of the Yukawa couplings equal to some numbers (often, but not always, integers) which are not so far from 1. That is why, looking at Fig. \ref{Figure_X_Dependence}, one can try to suggest certain values of parameters for which the ratios of the Yukawa couplings are, e.g., integers. In what follows we will emphasize two such possibilities.

However, before doing this, let us first estimate the ratio $|(Y_E)_{22}/(Y_D)_{22}|_{M_x}$ with the help of the RGI $\widetilde I_3$. Namely, using Eq. (\ref{I3_Tilde_Estimate}) and setting $\widetilde \Delta_3\to 1$, for $M_{\mbox{\scriptsize SUSY}} = 4\,\mbox{TeV}$ we obtain

\begin{equation}
\bigg|\frac{(Y_E)_{22}}{(Y_D)_{22}}\bigg|_{M_X} \approx 4.16
\end{equation}

\noindent
This value is in rather good agreement with the result of the numerical analysis for a wide range of $\mbox{tg}\,\beta$ (solving the renormalization group equations we obtained the value $4.26$, while a more precise calculation made in \cite{Antusch:2025fpm} gives approximately $4.33$). Given that $y\approx 0.305$, it is possible to estimate the analogous ratio for the third generation,

\begin{equation}
\bigg|\frac{(Y_E)_{33}}{(Y_D)_{33}}\bigg|_{M_X} = y\cdot \bigg|\frac{(Y_E)_{22}}{(Y_D)_{22}}\bigg|_{M_X} \approx 1.3.
\end{equation}

\noindent
Let us again note that here we use the unique supersymmetric threshold. Choosing the parameters which determine masses of superpartners one can change this ratio in a certain range. By a proper choice of these parameters, one can, in particular, achieve the top-bottom-tau unification for the third generation following from the invariant $\,\xbar{16}\,\times\,\xbar{16}\,\times 10$ of the group $SO(10)$, see, e.g., \cite{Hall:1993gn,Blazek:2001sb,Tobe:2003bc,Auto:2003ys}.

\begin{figure}[h]
\begin{picture}(0,7.2)
\put(4.7,0.6){\includegraphics[scale=0.34]{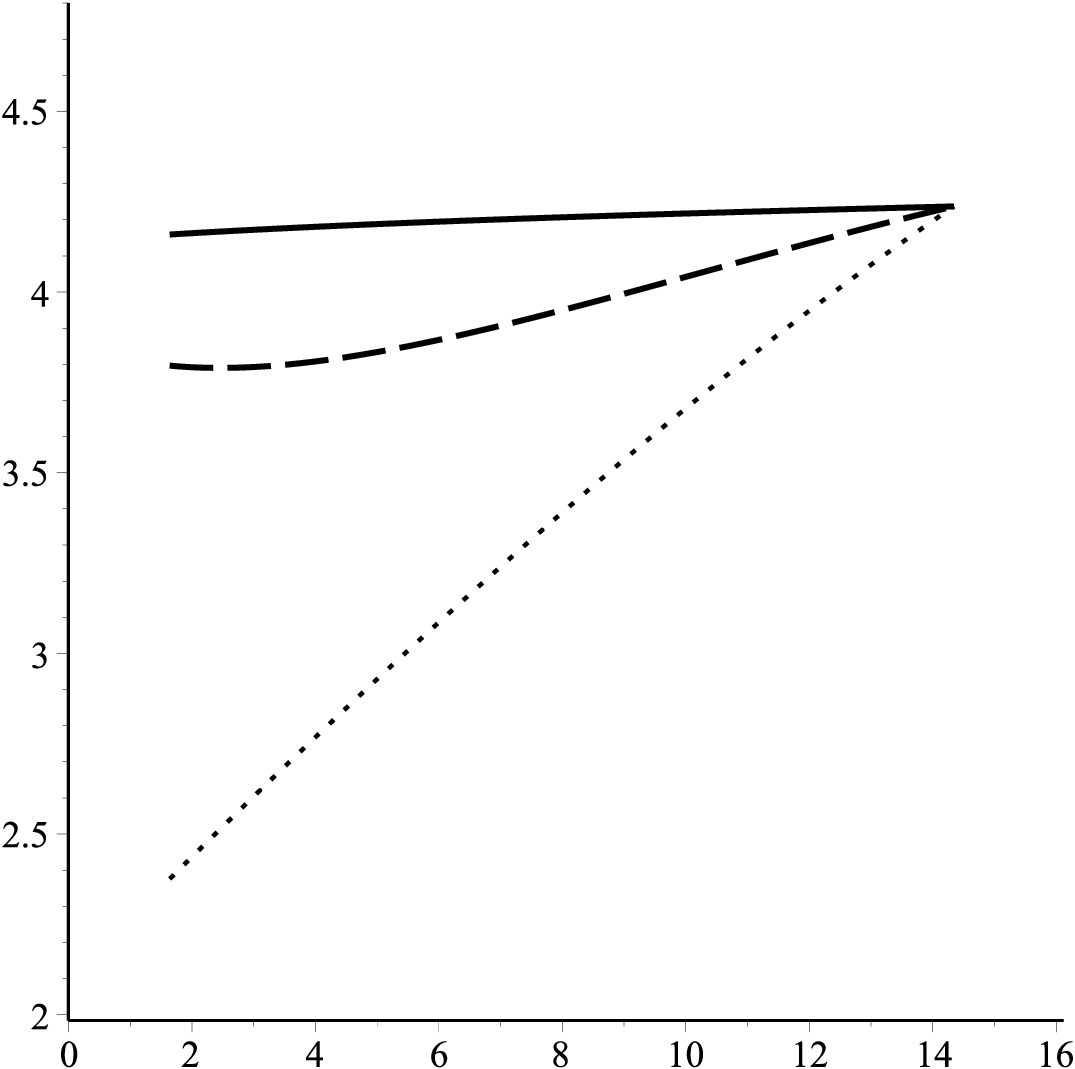}}
\put(10.2,0){$\log_{10}\frac{\mu}{m_Z}$}
\put(6.2,5.7){$\widetilde I_3$}
\put(6.2,4.1){$I_3$}
\put(6.7,2.2){${\displaystyle \Big|\frac{(Y_E)_{22}}{(Y_D)_{22}}\Big|}$}
\end{picture}
\caption{Scale dependence of the ratio $|(Y_E)_{22}/(Y_D)_{22}|$ (dotted line) and of the RGIs $I_3$ (dashed line) and $\widetilde I_3$ (solid line) for the MSSM with $\mbox{tg}\,\beta=28$ and $M_{\mbox{\scriptsize SUSY}}=4\,\mbox{TeV}$}\label{Figure_I3_Scale_Dependence}
\end{figure}

The plots of the RGIs $\widetilde I_3$, $I_3$, and of the ratio $|(Y_E)_{22}/(Y_D)_{22}|$ as functions of scale $\log_{10}\mu/m_Z$ are presented in Fig. \ref{Figure_I3_Scale_Dependence} for $\mbox{tg}\,\beta=28$ and $M_{\mbox{\scriptsize SUSY}}=4\,\mbox{TeV}$. We see that $I_3$ slightly but essentially depends on scale ($\Delta_3\approx 1.06$), while the expression $\widetilde I_3$ remains almost constant throughout the whole renormalization group evolution ($\widetilde \Delta_3\approx 1.02$).

Next, let us consider two interesting special cases corresponding to certain particular choices of the parameter $x= |(Y_U)_{33}/(Y_D)_{33}|_{M_X}$.

1. The first possibility is $x=1$, corresponding to the top-bottom Yukawa unification. This value of $x$ follows, e.g., from the simplest $SO(10)$ invariant $\,\xbar{16}\,\times\,\xbar{16}\,\times 10$. However, according to Fig. \ref{Figure_X_Dependence}, in this case the ratio $|(Y_D)_{22}/(Y_U)_{22}|_{M_X}$ appears to be rather large, about $5.4$. Therefore, (for $x=1$) at the unification scale the Yukawa couplings should satisfy the approximate relations

\begin{equation}\label{Yukawa_Unification_1}
\left|(Y_U)_{33}\right| = \left|(Y_D)_{33}\right| \approx (1.3)^{-1}\left|(Y_E)_{33}\right|;\qquad 5.4\cdot\left|(Y_U)_{22}\right| = \left|(Y_D)_{22}\right| \approx (4.3)^{-1} \left|(Y_E)_{22}\right|.
\end{equation}

\begin{figure}[!h]
\hspace*{-4mm}
\begin{picture}(0,15)
\put(1.1,8.0){\includegraphics[scale=0.34]{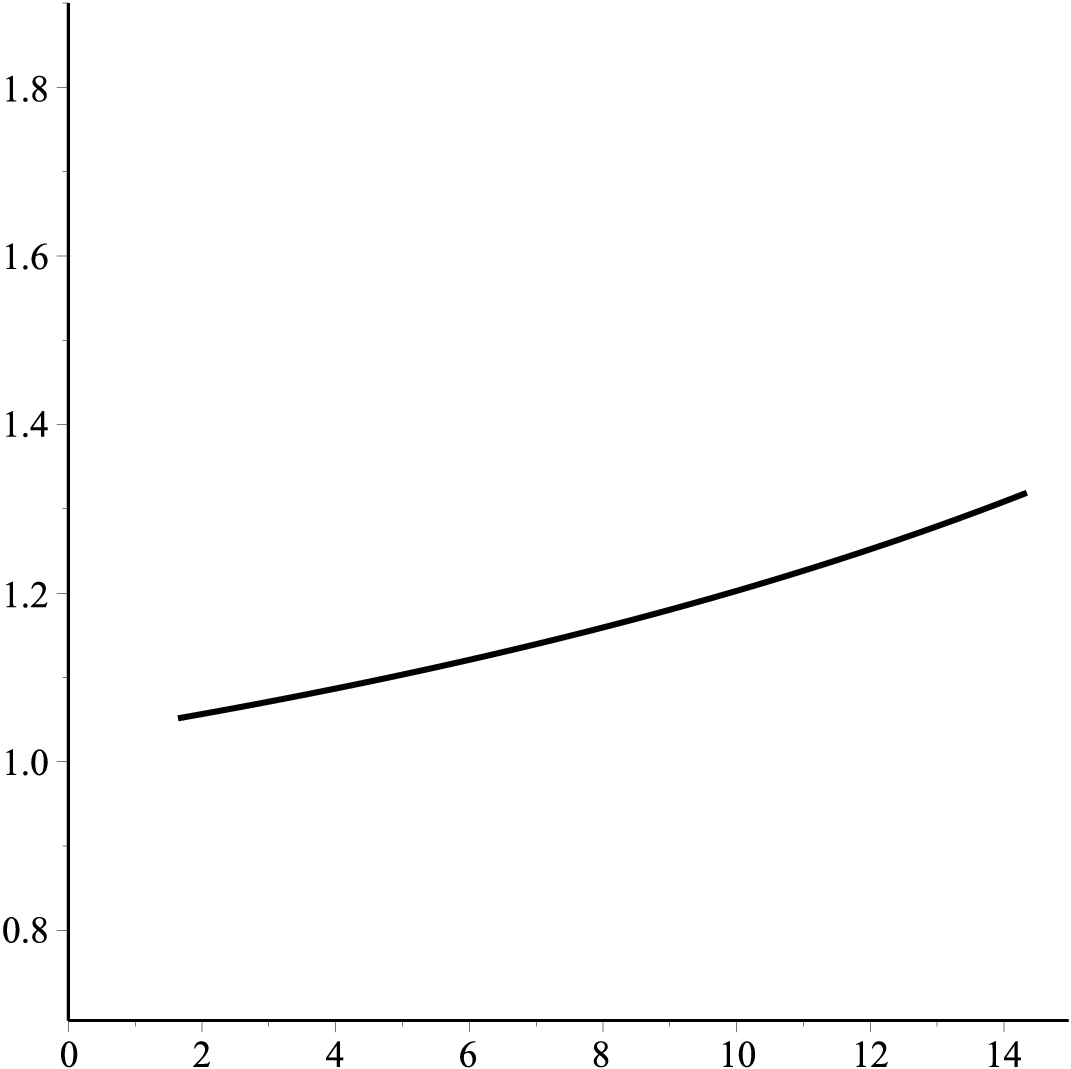}}
\put(8.75,8.0){\includegraphics[scale=0.34]{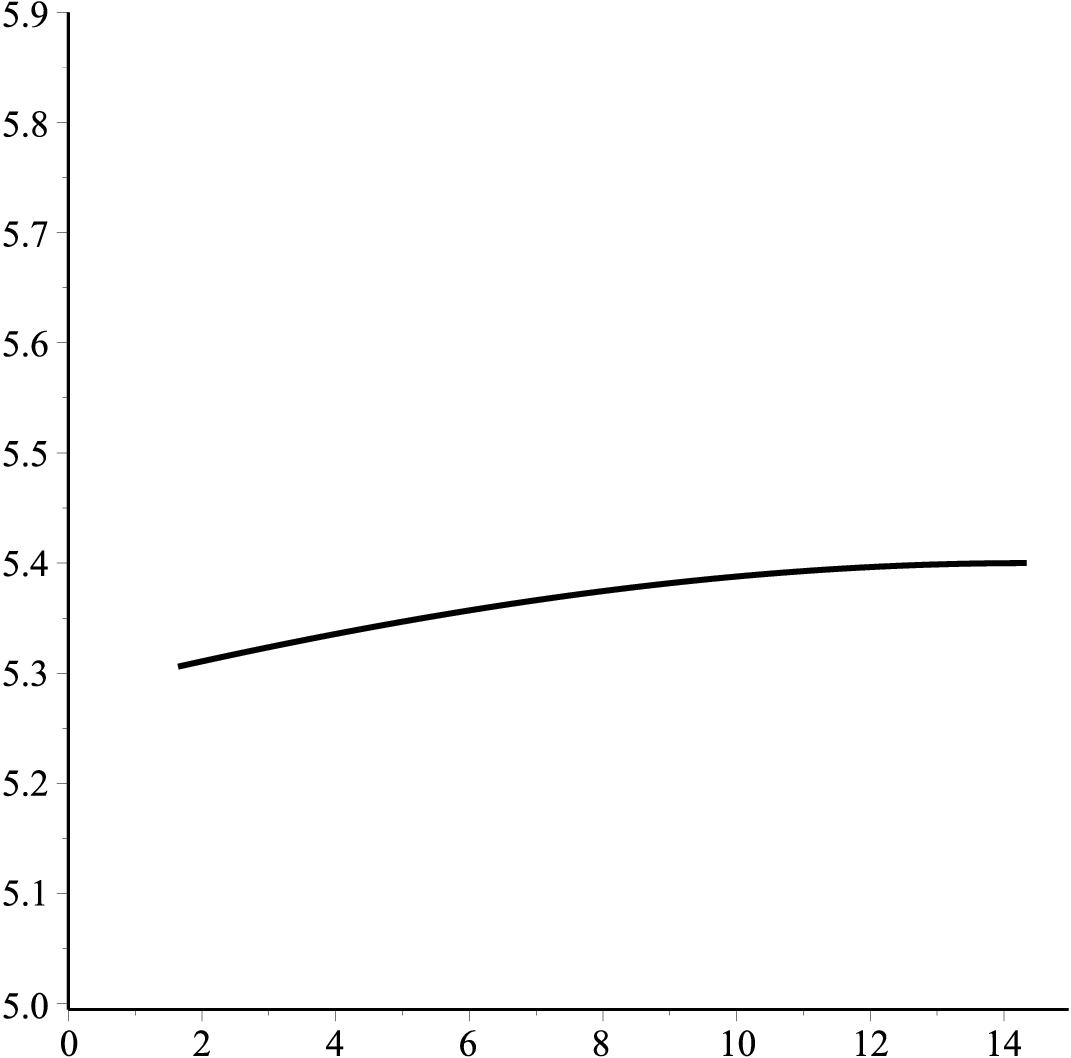}}
\put(0,14.0){${\displaystyle \frac{I_1}{\mbox{tg}^2\beta}}$}
\put(6.4,7.5){$\log_{10}\frac{\mu}{m_Z}$}
\put(8.2,14.0){$I_2$}
\put(14,7.5){$\log_{10}\frac{\mu}{m_Z}$}
\put(1.0,0.5){\includegraphics[scale=0.34]{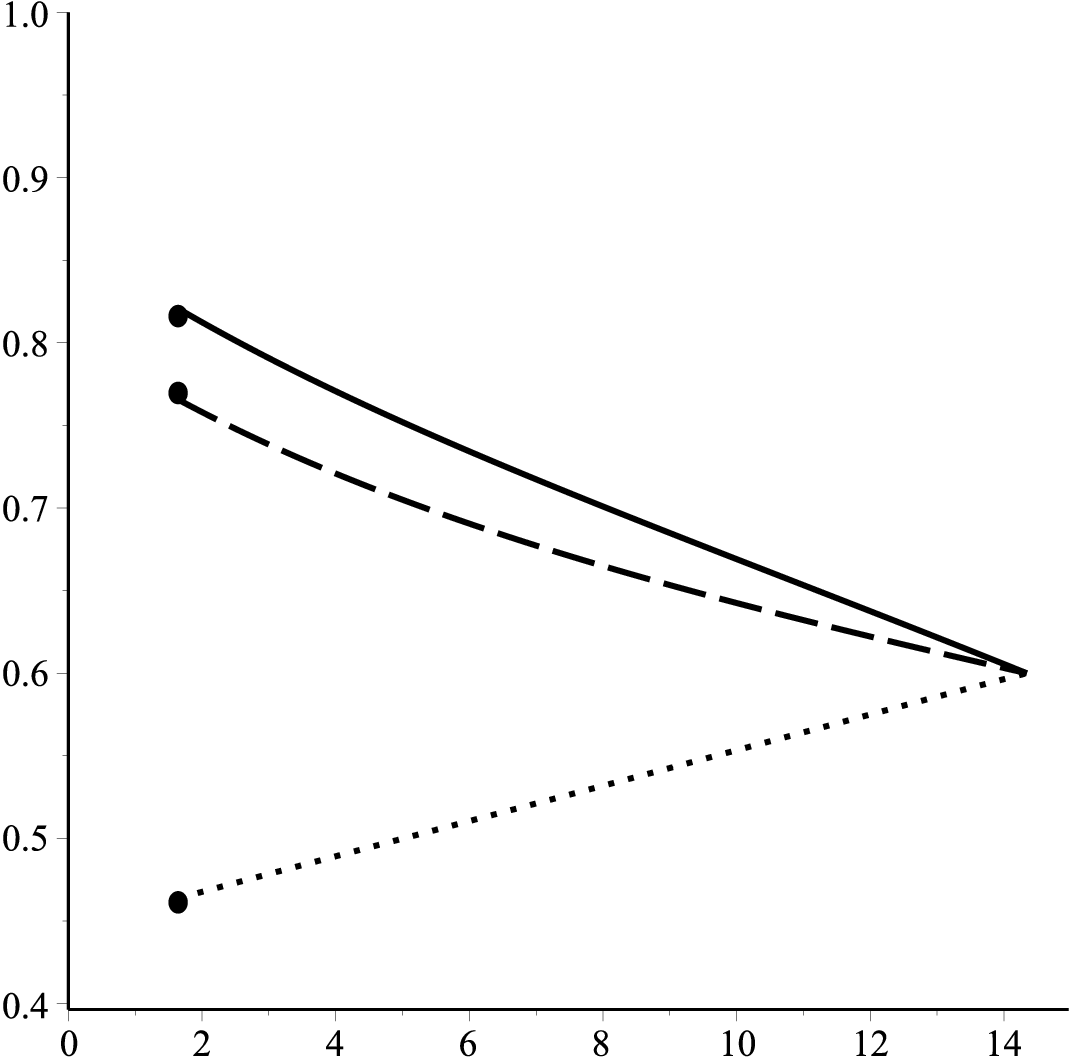}}
\put(8.6,0.5){\includegraphics[scale=0.34]{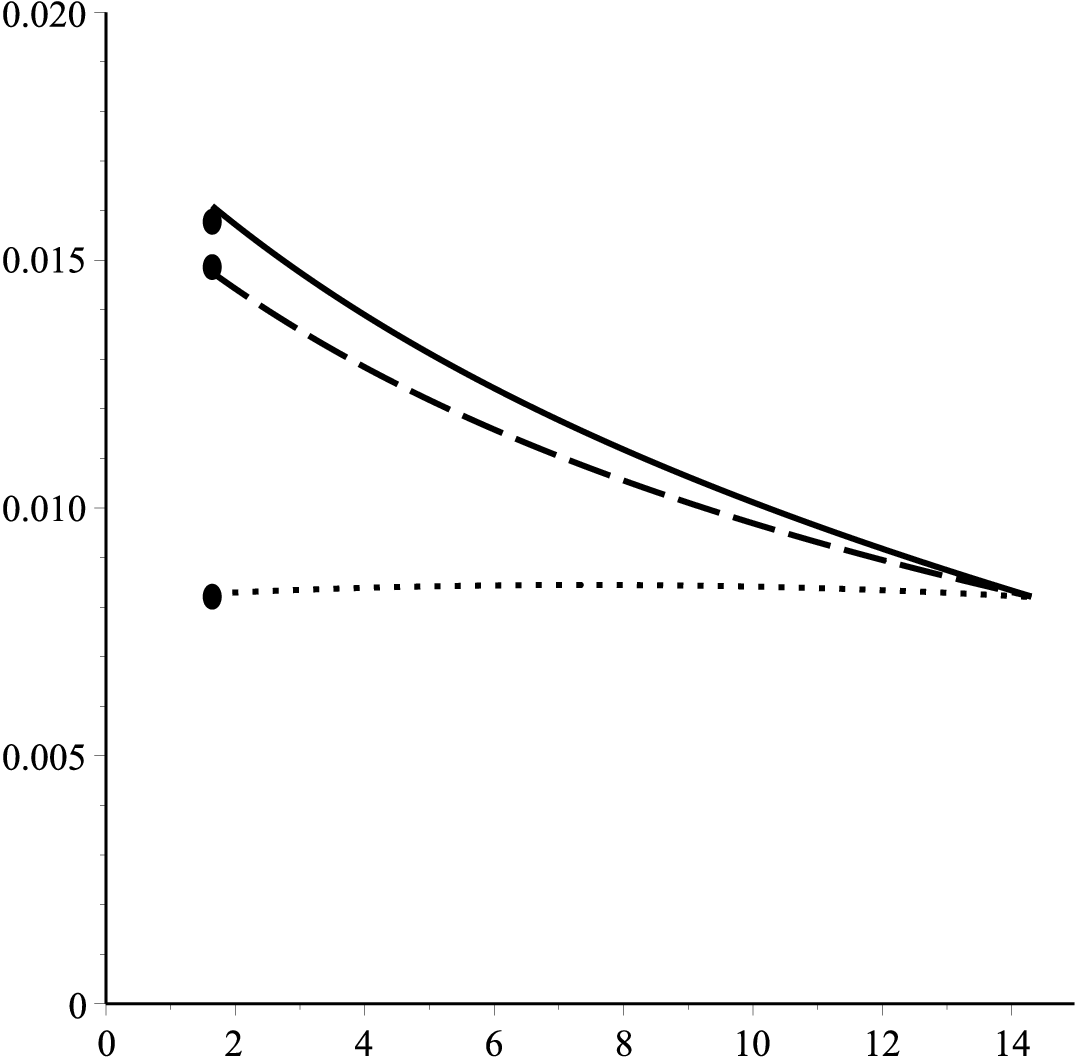}}
\put(0.4,6.5){$y_{33}$}
\put(6.4,0){$\log_{10}\frac{\mu}{m_Z}$}
\put(8.1,6.5){$y_{22}$}
\put(13.8,0){$\log_{10}\frac{\mu}{m_Z}$}
\end{picture}
\caption{The renormalization group running of the approximate RGIs and Yukawa couplings (of the third and second generations) for the MSSM with $\mbox{tg}\,\beta = 59$ (corresponding to $x=1$). The solid, dashed, and dotted lines correspond to $y_U$, $y_D$, and $y_E$ (given by Eq. (\ref{New_Yukawa_Option1})), respectively. The circles mark the values of the relevant couplings (\ref{Boundary}) at the scale $M_{\mbox{\scriptsize SUSY}}=4\,\mbox{TeV}$ obtained from the renormalizarion group equations for the Standard Model.}\label{Figure_Yukawa_Running_MSSM_Option1}
\end{figure}

\noindent
Three coefficients here are not equal to integers, and, moreover, the corresponding value of $\mbox{tg}\,\beta$ is very large, about $63$. (Certainly, the precision of this estimate is limited by the presence of $\Delta_1$ and $\Delta_2$ in Eq. (\ref{Tg_Beta}), which, as we will see later, somewhat reduce this value.) For such large values of $\mbox{tg}\,\beta$ the solutions of the renormalization group equations are unstable, and at low energies they tend to the infrared quasi-fixed point. That is why in this case it is convenient to impose initial conditions at the unification scale. The corresponding plots of the invariants $I_1$ and $I_2$ are presented in Fig. \ref{Figure_Yukawa_Running_MSSM_Option1} at the top left and at the top right, respectively, for $\mbox{tg}\,\beta=59.$\footnote{Let us remind that here we do not take into account the threshold effects, which can be very essential.} Note that for $x=1$ the expression for the RGI $I_2$ takes the form

\medskip

\begin{equation}\label{Yukawa_RGI2_Option1}
I_2 \Big|_{x=1} = \bigg|\frac{(Y_U)_{33}}{(Y_D)_{33}}\cdot \frac{(Y_D)_{22}}{(Y_U)_{22}}\bigg|.
\end{equation}

\noindent
We see that the expression $I_1$ in this case depends on scale essentially (because for large $\mbox{tg}\,\beta$ the Yukawa coupling $(Y_E)_{33}$ is rather large), but not very strongly. The expression $I_2$ depends on scale much less and remains almost constant throughout the whole renormalization group evolution starting from $M_{\mbox{\scriptsize SUSY}}$ up to the unification scale $M_X$. More precisely, the values of $\Delta_1$ and $\Delta_2$ defined by Eq. (\ref{Y_ED_Relations}) and of the expression $\Delta_1^{-1/2}\Delta_2^{5/2}$ present in Eq. (\ref{Tg_Beta}) are

\begin{equation}\label{Delta12_Values}
\Delta_1 \approx 1.25;\qquad \Delta_2\approx 1.02;\qquad \Delta_1^{-1/2}\Delta_2^{5/2}\approx 0.93.
\end{equation}

\noindent
Due to the last equality, the value of $\mbox{tg}\,\beta$ appears a little smaller than the original estimate and is approximately equal to 59.

Drawing the plots of the Yukawa couplings, we find it convenient to redefine them by introducing the values $(y_i)_{33}$ and $(y_i)_{22}$ (where $i = U, D, E$) such that the unification conditions take the form

\begin{equation}
(y_U)_{33} = (y_D)_{33} = (y_E)_{33};\qquad (y_U)_{22} = (y_D)_{22} = (y_E)_{22}.
\end{equation}

\noindent
(Note that the definitions of $(y_i)_{33}$ and $(y_i)_{22}$ are different for different options considered below.) In particular, according to Eq. (\ref{Yukawa_Unification_1}), for $x=1$ it is possible to define the new Yukawa couplings by the equations

\begin{eqnarray}\label{New_Yukawa_Option1}
&& (y_U)_{33} \equiv \left|(Y_U)_{33}\right|;\qquad\qquad (y_D)_{33} \equiv \left|(Y_D)_{33}\right|;\qquad (y_E)_{33} \equiv (1.3)^{-1} \left|(Y_E)_{33}\right|;\qquad\nonumber\\
&& (y_U)_{22} \equiv 5.4\cdot\left|(Y_U)_{22}\right|;\qquad (y_D)_{22} \equiv \left|(Y_D)_{22}\right|;\qquad (y_E)_{22} \equiv (4.3)^{-1} \left|(Y_E)_{22}\right|.
\end{eqnarray}

In Fig. \ref{Figure_Yukawa_Running_MSSM_Option1} we also present the renormalization group running of the Yukawa couplings $y_i$ for the third (at the bottom left) and second (at the bottom right) generations.\footnote{Certainly, we consider only the components $(Y_i)_{33}$ and $(Y_i)_{22}$ of the Yukawa matrices assuming that the other elements are much smaller.} These plots were obtained by imposing the boundary conditions

\begin{equation}
(y_i)_{33}\Big|_{M_X} = 0.6;\qquad (y_i)_{22}\Big|_{M_X} = 0.0082
\end{equation}

\noindent
at the unification scale. We see that the values of the Yukawa couplings at $M_{\mbox{\scriptsize SUSY}}$ are in rather good (but not exact) agreement with the corresponding values (marked by circles) obtained in the Standard Model by solving the non-supersymmetric renormalization group equations. This confirms that the values of the normalization constants in Eq. (\ref{New_Yukawa_Option1}), derived with the help of approximate RGIs, are in a rather good agreement with the values obtained from the renormalization group analysis. However, it seems at least difficult to obtain the values of the ratios $Y_E/Y_D$ from the group theory. Nevertheless, these ratios can be somewhat altered by a special tuning the supersymmetric thresholds. In particular, in this way it is even possible to achieve the complete top-bottom-tau unification \cite{Hall:1993gn,Blazek:2001sb,Tobe:2003bc,Auto:2003ys}. However, we consider only the situation when all thresholds coincides. In this case top-bottom-tau unification does not occur, but we notice that the curves for $(y_U)_{33} = x^{-1} |(Y_U)_{33}|$ and $(y_D)_{33}= |(Y_D)_{33}|$ are really close to each other, so that the approximation $|(Y_U)_{33}| \approx x |(Y_D)_{33}|$ used to derive the expression (\ref{Yukawa_RGI2}) is really satisfied.

\medskip

2. Looking at Fig. \ref{Figure_X_Dependence}, one can observe that the point

\begin{equation}
x=3;\qquad \Big|\frac{(Y_D)_{22}}{(Y_U)_{22}}\Big|_{M_X} = 2,
\end{equation}

\noindent
falls almost exactly on the left plot. In this case the Yukawa relations suggested by the analysis of the RGIs take the form

\begin{equation}\label{Yukawa_Unification_2}
\frac{1}{3} \left|(Y_U)_{33}\right| = \left|(Y_D)_{33}\right| \approx (1.3)^{-1}\left|(Y_E)_{33}\right|;\qquad 2 \left|(Y_U)_{22}\right| = \left|(Y_D)_{22}\right| \approx (4.3)^{-1} \left|(Y_E)_{22}\right|.
\end{equation}

\begin{figure}[!h]
\hspace*{-4mm}
\begin{picture}(0,15)
\put(1.1,8){\includegraphics[scale=0.34]{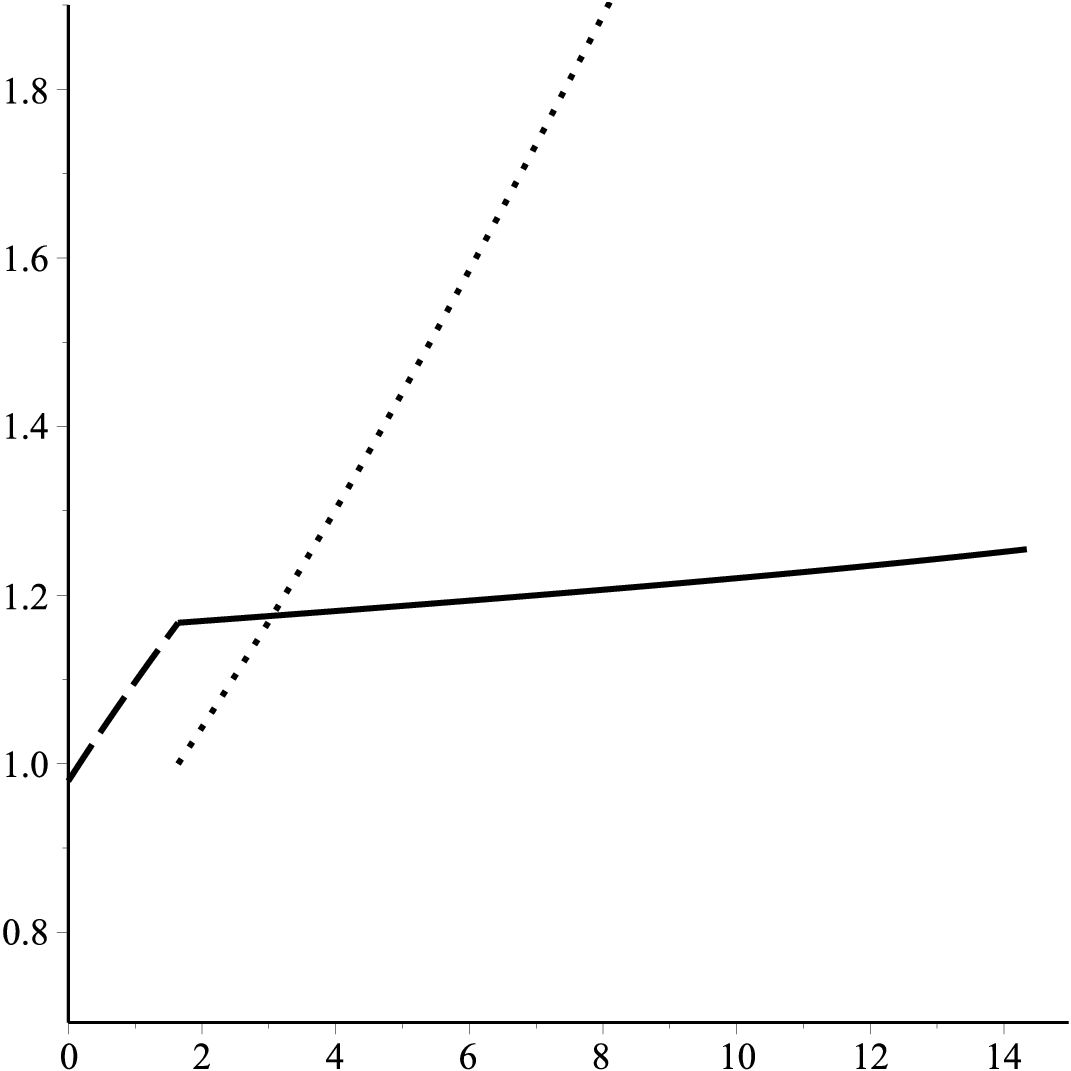}}
\put(8.8,8){\includegraphics[scale=0.34]{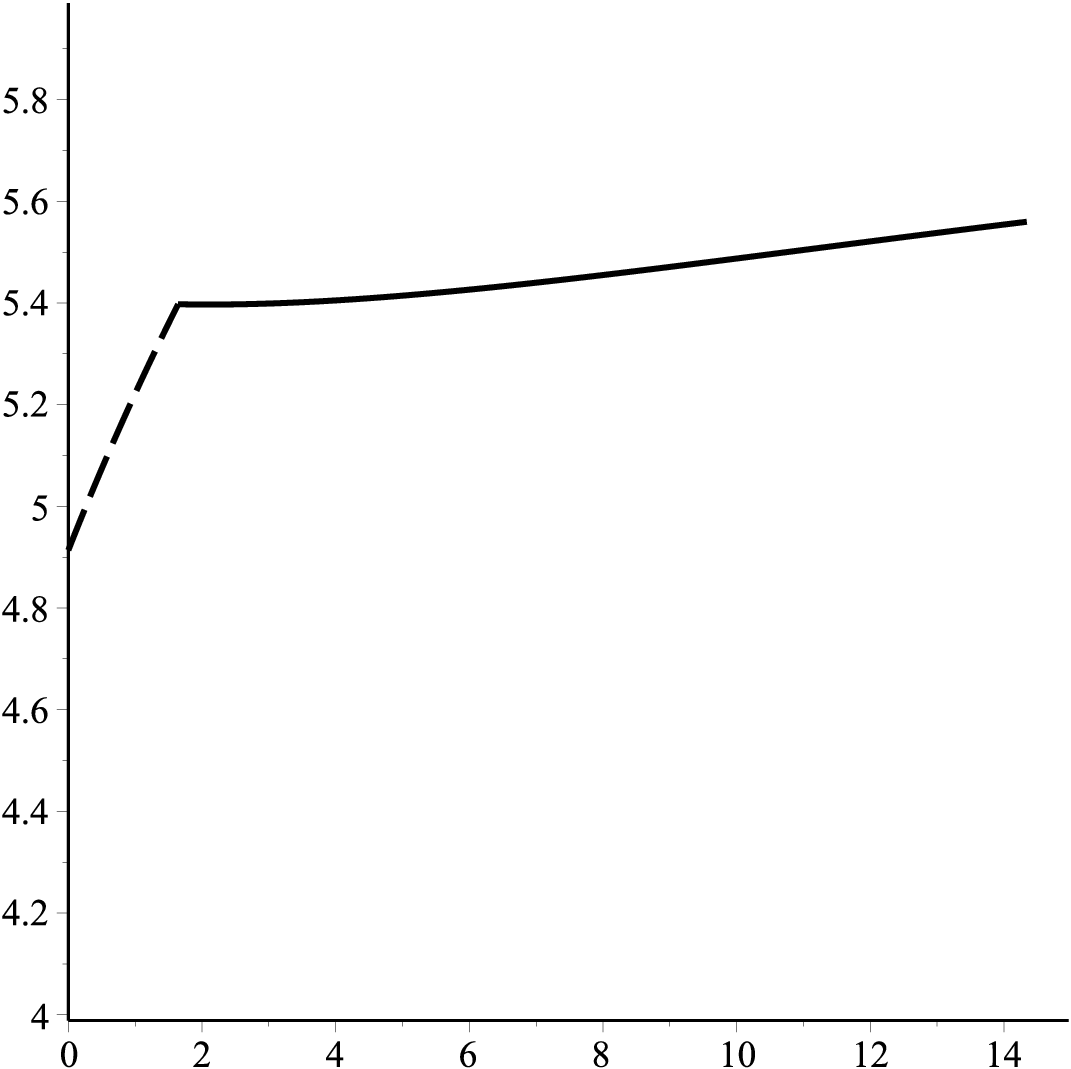}}
\put(0.1,14.0){${\displaystyle \frac{I_1}{\mbox{tg}^2\beta}}$}
\put(6.4,7.5){$\log_{10}\frac{\mu}{m_Z}$}
\put(7.2,14){$(3.05)^{4/3} I_2$}
\put(14,7.5){$\log_{10}\frac{\mu}{m_Z}$}
\put(1.0,0.5){\includegraphics[scale=0.34]{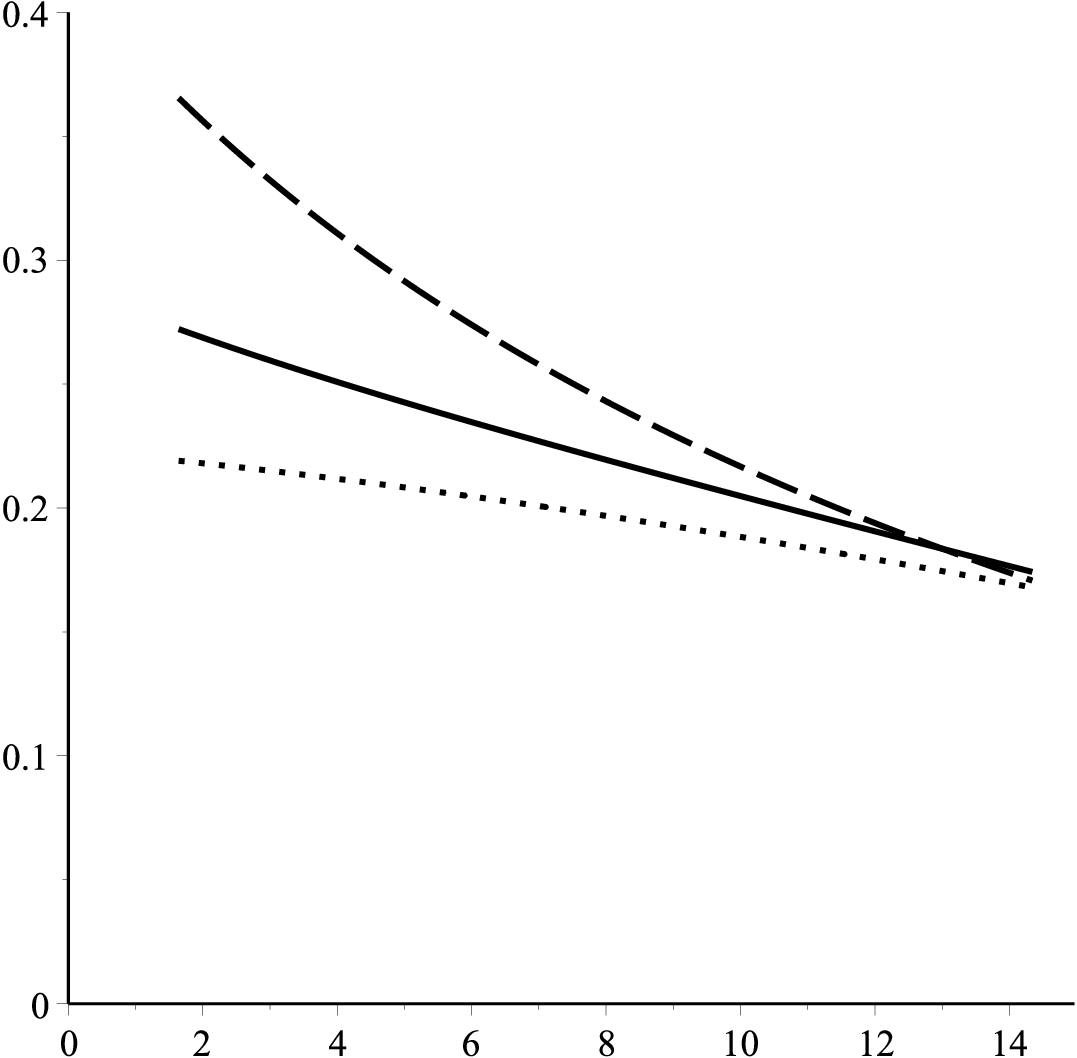}}
\put(8.6,0.5){\includegraphics[scale=0.34]{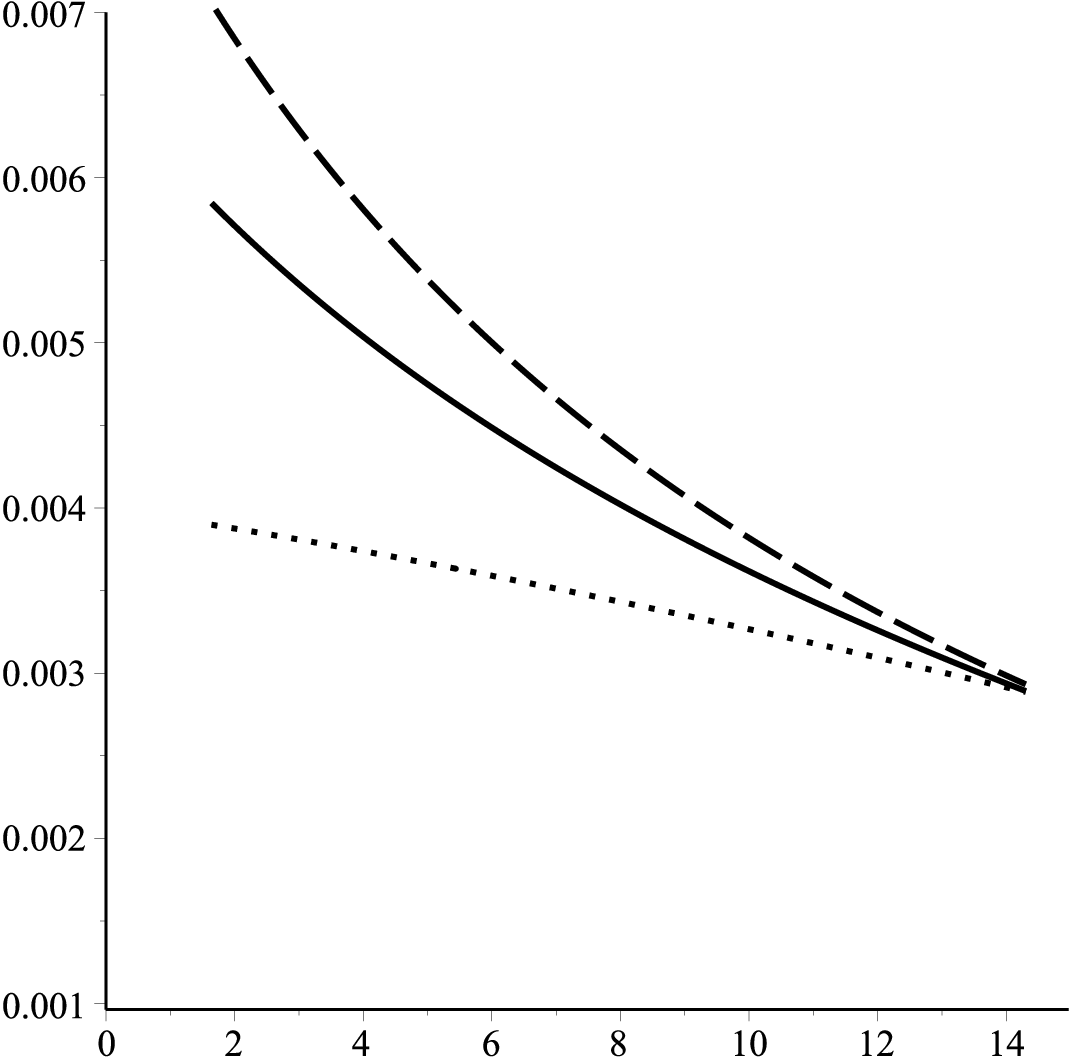}}
\put(0.4,6.5){$y_{33}$}
\put(6.4,0){$\log_{10}\frac{\mu}{m_Z}$}
\put(8.0,6.5){$y_{22}$}
\put(13.8,0){$\log_{10}\frac{\mu}{m_Z}$}
\end{picture}
\caption{The renormalization group running of the approximate RGIs and Yukawa couplings (of the third and second generations) for the MSSM with $\mbox{tg}\,\beta = 28$ (corresponding to $x=3$). The solid, dashed, and dotted lines correspond to $y_U$, $y_D$, and $y_E$ (defined by Eq. (\ref{New_Yukawa_Option2})), respectively.}\label{Figure_Yukawa_Running_MSSM_Option2}
\end{figure}

\noindent
Although in this case $Y_U$ and $Y_D$ are related in a rather simple way, it does not seem that the Yukawa relations for $Y_E$ could be easily derived from group theory. The value of $\mbox{tg}\,\beta$ obtained in this case from the right plot in Fig. \ref{Figure_X_Dependence} is about $26.9$. For such values of $\mbox{tg}\,\beta$ the solutions of the renormalization group equations are stable, so that the initial conditions can be imposed at low energies, namely, at the scale $M_{\mbox{\scriptsize SUSY}}$. The renormalization group analysis demonstrates that the relations (\ref{Yukawa_Unification_2}) hold better for $\mbox{tg}\,\beta=28.$ That is why the plots in Fig. \ref{Figure_Yukawa_Running_MSSM_Option2} are presented for this value. Namely, the plot of the expression $I_1/\mbox{tg}^2\beta$ is presented at the top left of this figure by the solid line. Below the supersymmetric threshold by dashes we also plotted the function

\medskip

\begin{equation}
\Bigg|\bigg(\frac{(Y_U^{\mbox{\scriptsize (SM)}})_{33}}{(Y_D^{\mbox{\scriptsize (SM)}})_{33}}\bigg)^3 \bigg(\frac{(Y_D^{\mbox{\scriptsize (SM)}})_{22}}{(Y_U^{\mbox{\scriptsize (SM)}})_{22}}\bigg)^5\Bigg|,
\end{equation}

\noindent
which depends on scale much more strongly. For comparison, by dots we also presented the plot of the function

\begin{equation}\
\frac{1}{\mbox{tg}^2\beta} \Bigg|\bigg(\frac{(Y_U)_{33}}{(Y_D)_{33}}\bigg)^5 \bigg(\frac{(Y_D)_{22}}{(Y_U)_{22}}\bigg)^3\Bigg|,
\end{equation}

\noindent
which was obtained from $I_1/\mbox{tg}^2\beta$ by the formal replacement of powers $3\leftrightarrow 5$. Comparing the growth of this function with the growth of $I_1/\mbox{tg}^2\beta$, we see that $I_1$ can really be considered as an approximate RGI ($\Delta_1\approx 1.07$).

For $x=3$ the RGI $I_2$ takes the form

\begin{equation}\label{Yukawa_RGI2_Option2}
I_2\Big|_{x=3} = \bigg|\frac{(Y_U)_{33}}{(Y_D)_{33}}\cdot \frac{(Y_D)_{22}}{(Y_U)_{22}}\cdot \bigg(\frac{(Y_E)_{33}}{(Y_D)_{33}}\cdot \frac{(Y_D)_{22}}{(Y_E)_{22}}\bigg)^{4/3}\bigg|.
\end{equation}

\noindent
The plot of the function $(3.05)^{4/3} I_2$ (above the supersymmetric threshold) is presented in Fig. \ref{Figure_Yukawa_Running_MSSM_Option2} at the top right by a solid line. Below the supersymmetric threshold by dashes we plot the function

\begin{equation}\label{SM_Analog_Of_I2}
\Bigg|\frac{(Y_U^{\mbox{\scriptsize (SM)}})_{33}}{(Y_D^{\mbox{\scriptsize (SM)}})_{33}}\cdot \frac{(Y_D^{\mbox{\scriptsize (SM)}})_{22}}{(Y_U^{\mbox{\scriptsize (SM)}})_{22}}\cdot \bigg(3.05\cdot \frac{(Y_E^{\mbox{\scriptsize (SM)}})_{33}}{(Y_D^{\mbox{\scriptsize (SM)}})_{33}}\cdot \frac{(Y_D^{\mbox{\scriptsize (SM)}})_{22}}{(Y_E^{\mbox{\scriptsize (SM)}})_{22}}\bigg)^{4/3}\Bigg|.
\end{equation}

\noindent
We see that the expression $I_2$ really remains almost constant above the supersymmetric threshold ($\Delta_2\approx 1.03$), while below this threshold the expression (\ref{SM_Analog_Of_I2}) strongly depends on scale. The normalization constant is chosen in such a way that at the unification scale the resulting function is equal to the value of the ratio

\begin{equation}\label{Y_UD_Ratio}
\bigg|\frac{(Y_U)_{33}}{(Y_D)_{33}}\cdot \frac{(Y_D)_{22}}{(Y_U)_{22}}\Bigg|.
\end{equation}

At the bottom left and at the bottom right of Fig. \ref{Figure_Yukawa_Running_MSSM_Option2} we present the plots of the running Yukawa couplings for the third and second generations, respectively. For convenience, they are redefined as

\begin{eqnarray}\label{New_Yukawa_Option2}
&& (y_U)_{33} \equiv \frac{1}{3} \left|(Y_U)_{33}\right|;\qquad (y_D)_{33} \equiv \left|(Y_D)_{33}\right|;\qquad (y_E)_{33} \equiv (1.3)^{-1} \left|(Y_E)_{33}\right|;\qquad\nonumber\\
&& (y_U)_{22} \equiv 2 \left|(Y_U)_{22}\right|;\qquad\, (y_D)_{22} \equiv \left|(Y_D)_{22}\right|;\qquad (y_E)_{22} \equiv (4.3)^{-1} \left|(Y_E)_{22}\right|,
\end{eqnarray}

\noindent
so that the fulfillment of the condition (\ref{Yukawa_Unification_2}) corresponds to meeting of these couplings at a single point. This implies that the Yukawa couplings $y_i$ in Fig. \ref{Figure_Yukawa_Running_MSSM_Option2} meet in a single point due to special tuning of the normalization constants for $y_E$ in Eq. (\ref{New_Yukawa_Option2}). However, the integers appearing in the normalization constants for $y_U$ were obtained from the analysis based on approximate RGIs. The plots presented in Fig. \ref{Figure_Yukawa_Running_MSSM_Option2} also demonstrate that the curves of $(y_U)_{33} = x^{-1} |(Y_U)_{33}|$ and $(y_D)_{33} = |(Y_D)_{33}|$ are again rather close to each other, so that the approximation used for deriving the RGI $I_2$ is also satisfied in this case.

\section{Renormalization group running of the Yukawa couplings for theories with additional matter superfields}
\hspace*{\parindent}\label{Section_Yukawa_With_Extra_Matter}

To achieve the unification of the Yukawa couplings, one can try either to tune the thresholds or to modify the theory in such a way as to achieve the Yukawa unification without spoiling the unification of the gauge couplings. In the latter case the gauge coupling unification is preserved (at least, in the one-loop approximation) if the superfields added to the MSSM field content form complete $SU(5)$ multiplets and their masses are not too different \cite{Ghilencea:1997mu,Amelino-Camelia:1998xgn}. The number and representations of these exotic superfields are limited by the requirement that a (single) value of the inverse gauge couplings at the unification scale should be positive. Here we will try to understand how the Yukawa relations will change upon the addition of certain extra superfields forming complete $SU(5)$ multiplets under the assumption that their Yukawa interaction is relatively small. Namely, we will consider two examples of theories which contain different numbers of additional fields and different values of their masses. These parameters were chosen in such a way that the resulting Yukawa relations contain not very complex fractions or square roots.

1. First, let us add 2 additional superfields forming the representation $5+\,\xbar{5}\,$ of the group $SU(5)$. (An even number of superfields is needed for the anomaly cancellation.) We choose their masses equal to $4\cdot 10^{4}$\, GeV. In this case we obtained that at the unification scale $|(Y_E)_{22}/(Y_D)_{22}| \approx 9/2$ (instead of $4.3$ for the MSSM) for a rather large range of $\mbox{tg}\,\beta$. According to \cite{Antusch:2009gu,Antusch:2013rxa}, this value may follow from a certain quartic $SU(5)$ invariant obtained from $5 \times 24\times \,\xbar{5}$ and $\,\xbar{10}\,\times 45\times 5$ after integrating out the massive superfields in representations $\,\xbar{5}\,$ and $5$.

\begin{figure}[!h]
\hspace*{-4mm}
\begin{picture}(0,8)
\put(1.1,0.5){\includegraphics[scale=0.34]{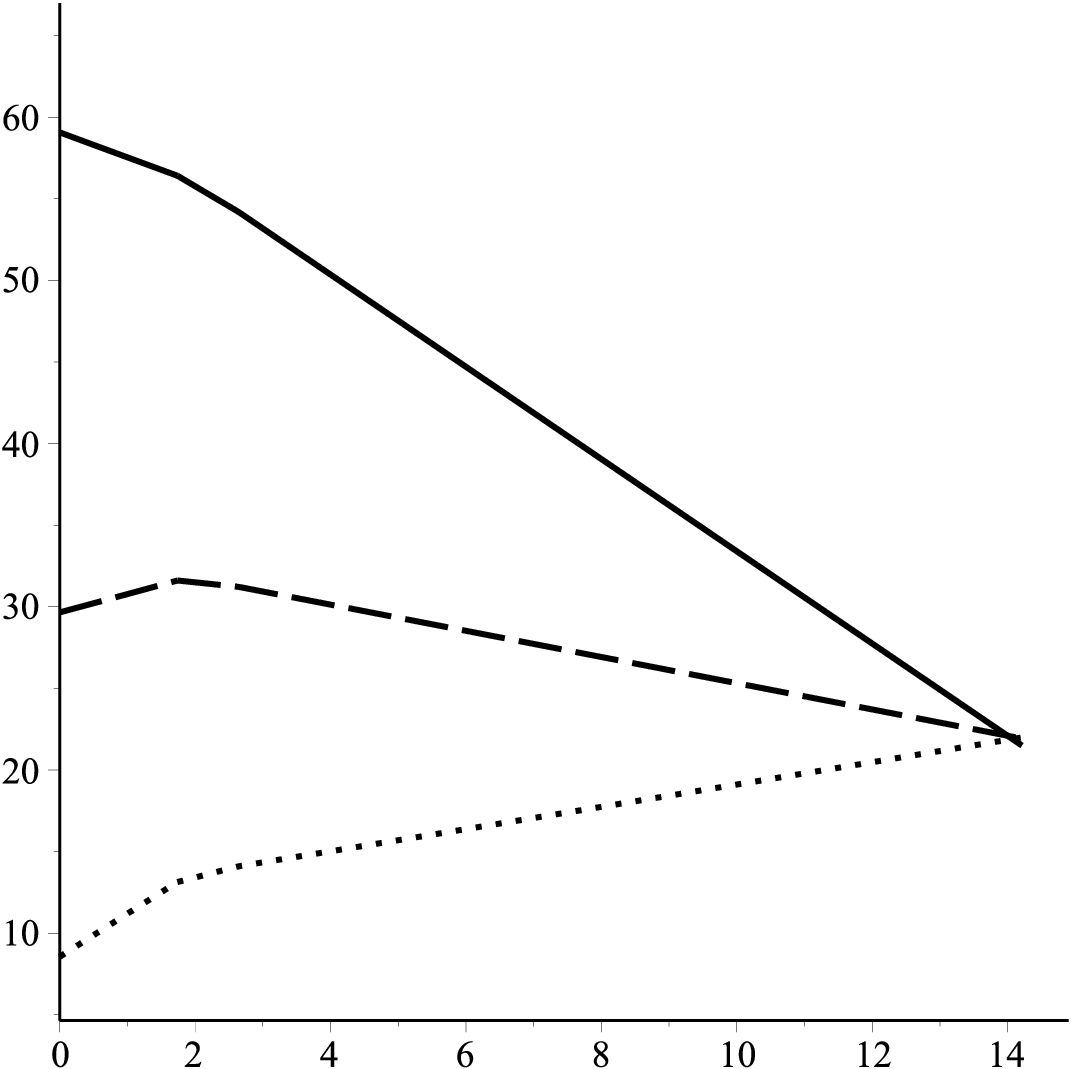}}
\put(8.8,0.5){\includegraphics[scale=0.34]{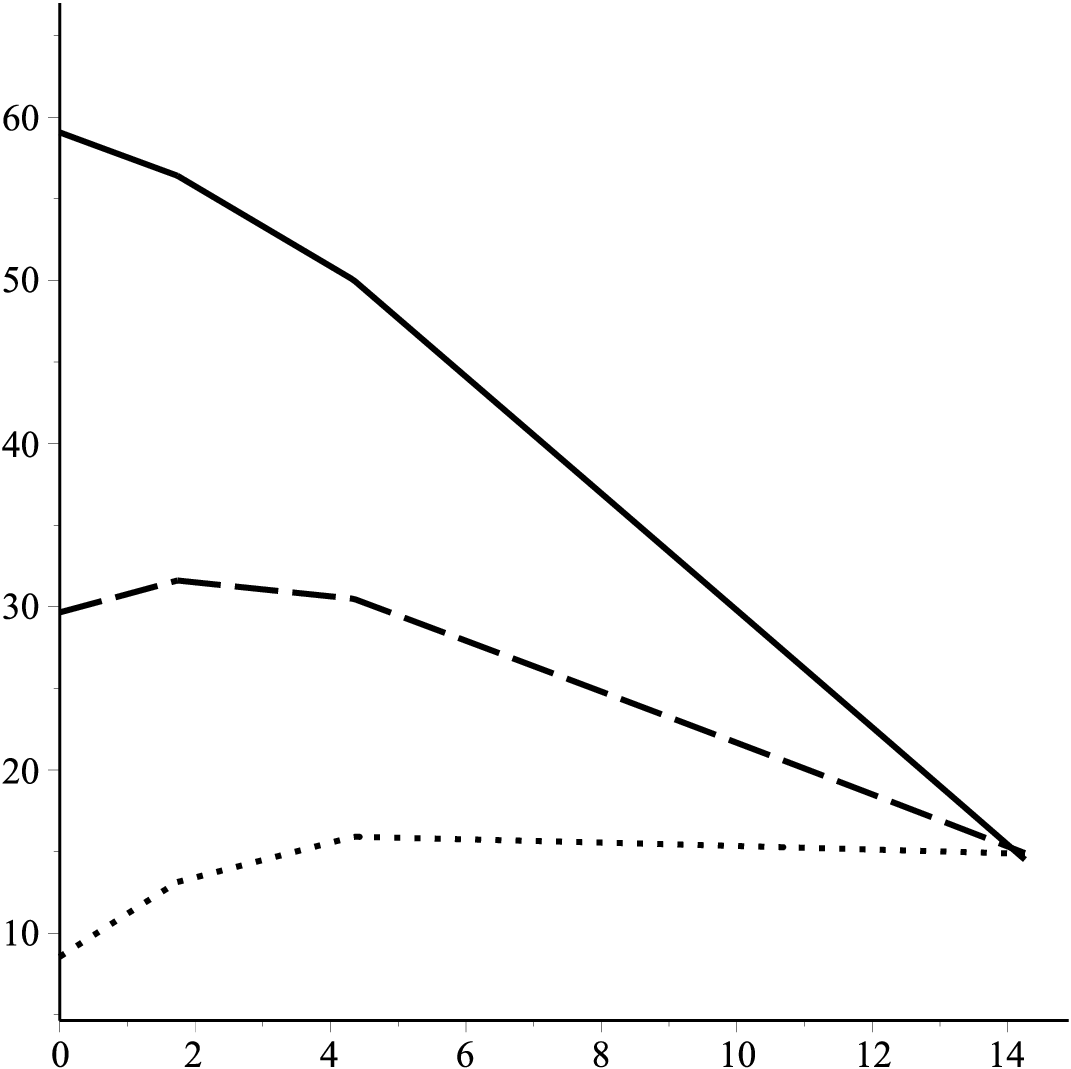}}
\put(0.4,6.5){$1/\alpha$}
\put(8.1,6.5){$1/\alpha$}
\put(6.4,0){$\log_{10}\frac{\mu}{m_Z}$}
\put(13.8,0){$\log_{10}\frac{\mu}{m_Z}$}
\end{picture}
\caption{The renormalization group running of the gauge couplings for the theories with additional superfields described in the text. The left plot corresponds to the theory with $n_5=2$, $M=4\cdot 10^4$ GeV, while the right plot is drawn for $n_5=6$, $M=2\cdot 10^6$\, GeV.}\label{Figure_Gauge_Couplings}
\end{figure}

\begin{figure}[!h]
\hspace*{-4mm}
\begin{picture}(0,8)
\put(1.1,0.5){\includegraphics[scale=0.34]{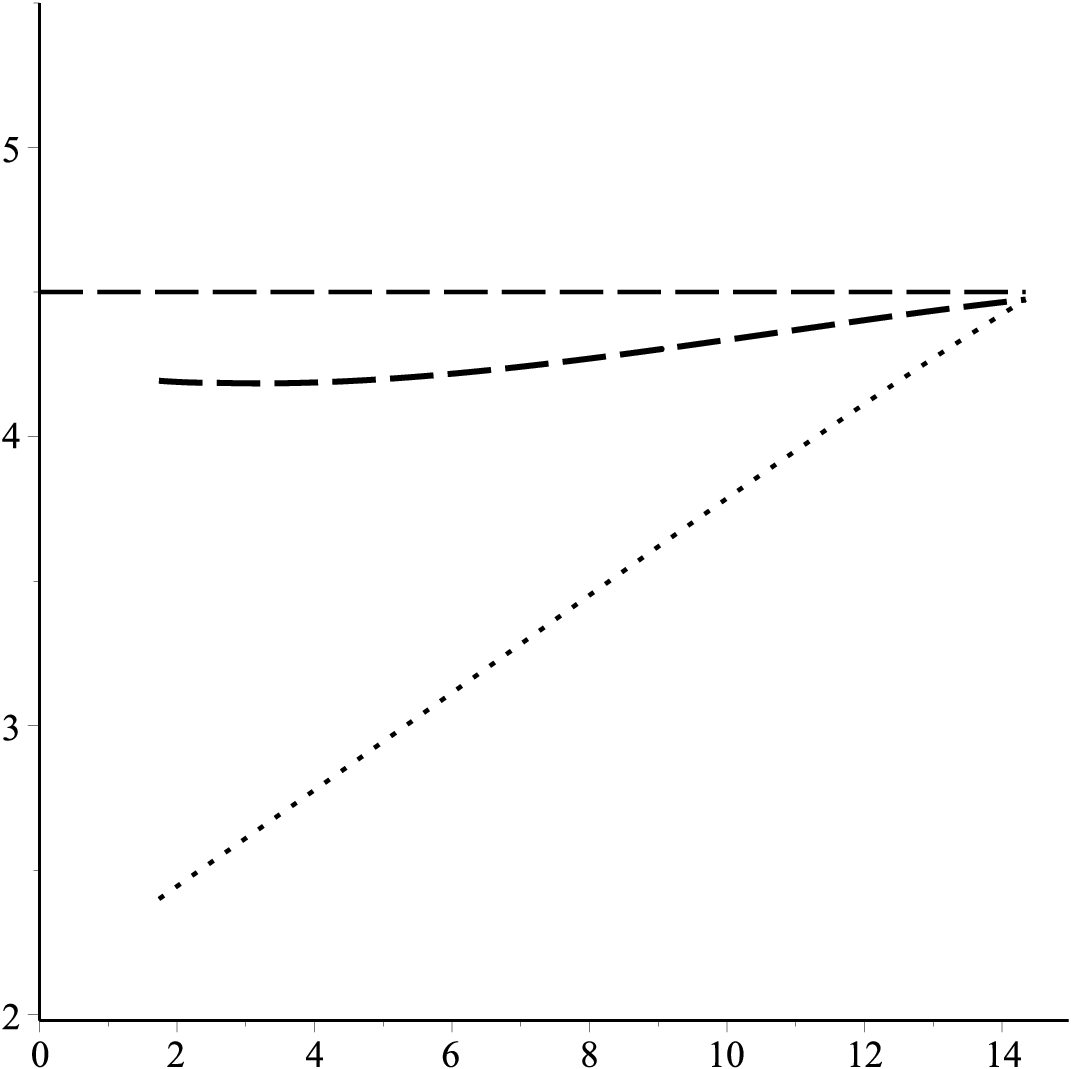}}
\put(8.8,0.5){\includegraphics[scale=0.34]{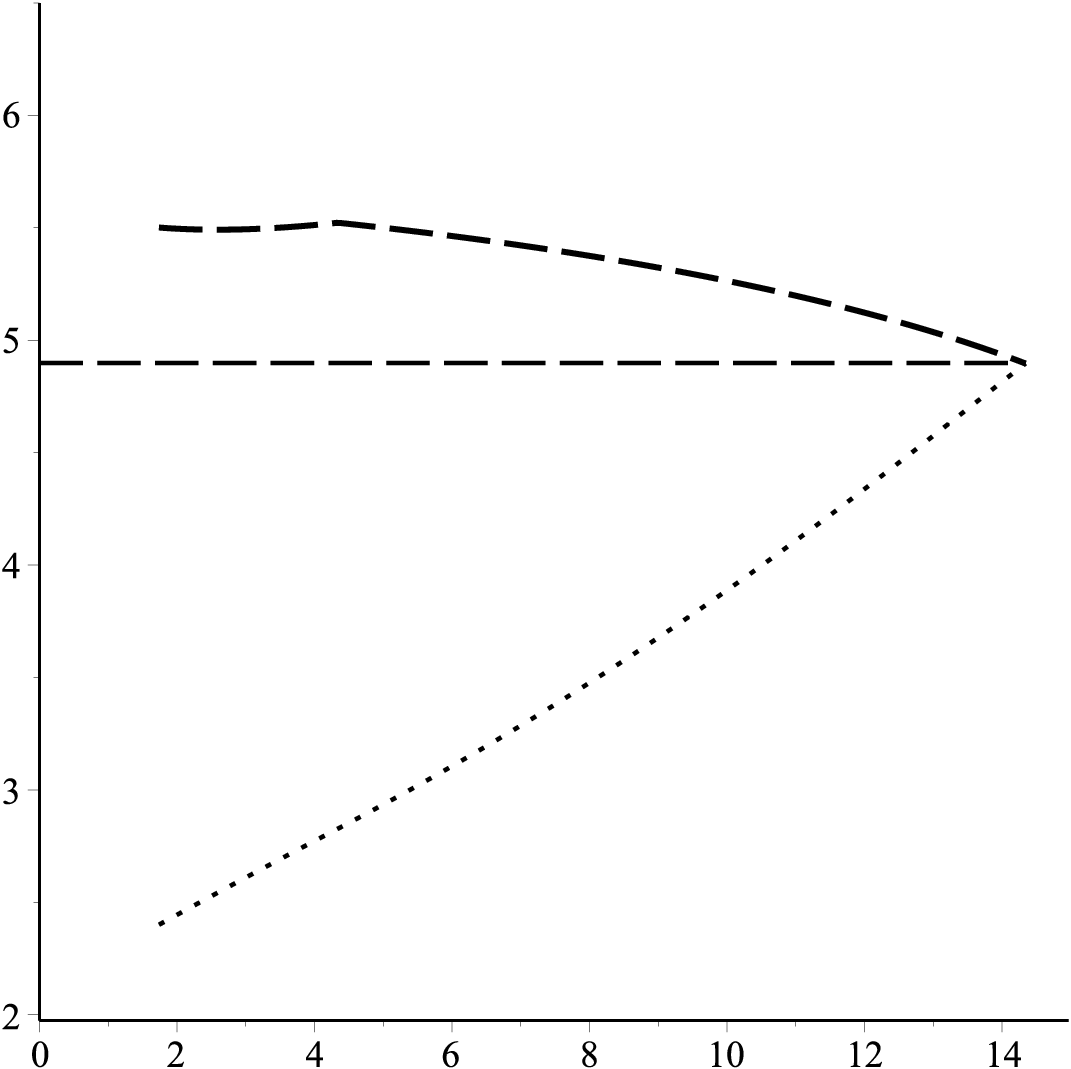}}
\put(2.4,4.0){$I_3$}
\put(10.1,5.6){$I_3$}
\put(0.45,4.85){$9/2$}
\put(7.95,4.45){$\sqrt{24}$}
\put(3.9,2.2){${\displaystyle \Big|\frac{(Y_E)_{22}}{(Y_D)_{22}}\Big|}$}
\put(10.0,2.7){${\displaystyle \Big|\frac{(Y_E)_{22}}{(Y_D)_{22}}\Big|}$}
\put(6.4,0){$\log_{10}\frac{\mu}{m_Z}$}
\put(13.8,0){$\log_{10}\frac{\mu}{m_Z}$}
\end{picture}
\caption{The renormalization group running of the approximate RGI $I_3$ for two theories containing some extra superfields. The left plot corresponds to the theory with $n_5=2$, $M=4\cdot 10^4$ GeV, while the right plot is drawn for $n_5=6$, $M=2\cdot 10^6$\, GeV.}\label{Figure_I3_Exotics}
\end{figure}

The plot of the running gauge couplings for this theory is presented in Fig. \ref{Figure_Gauge_Couplings} on the left, and the renormalization group evolution of the ratio $|(Y_E)_{22}/(Y_D)_{22}|$ and of the expression $I_3$ (given by Eq. (\ref{I3_RGI})) is shown in Fig. \ref{Figure_I3_Exotics} on the left. In particular, we see that, comparing with the MSSM, the value of this ratio at the unification scale is larger due to the presence of the extra superfields, and the RGI $I_3$ allows to estimate it (although very roughly). In particular, the existence of the RGI $I_3$ indicates that the ratio $|(Y_E)_{22}/(Y_D)_{22}|$ slightly depends on $\mbox{tg}\,\beta$ due to Eq. (\ref{Y22_Ration_From_I3}). Note that in this case when solving the renormalization group equations the initial conditions are imposed at the scale $M_{\mbox{\scriptsize SUSY}}= 4\,\mbox{TeV}$ using the solution of the renormalization group equations for the Standard Model and Eq. (\ref{Boundary}).

For the value $\mbox{tg}\beta = 33$ we obtained that at the unification scale the Yukawa couplings (approximately) satisfy the relations

\begin{equation}\label{Unification_Exotics_Option1}
\frac{1}{\sqrt{6}} (Y_U)_{33} = (Y_{D})_{33} = \frac{3}{4} (Y_{E})_{33};\qquad \sqrt{6} (Y_U)_{22} = (Y_{D})_{22} = \frac{2}{9} (Y_{E})_{22}
\end{equation}

\begin{figure}[!h]
\hspace*{-4mm}
\begin{picture}(0,15)
\put(1.1,8){\includegraphics[scale=0.34]{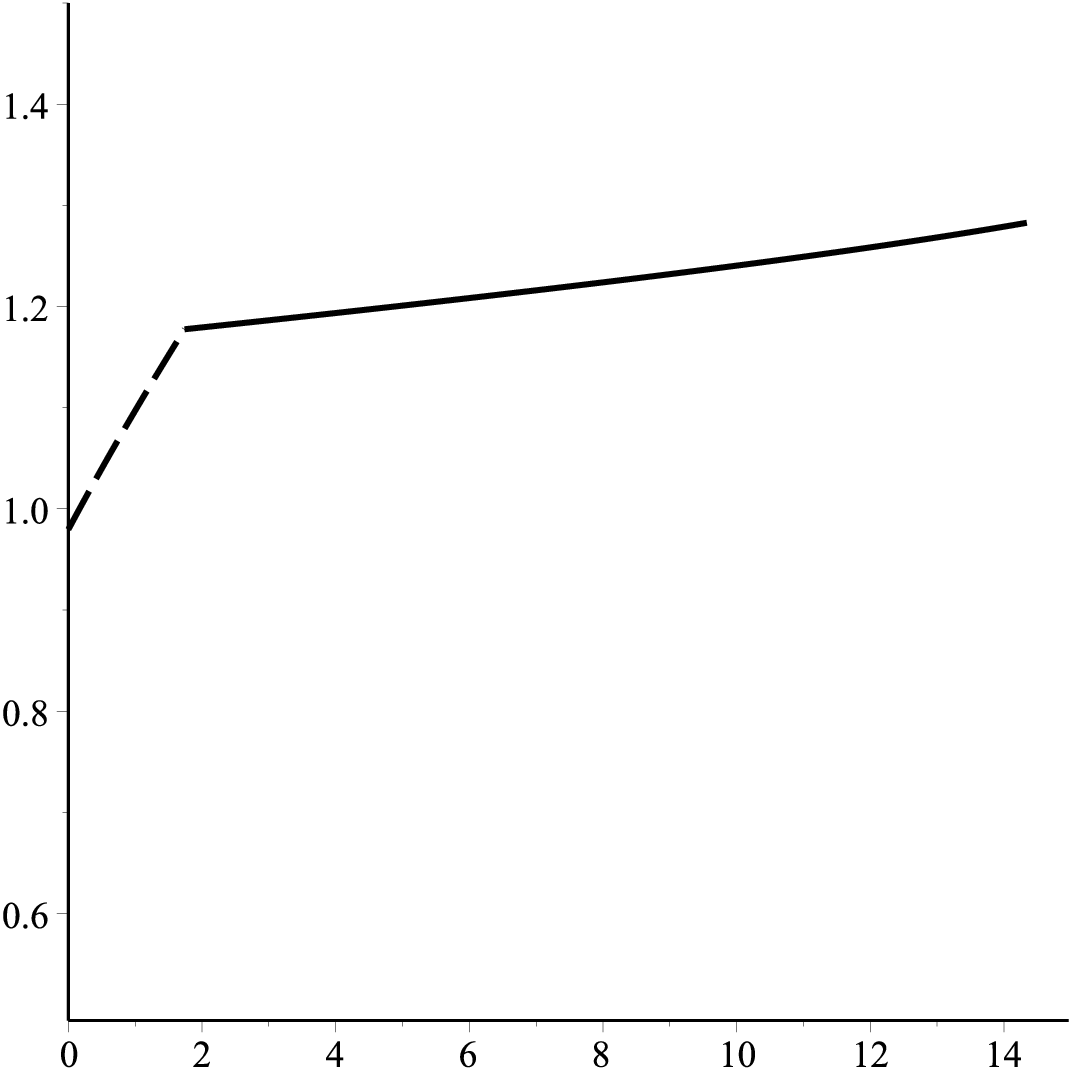}}
\put(8.8,8){\includegraphics[scale=0.34]{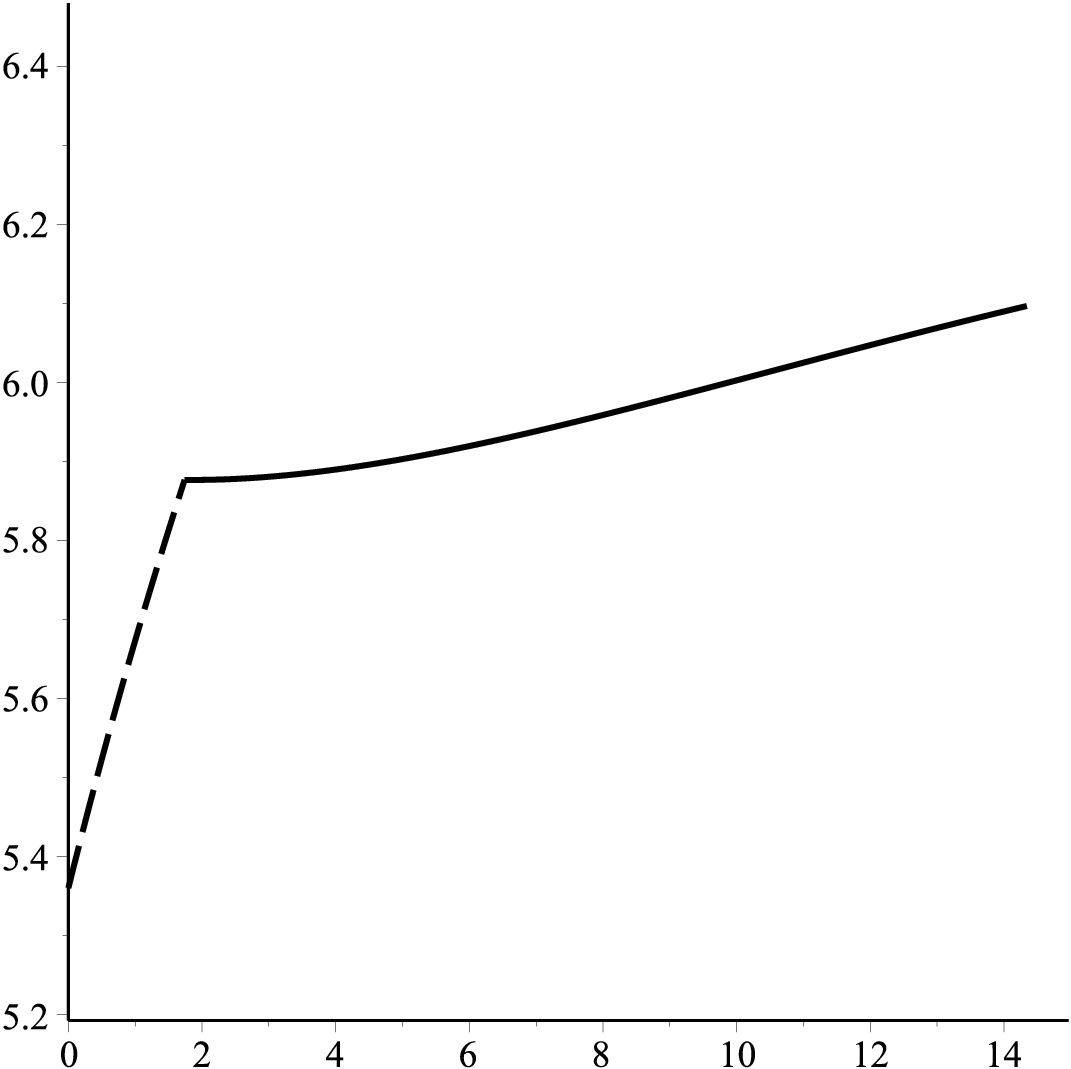}}
\put(0,14.0){${\displaystyle \frac{I_1}{\mbox{tg}^2\beta}}$}
\put(6.4,7.5){$\log_{10}\frac{\mu}{m_Z}$}
\put(7.0,14){${\displaystyle (3.28)^{10/9} I_2}$}
\put(14,7.5){$\log_{10}\frac{\mu}{m_Z}$}
\put(1.0,0.5){\includegraphics[scale=0.34]{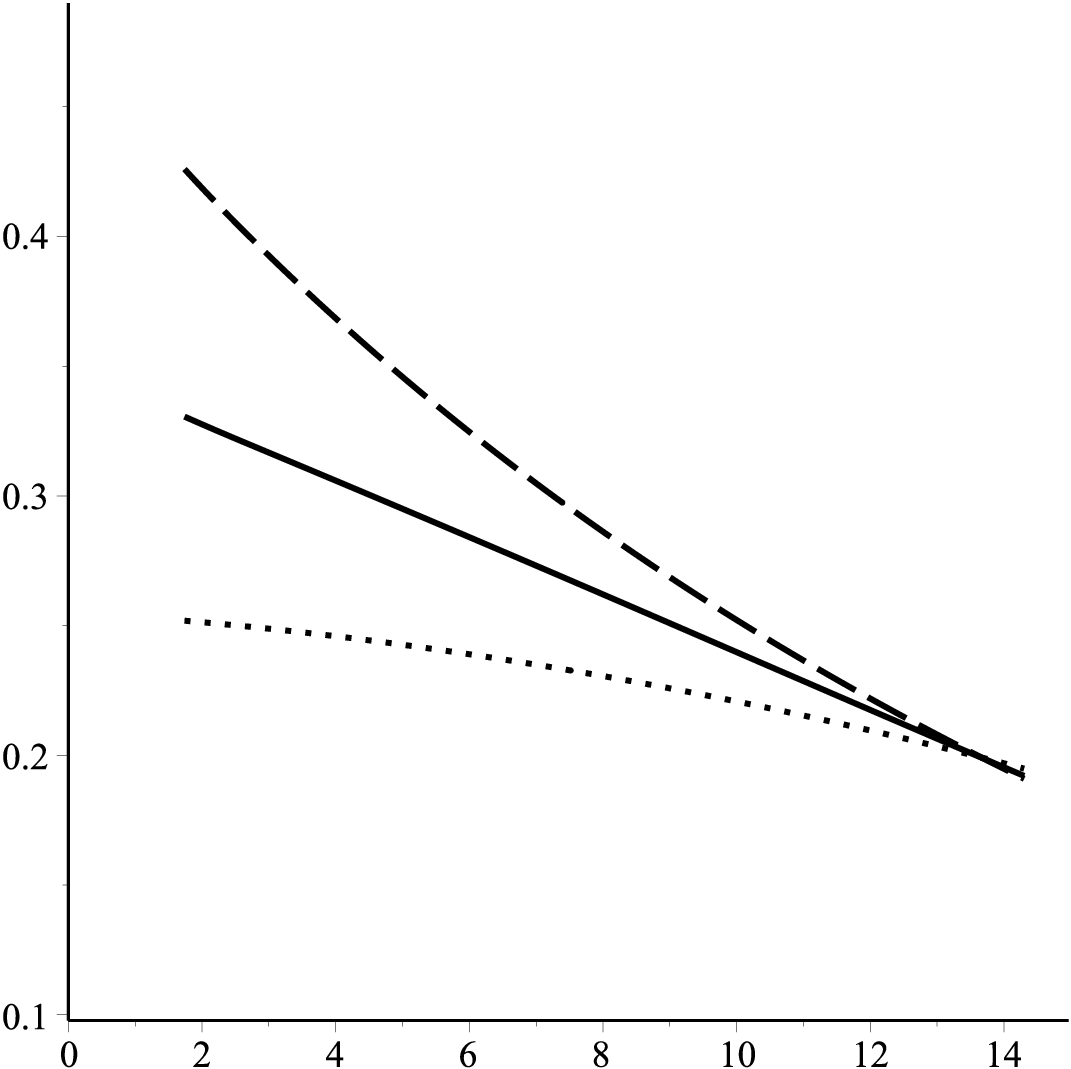}}
\put(8.6,0.5){\includegraphics[scale=0.34]{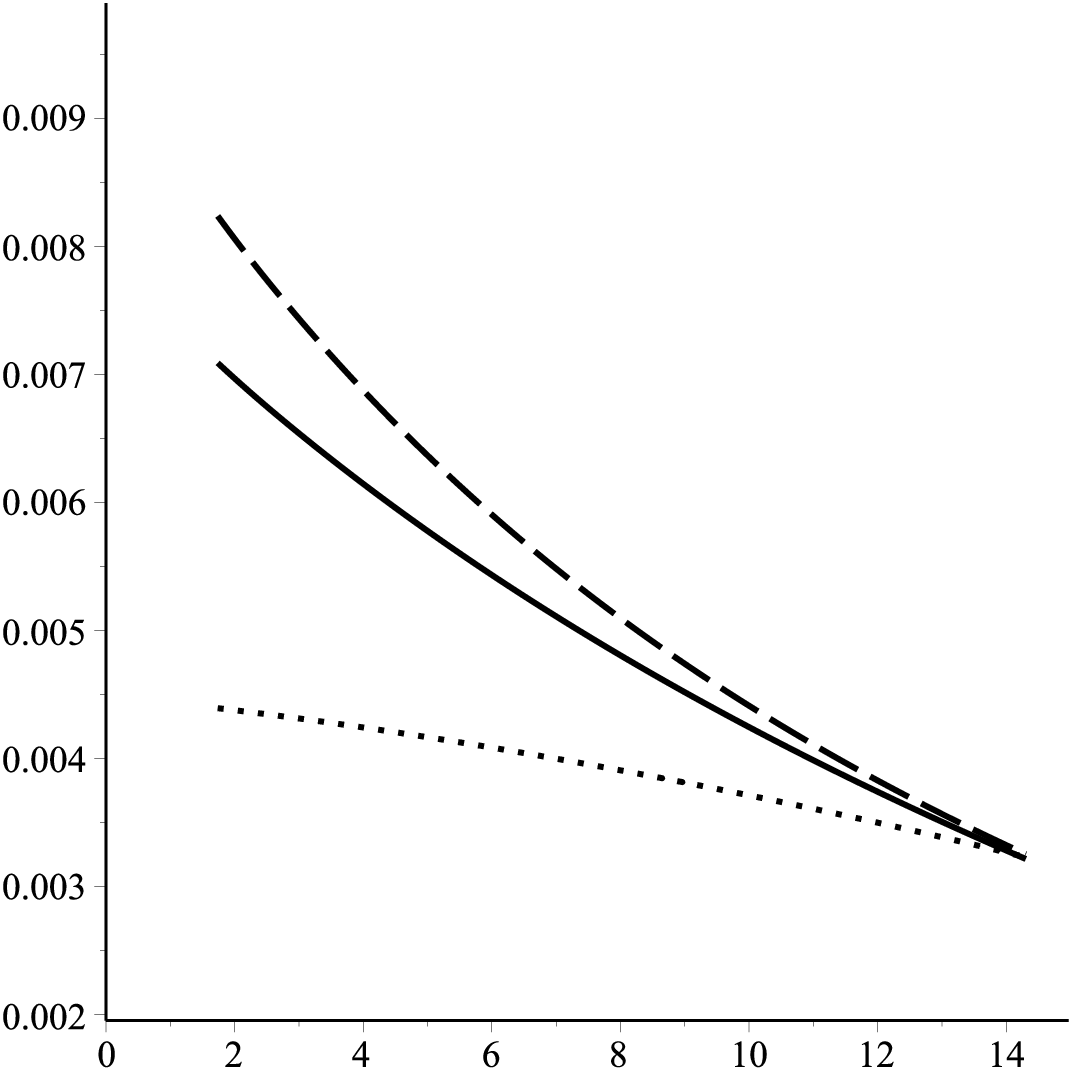}}
\put(0.4,6.5){$y_{33}$}
\put(6.4,0){$\log_{10}\frac{\mu}{m_Z}$}
\put(8.1,6.5){$y_{22}$}
\put(13.8,0){$\log_{10}\frac{\mu}{m_Z}$}
\end{picture}
\caption{The renormalization group running of the approximate RGIs and Yukawa couplings (for the third and second generations) for the extension of the MSSM with $n_5=2$, $M=4\cdot 10^4$ GeV, $\mbox{tg}\,\beta = 33$ and $x=\sqrt{6}$. The solid, dashed, and dotted lines correspond to $y_U$, $y_D$, and $y_E$ (given by Eq. (\ref{New_Yukawa_Exotics_Option1})), respectively. The apparent Yukawa unification implies that Eq. (\ref{Unification_Exotics_Option1}) holds at the unification scale.}\label{Figure_Yukawa_Running_Exotics_Option1}
\end{figure}

\noindent
(corresponding to $y = 8/27$). The plots of the approximate RGIs $I_1$ and $I_2$ (the latter is normalized in such a way that at the unification scale we obtain the value of the ratio (\ref{Y_UD_Ratio})) for this case are presented in Fig. \ref{Figure_Yukawa_Running_Exotics_Option1} at the top left and top right, respectively. We see that they really remain almost constant throughout the evolution from $M_{\mbox{\scriptsize SUSY}}$ to $M_X$. More precisely, the values of $\Delta_i=I_i(M_X)/I_i(M_{\mbox{\scriptsize SUSY}})$  appeared to be

\begin{equation}\label{Delta_Exotics_Option1}
\Delta_1\approx 1.09;\qquad \Delta_2\approx 1.13;\qquad \Delta_3\approx 1.07,
\end{equation}

\noindent
while the solution of the renormalization group equations gives $y\approx 3.28$. As earlier, we introduce new Yukawa couplings

\begin{eqnarray}\label{New_Yukawa_Exotics_Option1}
&& (y_U)_{33} \equiv \frac{1}{\sqrt{6}} \left|(Y_U)_{33}\right|;\qquad (y_D)_{33} \equiv \left|(Y_D)_{33}\right|;\qquad (y_E)_{33} \equiv \frac{3}{4} \left|(Y_E)_{33}\right|;\qquad\nonumber\\
&& (y_U)_{22} \equiv \sqrt{6}\,\left|(Y_U)_{22}\right|;\qquad\, (y_D)_{22} \equiv \left|(Y_D)_{22}\right|;\qquad (y_E)_{22} \equiv \frac{2}{9} \left|(Y_E)_{22}\right|,
\end{eqnarray}

\noindent
in such a way that the validity of Eq. (\ref{Unification_Exotics_Option1}) corresponds to the equality of these couplings. The renormalization group running of these couplings is presented at the bottom of Fig. \ref{Figure_Yukawa_Running_Exotics_Option1}. The Yukawa couplings for the third and second generations are presented in the left and right plots, respectively. From these plots we see that it is possible to achieve the validity of Eq. (\ref{Unification_Exotics_Option1}) at the unification scale. Moreover, the plots of $y_u$ and $y_d$ are really close to each other for the third generation, as is necessary to derive the approximate RGI (\ref{Yukawa_RGI2}). However, the value of the ratio (\ref{Y_UD_Ratio}) appears to be rather large and, for a large range of $\mbox{tg}\,\beta$ values, is about 6. At present, we do not know whether it is possible to derive such a ratio from certain group invariants.

\medskip

2. As another example, let us consider a theory containing 6 superfields forming the representation $3\times(5+\,\xbar{5}\,)$ of the group $SU(5)$. This choice is motivated by the fact that the representation $\,\xbar{27}\,$ of the group $E_6$ produces the representations $5$ and $\,\xbar{5}\,$ (together with a singlet) in addition to the usual $1+5+\,\xbar{10}\,$ forming a particle content of a single generation. In this case the masses are taken equal to $2\cdot 10^{6}$\, GeV, because in this case the value of the ratio $|(Y_E)_{22}/(Y_D)_{22}|$ at the unification scale is rather close to $\sqrt{24}$. The renormalization group evolution of the gauge couplings for this case is shown in Fig. \ref{Figure_Gauge_Couplings} on the right, while the running of the ratio $|(Y_E)_{22}/(Y_D)_{22}|$ together with the plot of the RGI $I_3$ are presented in Fig. \ref{Figure_I3_Exotics} on the right. We see that the RGI $I_3$ may be used for constructing a rough estimate of this ratio. Again, the initial conditions were imposed at $M_{\mbox{\scriptsize SUSY}}=4\,\mbox{GeV}$. The ratio (\ref{Y_UD_Ratio}) appears to be close to 6 for a wide range of $\mbox{tg}\,\beta$. We chose the value $\mbox{tg}\,\beta=27.2$, because in this case the Yukawa relations at the unification scale resemble the equations

\begin{equation}\label{Unification_Exotics_Option2}
\frac{1}{3} (Y_U)_{33} = (Y_{D})_{33} = \frac{2}{3} (Y_{E})_{33};\qquad 2\,(Y_U)_{22} = (Y_{D})_{22} = \frac{1}{\sqrt{24}} (Y_{E})_{22}.
\end{equation}

\noindent
Therefore, in this case it is convenient to introduce the redefined Yukawa couplings as

\begin{eqnarray}\label{New_Yukawa_Exotics_Option2}
&& (y_U)_{33} \equiv \frac{1}{3} \left|(Y_U)_{33}\right|;\qquad (y_D)_{33} \equiv \left|(Y_D)_{33}\right|;\qquad (y_E)_{33} \equiv \frac{2}{3} \left|(Y_E)_{33}\right|;\qquad\nonumber\\
&& (y_U)_{22} \equiv 2\,\left|(Y_U)_{22}\right|;\qquad\, (y_D)_{22} \equiv \left|(Y_D)_{22}\right|;\qquad (y_E)_{22} \equiv \frac{1}{\sqrt{24}} \left|(Y_E)_{22}\right|.
\end{eqnarray}

\begin{figure}[!h]
\hspace*{-4mm}
\begin{picture}(0,15)
\put(1.1,8){\includegraphics[scale=0.34]{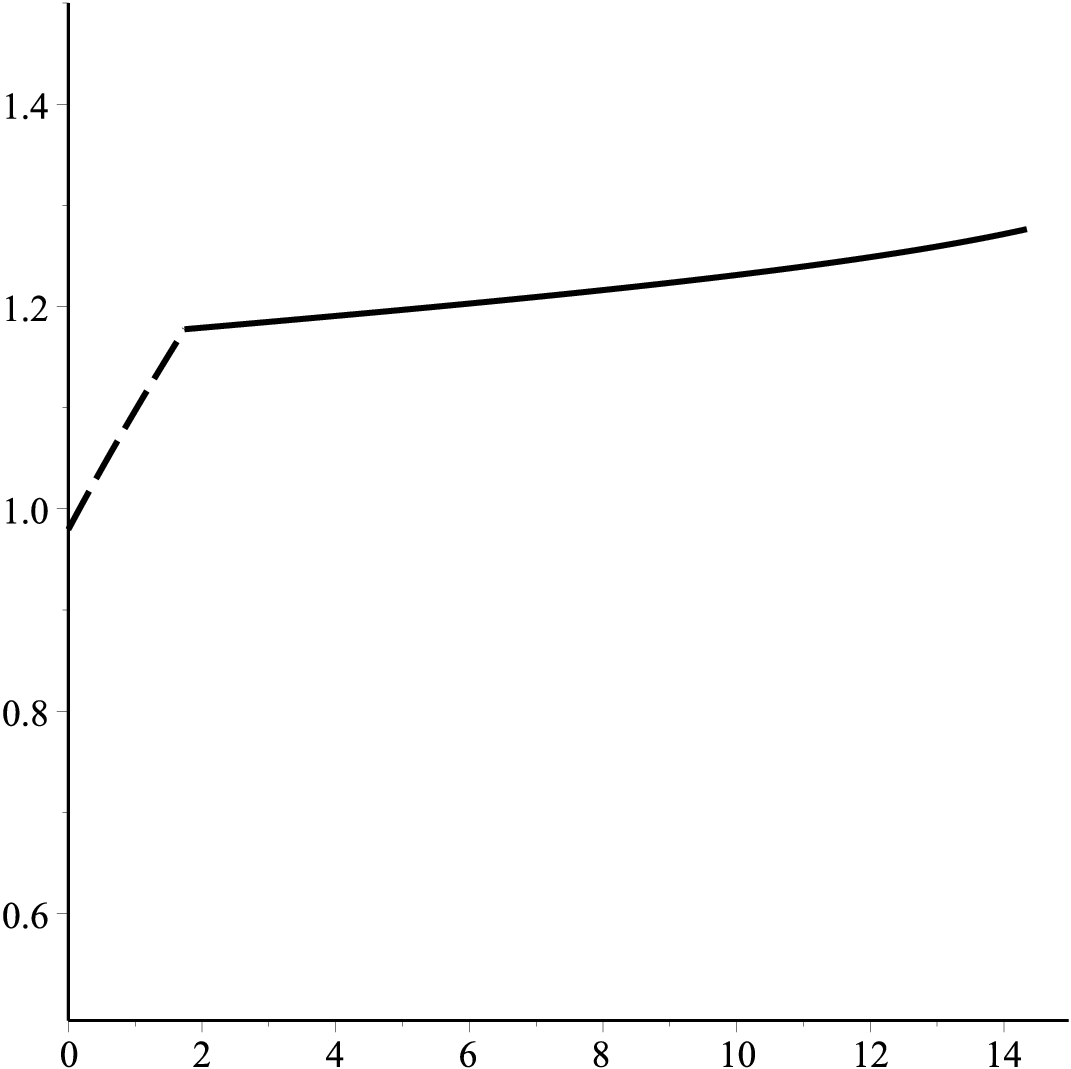}}
\put(8.8,8){\includegraphics[scale=0.34]{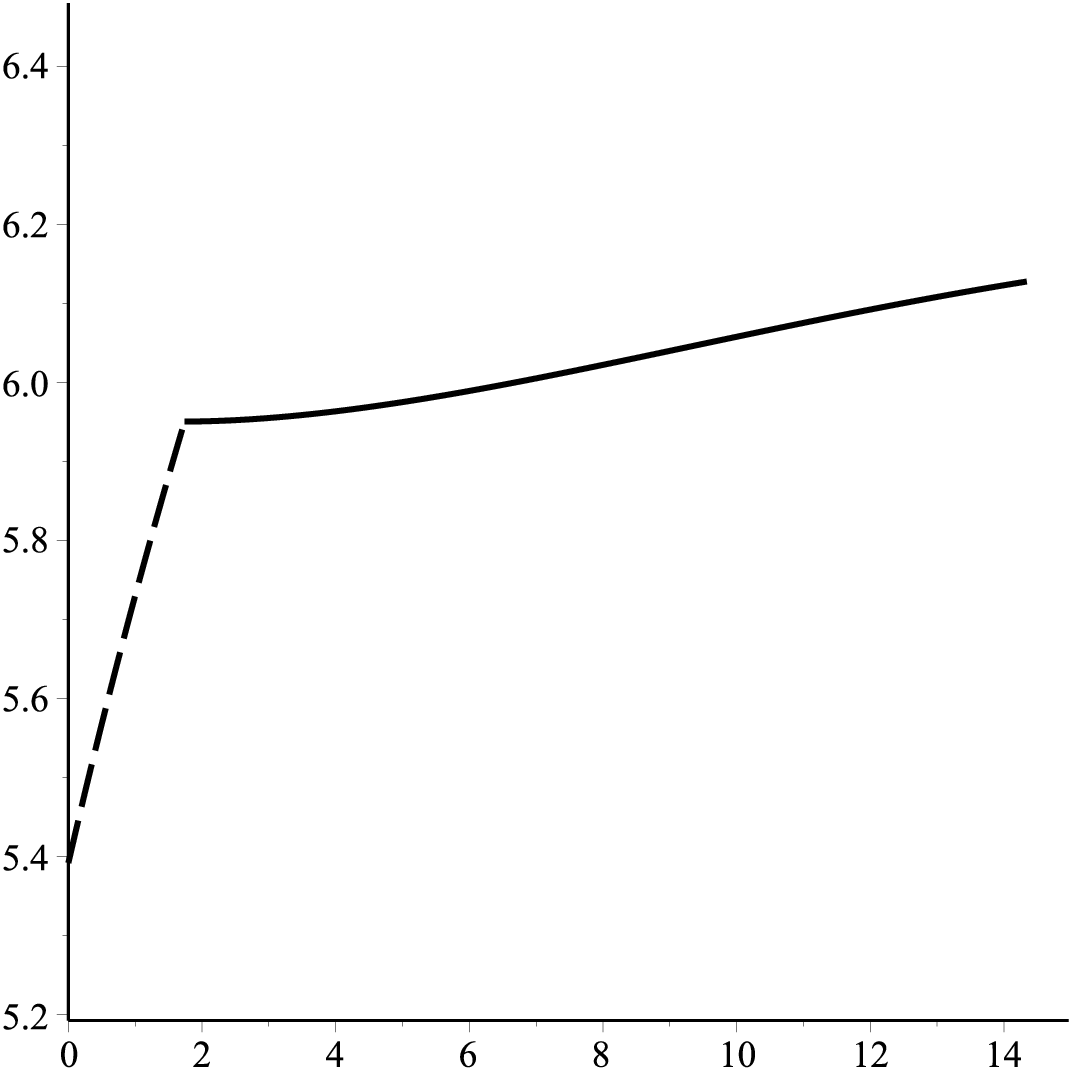}}
\put(0,14.0){${\displaystyle \frac{I_1}{\mbox{tg}^2\beta}}$}
\put(6.4,7.5){$\log_{10}\frac{\mu}{m_Z}$}
\put(7.0,14){${\displaystyle (3.27)^{4/3} I_2}$}
\put(14,7.5){$\log_{10}\frac{\mu}{m_Z}$}
\put(1.0,0.5){\includegraphics[scale=0.34]{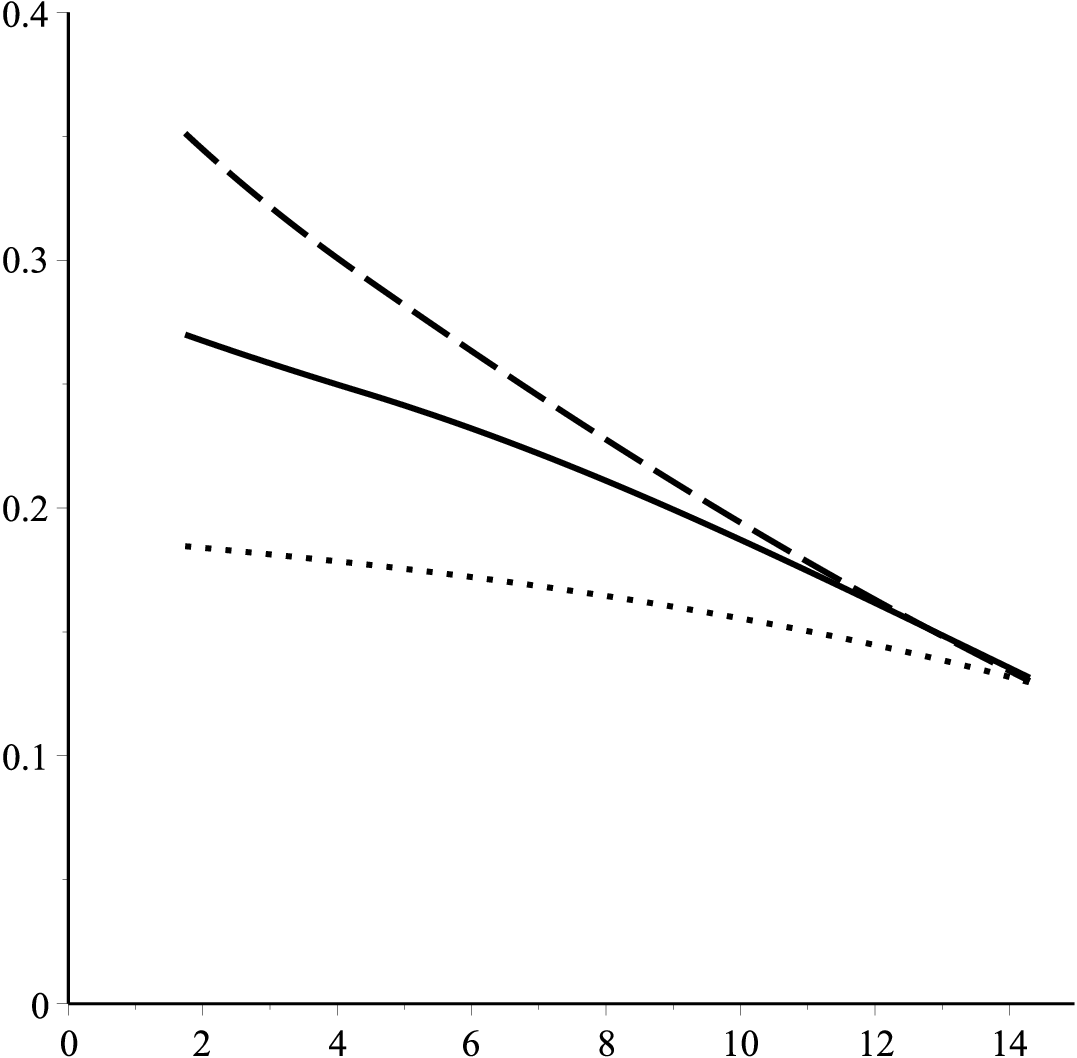}}
\put(8.6,0.5){\includegraphics[scale=0.34]{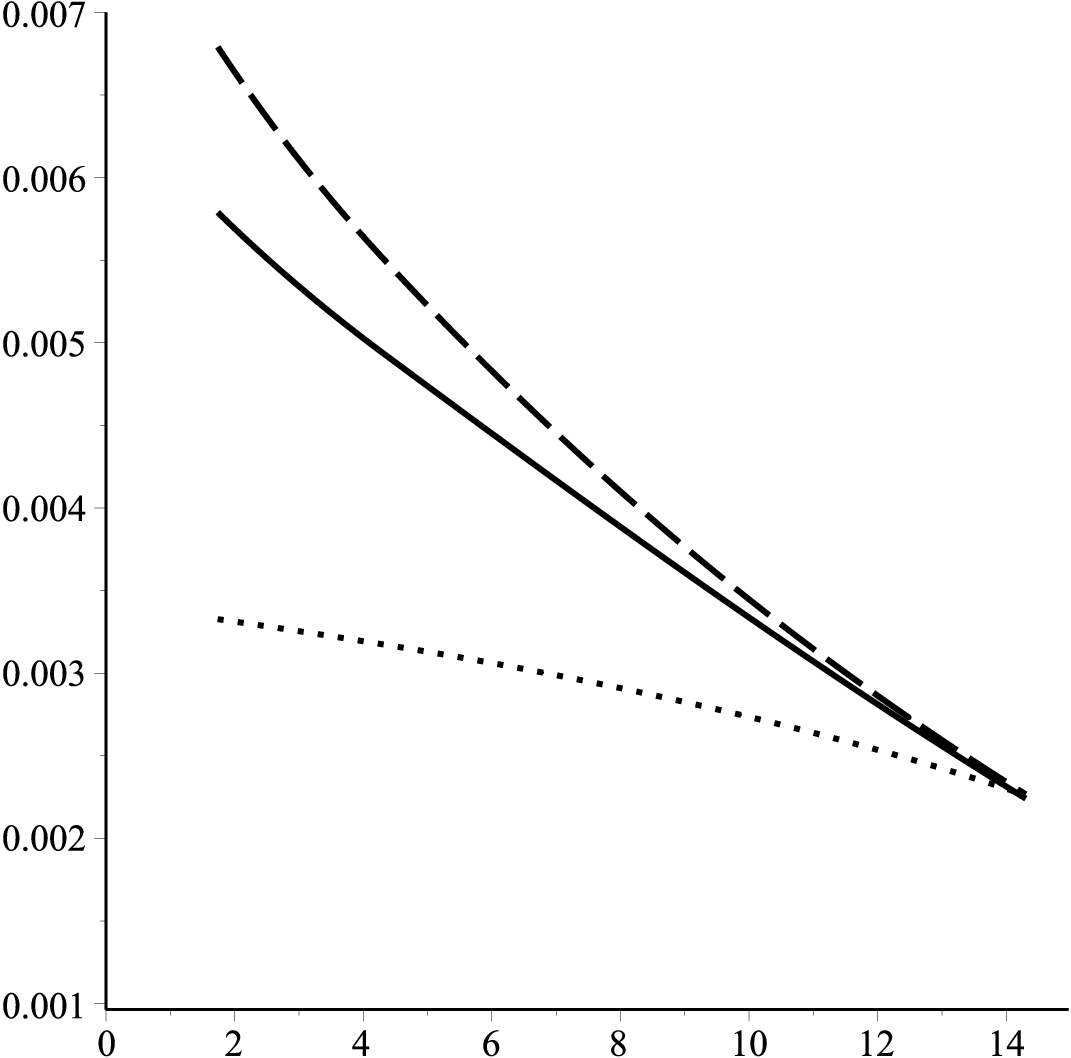}}
\put(0.4,6.5){$y_{33}$}
\put(6.4,0){$\log_{10}\frac{\mu}{m_Z}$}
\put(8.1,6.5){$y_{22}$}
\put(13.8,0){$\log_{10}\frac{\mu}{m_Z}$}
\end{picture}
\caption{The renormalization group running of the approximate RGIs and Yukawa couplings (for the third and second generations) for the extension of the MSSM with $n_5=6$, $M=2\cdot 10^6$ GeV, $\mbox{tg}\,\beta = 27.2$ and $x=3$. The solid, dashed, and dotted lines correspond to $y_U$, $y_D$, and $y_E$ (given by Eq. (\ref{New_Yukawa_Exotics_Option2})), respectively. The apparent Yukawa unification implies that Eq. (\ref{Unification_Exotics_Option2}) holds at the unification scale.}\label{Figure_Yukawa_Running_Exotics_Option2}
\end{figure}

\medskip

The plots of the invariants $I_1$ and $I_2$ (again, the latter is normalized in such a way that at the unification scale we obtain the value of the ratio (\ref{Y_UD_Ratio})) are presented in Fig. \ref{Figure_Yukawa_Running_Exotics_Option2} at the top left and right, respectively. The scale dependence of the approximate RGIs is characterized by the values $\Delta_i$, which in this case take the values

\begin{equation}\label{Delta_Exotics_Option2}
\Delta_1\approx 1.08;\qquad \Delta_2\approx 1.13;\qquad \Delta_3\approx 0.89.
\end{equation}

The renormalization group running of the Yukawa couplings (\ref{New_Yukawa_Exotics_Option2}) of the third and second generations is presented in Fig. \ref{Figure_Yukawa_Running_Exotics_Option2} at the bottom left and right, respectively. The apparent unification implies that Eq. (\ref{Unification_Exotics_Option2}) is satisfied at the unification scale. However, it is at least very difficult to construct a theory which produces numerical constants present in the corresponding Yukawa relations. Presumably, such relations may be obtained if the superpotential is constructed from sufficiently small representations, but includes some non-renormalizable terms of the fourth or higher degree in the chiral matter superfields. Such superpotentials were often considered in the literature, see, e.g., \cite{Altarelli:2000fu,Antusch:2009gu,Antusch:2013rxa,Bork:2021mmm,Lakhal:2025nbh} and \cite{Raby:2017ucc} for more references.

In fact, analyzing the running of the Yukawa couplings, we see that the expressions

\begin{equation}\label{Y_UD_Ratio_2}
\bigg|\frac{(Y_{U})_{33}}{(Y_D)_{33}}\cdot \frac{(Y_{D})_{22}}{(Y_U)_{22}}\bigg|\qquad \mbox{and}\qquad \bigg|\frac{(Y_E)_{22}}{(Y_D)_{22}}\bigg|
\end{equation}

\noindent
seem to be rather large (about $6$ and $4.5$, respectively), at least, if the thresholds and the Yukawa interaction of exotic additional superfields is neglected. Possibly, such values could somehow be derived from certain group invariants. However, at present we do not know how to do this. Nevertheless, it is possible to present an example of an invariant which produces similar relations with smaller values of these ratios. In particular, in the next section we demonstrate that similar Yukawa relations can be derived from the $E_6$ invariant $\xbar{27}\,\times \,\xbar{27}\,\times \,\xbar{351}\,'$, separately for the third and second generations.

\section{Derivation of the Yukawa relations with large ratios (\ref{Y_UD_Ratio_2}) from the $E_6$ invariant $\,\xbar{27}\,\times \,\xbar{27}\,\times \,\xbar{351}\,'$}
\hspace*{\parindent}\label{Section_Yukawa_Relation}

Certainly, it would be desirable to present a satisfactory model producing the Yukawa relations similar to those discussed above. At present, we did not managed to do this. Nevertheless, it would be interesting to investigate if these or similar relations can in principle be derived starting from the group theory arguments. In this section we demonstrate that relations similar to, e.g., (\ref{Unification_Exotics_Option1}) can be obtained {\it separately} for the third and second generations from the $E_6$ invariant $\,\xbar{27}\,\times \,\xbar{27}\,\times \,\xbar{351}\,'$, but the ratios of the Yukawa couplings have different values. Moreover, the Higgs superfields will be different for the third and second generations, so that the results cannot be directly applied for the phenomenological purposes. However, here we are not yet trying to construct a model describing physics beyond the Standard Model. The purpose is simply to demonstrate the very existence of an invariant which is able to produce, at least, qualitatively, equations similar to those which follows from the renormalization group analysis.

Let us consider a theory with the gauge group $E_6$ and two its invariants $\,\xbar{27}\,\times \,\xbar{27}\,\times \,\xbar{351}\,'$ for the third and second generations, with the corresponding Yukawa couplings $Y_3$ and $Y_2$, respectively. The main information about the group $E_6$ can be found in \cite{Slansky:1981yr}, while the explicit form of the commutation relations and generators of the fundamental representation are presented in, e.g., \cite{Stepanyantz:2023vat,Stepanyantz:2024ukh}. We will assume the simplest symmetry breaking pattern

\begin{equation}
E_6\ \to\ SO(10)\ \to\ SU(5)\ \to\ SU(3)\times SU(2)\times U(1).
\end{equation}

\noindent
In what follows we will need some branching rules starting from the representations $\,\xbar{27}\,$ and $\,\xbar{351}\,'$.  They can be written as

\begin{eqnarray}\label{Branchings_E_6_27}
&&\hspace*{-5mm} \,\xbar{27}\,\Big|_{E_6} = 1(-4) + 10(2) + \,\xbar{16}\,(-1)\Big|_{SO(10)\times U(1)};\\
\label{Branchings_E_6_351'}
&&\hspace*{-5mm} \,\xbar{351}\,'\Big|_{E_6} = 1(8) + 10(2) + 16(5) + 54(-4) + 126(2) + \,\xbar{144}\,(-1)\Big|_{SO(10)\times U(1)};\\
&&\vphantom{1}\nonumber\\
\label{Branchings_SO(10)_10}
&&\hspace*{-5mm} 10\Big|_{SO(10)} = 5(2) + \,\xbar{5}\,(-2)\Big|_{SU(5)\times U(1)};\\
\label{Branchings_SO(10)_16}
&&\hspace*{-5mm} \,\xbar{16}\,\Big|_{SO(10)} = 1(5)+5(-3)+\,\xbar{10}\,(1)\Big|_{SU(5)\times U(1)};\\
\label{Branchings_SO(10)_126}
&&\hspace*{-5mm} 126\Big|_{SO(10)} = 1(-10)+\,\xbar{5}\,(-2)+10(-6)+\,\xbar{15}\,(6)+45(2)+\,\xbar{50}\,(-2)\Big|_{SU(5)\times U(1)};\\
\label{Branchings_SO(10)_144}
&&\hspace*{-5mm} \,\xbar{144}\,\Big|_{SO(10)} = 5(-3)+\,\xbar{5}\,(-7)+\,\xbar{10}\,(1)+\,\xbar{15}\,(1)+24(5)+\,\xbar{40}\,(1)+45(-3)\Big|_{SU(5)\times U(1)};\\
&&\vphantom{1}\nonumber\\
\label{Branchings_SU(5)_5}
&&\hspace*{-5mm} 5\Big|_{SU(5)} = (1,2)(3)+(3,1)(-2)\Big|_{SU(3)\times SU(2)\times U(1)};\\
\label{Branchings_SU(5)_Bar5}
&&\hspace*{-5mm} \,\xbar{5}\,\Big|_{SU(5)} = (1,2)(-3)+(\,\xbar{3}\,,1)(2)\Big|_{SU(3)\times SU(2)\times U(1)};\\
\label{Branchings_SU(5)_10}
&&\hspace*{-5mm} \,\xbar{10}\,\Big|_{SU(5)} = (1,1)(-6)+(3,1)(4)+(\,\xbar{3}\,,2)(-1)\Big|_{SU(3)\times SU(2)\times U(1)};\\
&&\hspace*{-5mm} 45\Big|_{SU(5)} = (1,2)(3)+(3,1)(-2)+(3,3)(-2)+(\,\xbar{3}\,,1)(8)
\nonumber\\
&&\hspace*{-5mm}\qquad\qquad\qquad\qquad\qquad\qquad\ \ +(\,\xbar{3}\,,2)(-7)+(\,\xbar{6}\,,1)(-2)+(8,2)(3)\Big|_{SU(3)\times SU(2)\times U(1)}.\qquad
\end{eqnarray}

The invariant under consideration written in terms of the $SU(3)\times SU(2)\times U(1)$ superfields contains some terms which can be interpreted as certain parts in the MSSM superpotential (\ref{Superpotential_For_MSSM}). Evidently, the corresponding coefficients can be expressed in terms of $Y_3$ and $Y_2$ for the third and second generations, respectively. The way for obtaining these terms is presented in Table \ref{Table_Branching3} for the third generation and in Table \ref{Table_Branching2} for the second generation. In particular, in these tables we present the expressions for coefficients of various intermediate (for the subgroups $SO(10)$, $SU(5)$) and final (for the subgroup $SU(3)\times SU(2)\times U(1)$ corresponding to the MSSM) invariants in terms of the original couplings $Y_3$ and $Y_2$, which are the coefficients of the original $\,\xbar{27}\,\times \,\xbar{27}\,\times \,\xbar{351}\,'$ invariants.

\begin{table}[!h]
\hspace*{-4mm}
\begin{picture}(0,6.3)
\put(0.2,6.1){\line(1,0){15.5}}
\put(0.2,6.1){\line(0,-1){6}}
\put(1.8,6.1){\line(0,-1){6}}
\put(15.7,6.1){\line(0,-1){6}}
\put(0.7,5.2){$E_6$} \put(7,5.2){$Y_3\cdot\,\xbar{27}\,\times \,\xbar{27}\,\times \,\xbar{351}\,'$}
\put(0.2,4.6){\line(1,0){15.5}}
\put(4.4,4.6){\line(0,-1){4.5}}
\put(10.7,4.6){\line(0,-1){4.5}}
\put(0.35,3.7){$SO(10)$} \put(2.0,3.7){$\,\xbar{16}\,\times \,\xbar{16}\,\times 10$}  \put(5.9,3.7){${\displaystyle \frac{1}{4} Y_3\cdot\,\xbar{16}\,\times \,\xbar{16}\,\times 126}$} \put(11.3,3.7){$\sqrt{2}\, Y_3\cdot 10 \times \,\xbar{16}\,\times \,\xbar{144}\,$}
\put(0.2,3.1){\line(1,0){15.5}}
\put(6.7,3.1){\line(0,-1){3.0}}
\put(0.45,2.2){$SU(5)$} \put(2.0,2.2){$\,\xbar{10}\,\times \,\xbar{10}\,\times \,\xbar{5}\,$}
\put(4.6,2.2){$5 \times \,\xbar{10}\,\times 45$} \put(6.9,2.2){${\displaystyle -\frac{1}{\sqrt{3}} Y_3 \cdot\,\xbar{10}\,\times \,\xbar{10}\,\times \,\xbar{5}\,}$}
\put(11.3,2.2){${\displaystyle -\frac{1}{\sqrt{10}}\,Y_3 \cdot 5 \times \,\xbar{10}\,\times 5}$}
\put(0.2,1.6){\line(1,0){15.5}}
\put(0.4,0.7){MSSM} \put(7.2,0.7){${\displaystyle (Y_U)_{33} = -\frac{1}{\sqrt{3}} Y_3}$} \put(10.8,0.7){${\displaystyle (Y_E)_{33} = (Y_D)_{33} = -\frac{1}{\sqrt{10}} Y_3}$}
\put(0.2,0.1){\line(1,0){15.5}}
\end{picture}
\caption{The branchings used for producing superfields of the third generation}\label{Table_Branching3}
\end{table}

\begin{table}[!h]
\hspace*{-4mm}
\begin{picture}(0,6.3)
\put(0.2,6.1){\line(1,0){15.5}}
\put(0.2,6.1){\line(0,-1){6}}
\put(1.8,6.1){\line(0,-1){6}}
\put(15.7,6.1){\line(0,-1){6}}
\put(0.7,5.2){$E_6$} \put(7,5.2){$Y_2\cdot\,\xbar{27}\,\times \,\xbar{27}\,\times \,\xbar{351}\,'$}
\put(0.2,4.6){\line(1,0){15.5}}
\put(5.8,4.6){\line(0,-1){4.5}}
\put(13.1,4.6){\line(0,-1){4.5}}
\put(0.35,3.7){$SO(10)$} \put(2.0,3.7){${\displaystyle \frac{1}{4\sqrt{5}} Y_2\cdot \,\xbar{16}\,\times \,\xbar{16}\,\times 10}$}  \put(7.5,3.7){${\displaystyle \frac{1}{4} Y_2\cdot\,\xbar{16}\,\times \,\xbar{16}\,\times 126}$} \put(13.2,3.7){$10 \times \,\xbar{16}\,\times \,\xbar{144}\,$}
\put(0.2,3.1){\line(1,0){15.5}}
\put(10.8,3.1){\line(0,-1){3.0}}
\put(0.45,2.2){$SU(5)$} \put(1.9,2.2){${\displaystyle-\frac{1}{\sqrt{10}} Y_2\cdot \,\xbar{10}\,\times \,\xbar{10}\,\times \,\xbar{5}\,}$}
\put(6.9,2.2){$2Y_2\cdot 5 \times \,\xbar{10}\,\times 45$} \put(10.9,2.2){$\,\xbar{10}\,\times \,\xbar{10}\,\times \,\xbar{5}\,$}
\put(13.5,2.2){$5 \times \,\xbar{10}\,\times 5$}
\put(0.2,1.6){\line(1,0){15.5}}
\put(0.4,0.7){MSSM} \put(2.25,0.7){${\displaystyle (Y_U)_{22} = -\frac{1}{\sqrt{10}} Y_2}$} \put(5.9,0.7){${\displaystyle \frac{1}{3} (Y_E)_{22} = - (Y_D)_{22} =  \frac{1}{\sqrt{3}} Y_2}$}
\put(0.2,0.1){\line(1,0){15.5}}
\end{picture}
\caption{The branchings used for producing superfields of the second generation}\label{Table_Branching2}
\end{table}

\begin{table}[!h]
\begin{center}
\begin{tabular}{|c|c|c|c||c|c|c|c|}
\hline
\ \parbox{2cm}{The third\\ generation}\ & $\,SU(5)\,$ ${\displaystyle\vphantom{\int\limits_p^d}}$ & $SO(10)$ & $\quad E_6\quad $ & \parbox{2cm}{The second\\ generation} & $\,SU(5)\,$ & $SO(10)$ & $\quad E_6\quad $ \\
\hline
\hline
$D_3$,\ $L_3$ ${\displaystyle\vphantom{\frac{1}{2}}}$ & $5$ & $10$ & $\xbar{27}$ & $D_2$,\ $L_2$ ${\displaystyle\vphantom{\frac{1}{2}}}$ & $5$ & $\xbar{16}$ & $\xbar{27}$\\
\hline
$Q_3$,\,$U_3$,\,$E_3$ ${\displaystyle\vphantom{\frac{1}{2}}}$ & $\xbar{10}$ & $\xbar{16}$ & $\xbar{27}$ & $Q_2$,\,$U_2$,\,$E_2$ ${\displaystyle\vphantom{\frac{1}{2}}}$ & $\xbar{10}$ & $\xbar{16}$ & $\xbar{27}$\\
\hline
$(H_u)_3$ ${\displaystyle\vphantom{\frac{1}{2}}}$ & $\xbar{5}$ & $126$ & $\,\xbar{351}\,{}'$ & $(H_u)_2$ ${\displaystyle\vphantom{\frac{1}{2}}}$ & $\xbar{5}$ & $10$ & $\,\xbar{351}\,{}'$\\
\hline
$(H_d)_3$ ${\displaystyle\vphantom{\frac{1}{2}}}$ & $5$ & $\xbar{144}$ & $\,\xbar{351}\,{}'$ & $(H_d)_2$ ${\displaystyle\vphantom{\frac{1}{2}}}$ & $45$ & $126$ & $\,\xbar{351}\,{}'$\\
\hline
\end{tabular}
\end{center}
\caption{This table illustrates how we take various low energy superfields for deriving the Yukawa relations (\ref{Yukawa_Unification_2})}\label{Table_Superfields}
\end{table}

Table \ref{Table_Superfields} illustrates how we construct the superfields of the third and second generations in order to derive the Yukawa relations (\ref{Yukawa_Unification_2}). In particular, it should be noticed that
the Higgs superfields in this case are different for the third and second generations. However, in this section we only illustrate the very possibility of deriving the Yukawa relations (\ref{Yukawa_Unification_2}) using the group theory arguments without trying to construct any phenomenologically satisfactory model.

The $SO(10)$ invariants present in the Tables \ref{Table_Branching3} and \ref{Table_Branching2} are defined by the equations

\begin{eqnarray}\label{SO(10)_Normalization}
&& I_{10}\equiv \,\xbar{16}\,\times \,\xbar{16}\,\times 10 \equiv P^a (\Gamma_i B)_{ab} P^b A_i;\vphantom{\frac{1}{2}}\nonumber\\
&& I_{126}\equiv \,\xbar{16}\,\times \,\xbar{16}\,\times 126 \equiv \frac{1}{5!}\,P^a (\Gamma_{ijklm} B)_{ab} P^b A_{ijklm};\nonumber\\
&& I_{144}\equiv 10 \times \,\xbar{16}\,\times \,\xbar{144}\, \equiv P_i\,P^a A_{ia}.\vphantom{\frac{1}{2}}
\end{eqnarray}

\noindent
Here the representation $\,\xbar{27}\,$ is composed of the superfields $\big\{P,\,P_i,\,P^a\big\}$ in the representations $1$, $10$, and $\,\xbar{16}\,$ of the subgroup $SO(10)$. The index $i$ ranges from 1 to 10, while the index $a$ ranges from 1 to 16. The parts of the representation $\,\xbar{351}\,'$  corresponding to the representations $10$, $126$, and $144$ of the subgroup $SO(10)$ are denoted by $A_i$, $A_{ijklm}$, and $A_{ia}$, respectively. In our notation, the $D=10$ Euclidean $32\times 32$ gamma-matrices are denoted by $\bm{\Gamma}_i$. The details of their construction can be found in \cite{Stepanyantz:2023vat}. It is also useful to define the $32\times 32$ matrixes

\begin{equation}
\bm{\Gamma}_{11} \equiv -i\,\bm{\Gamma}_1\bm{\Gamma}_2\cdot\ldots\cdot\bm{\Gamma}_{10};\qquad \bm{B}\equiv \bm{\Gamma}_1\bm{\Gamma}_3\bm{\Gamma}_5\bm{\Gamma}_7\bm{\Gamma}_9,
\end{equation}

\noindent
which are analogs of $\gamma_5$ and the charge conjugation matrix $C$ for the usual $D=4$ case. The antisymmetrized products of $\bm{\Gamma}_i$ are denoted by $\bm{\Gamma}_{i_1 i_2\ldots i_k}$. Using them, we construct the $16\times 16$ matrices

\begin{eqnarray}
&& (\Gamma_{i_1i_2\ldots i_{2k}})_a{}^b \equiv \Big[\frac{1}{2}\bm{\Gamma}_{i_1i_2\ldots i_{2k}} (1+\bm{\Gamma}_{11})\Big]_a{}^{b};\qquad\nonumber\\
&& (\Gamma_{i_1i_2\ldots i_{2k+1}} B)_{ab} \equiv \Big[\frac{1}{2}\bm{\Gamma}_{i_1i_2\ldots i_{2k+1}} \bm{B} (1+\bm{\Gamma}_{11})\Big]_a{}^{b};\nonumber\\
&& (B\Gamma_{i_1i_2\ldots i_{2k+1}})^{ab} \equiv \Big[\frac{1}{2}\bm{B}\bm{\Gamma}_{i_1i_2\ldots i_{2k+1}} (1+\bm{\Gamma}_{11})\Big]_{a}{}^{b}.\qquad
\end{eqnarray}

\noindent
More details and the explicit form of the invariants (\ref{SO(10)_Normalization}) in terms of $SU(5)$ representations can be found in Appendix \ref{Appendix_Invariants}. In particular, Appendix \ref{Appendix_Invariants} contains the explanation of how the normalization is fixed for superfields in various representations of the groups $SO(10)$ and $SU(5)$.

The $SU(5)$ invariants present in Tables \ref{Table_Branching3} and \ref{Table_Branching2} are defined by the equation

\begin{eqnarray}\label{SU(5)_Normalization}
&& \,\xbar{10}\,\times \,\xbar{10}\,\times \,\xbar{5}\,\equiv \frac{1}{8}\varepsilon_{\mu\nu\alpha\beta\gamma} \phi^{\mu\nu} \phi^{\alpha\beta} \,\xbar{5}\,{}^\gamma;\nonumber\\
&& 5 \times \,\xbar{10}\,\times 45 \equiv \frac{1}{2} \phi_\alpha\, \phi^{\mu\nu} 45^\alpha_{\mu\nu};\nonumber\\
&& 5 \times \,\xbar{10}\,\times 5 \equiv \phi_\mu\, \phi^{\mu\nu}\, 5_\nu.\vphantom{\frac{1}{2}}
\end{eqnarray}

\noindent
In our notation, the superfields $\phi$, $\phi_\mu$, and $\phi^{\mu\nu}$ in the representations $1$, $5$, and $\,\xbar{10}\,$, respectively, form the representation $\,\xbar{16}\,$ of the group $SO(10)$. The superfields $\,\xbar{5}\,{}^\mu$, $45^\alpha_{\mu\nu}$, and $5_\mu$ are the relevant parts of the $E_6$ representation $\,\xbar{351}'\,$ which transform as $\,\xbar{5}\,$, $45$, and $5$ with respect to the subgroup $SU(5)$.

The (rather complicated) derivation of the Yukawa relations summarized in Tables \ref{Table_Branching3} and \ref{Table_Branching2} is described in detail in Appendix \ref{Appendix_Invariants}. Taking into account the arbitrariness in the choice of signs for various superfields, it is convenient to present the result in the form (see the last strings in the tables)

\begin{eqnarray}\label{Yukawa_Relation_351_3}
&& \left|(Y_E)_{33}\right| = \left|(Y_D)_{33}\right| = \sqrt{\frac{3}{10}} \left|(Y_U)_{33}\right| = \frac{1}{\sqrt{10}} \left|Y_3\right|;\\
\label{Yukawa_Relation_351_2}
&& \frac{1}{3} \left|(Y_E)_{22}\right| = \left|(Y_D)_{22}\right| = \sqrt{\frac{10}{3}} \left|(Y_U)_{22}\right| = \frac{1}{\sqrt{3}} \left|Y_2\right|.\qquad
\end{eqnarray}

\noindent
We see that these equations are similar to those which follows from the renormalization group analysis in a sense that they correctly indicate which Yukawa constants are larger and which are smaller. However, the numerical values of the Yukawa ratios at the unification scale do not agree with the ones follows from the analysis made in Sects.~\ref{Section_Yukawa_MSSM} and \ref{Section_Yukawa_With_Extra_Matter}. For instance, the ratios (\ref{Y_UD_Ratio_2}) are now equal to $10/3$ and 3 instead of (approximate) $6$ and $4.5$ suggested by the renormalization group analysis.

\section{Conclusion}
\hspace*{\parindent}

In this paper we have constructed some expressions which do not receive very large quantum corrections in the MSSM and some of its modifications. These expressions may possibly be used to analyse some aspects of Grand Unified Theories, for instance, the unification of the Yukawa couplings. Although for the MSSM one can construct expressions that are scale independent in all orders \cite{Rystsov:2024soq}, for the phenomenological purposes it seems more preferable to investigate approximate RGIs. They are simpler, but slightly depend on the renormalization point. The RGIs constructed in this paper are given by the expressions (\ref{Yukawa_RGI1}), (\ref{Yukawa_RGI2}), (\ref{I3_RGI}), and (\ref{I3_Tilde_RGI}). (The RGI (\ref{I3_Tilde_RGI}) was derived for the MSSM only.) The RGI $I_2$ has been constructed under the assumption that the ratio $|(Y_U)_{33}/(Y_D)_{33}|$ does not significantly deviate from its value at the unification scale denoted by $x$. The renormalization group analysis demonstrates that the expressions for the RGIs depend on scale, but this dependence is rather small. Ignoring this scale dependence, it is possible to estimate a certain combination of the Yukawa couplings at the unification scale and find an approximate relation between the values of $\mbox{tg}\,\beta$ and $x$ (given by Eq. (\ref{Tg_Beta})). The use of the approximate RGIs constructed in this paper allows one to guess some equations (see, e.g., Eqs. (\ref{Yukawa_Unification_1}) and (\ref{Yukawa_Unification_2})) relating Yukawa couplings at the unification scale. Presumably, these relations should follow from certain expressions invariant under the gauge group of a Grand Unified Theory. However, it is rather difficult to construct such expressions. Another possibility is to tune thresholds or add extra superfields to achieve better ratios between the Yukawa couplings. Note that, to avoid breaking of the gauge coupling unification, it is necessary for the new superfields to form complete $SU(5)$ multiplets. Under the assumption that the Yukawa interaction of the additional superfields can be neglected, we considered two variants in which the ratios of the Yukawa couplings at the unification scale are given by not so complex fractions or square roots. However, even in this case we did not manage to construct such group invariants that can produce the satisfactory Yukawa relations. Nevertheless, it is possible to present an invariant which is able to give the relations which qualitatively resemble the ones obtained by solving the renormalization group equations. However, the numerical values for the ratios of the Yukawa couplings obtained by the group theory methods do not agree with the ones obtained from the renormalization group analysis. Moreover, Higgs superfields are in this case different for different generations, so that it seems impossible to construct a phenomenologically satisfactory model based on this invariant.

Thus, the RGIs constructed in this paper can be used for analysing the Yukawa relations at the unification scale, but it is very difficult to construct a phenomenologically acceptable model. Probably, such a model may involve some extra superfields or include invariants more than cubic in chiral matter superfields. Such invariants break renormalizability, but can effectively appear in renormalizable theories after integrating out some massive superfields, see, e.g., \cite{Veltman:1980mj} or in supergravity theories \cite{Nilles:1983ge,Lahanas:1986uc}. Certainly, both threshold corrections and Yukawa interaction of extra superfields significantly affect running of the Yukawa couplings and can be crucial for the Yukawa unification. We hope to analyze these interesting issues in future research.

\section*{Acknowledgements}
\hspace*{\parindent}

K.S. is very grateful to S.S.Aleshin for valuable discussions.

The work of D.R. was supported by Foundation for Advancement of Theoretical Physics and Mathematics ``BASIS'', grant No. 25-2-1-36-1.

\appendix

\section*{Appendix}

\section{Two-loop $\beta$-functions and one-loop anomalous dimensions for the MSSM, its extensions and for the Standard Model}
\hspace*{\parindent}\label{Appendix_Lowest_RGFs}

The two-loop MSSM $\beta$-functions can be written in the form (see, e.g., \cite{Bjorkman:1985mi,Jack:2004ch})

\begin{eqnarray}\label{Beta1_2Loops}
&& \frac{\beta_1(\alpha,Y)}{\alpha_{1}^2} = - \frac{1}{2\pi}\cdot \frac{3}{5} \bigg\{ -11 -\frac{199\alpha_{1}}{60\pi} -\frac{9\alpha_{2}}{4\pi} -\frac{22\alpha_{3}}{3\pi}
\nonumber\\
&&\qquad\qquad\qquad + \frac{1}{8\pi^2} \mbox{tr}\Big(\frac{13}{3} Y_{U} Y_{U}^+ + \frac{7}{3} Y_{D} Y_{D}^+ + 3 Y_{E} Y_{E}^+ \Big)\bigg\} + O\Big(\alpha^2,\alpha Y^2, Y^4\Big);\qquad\vphantom{\frac{1}{2}}\\
\label{Beta2_2Loops}
&& \frac{\beta_2(\alpha,Y)}{\alpha_{2}^2} = - \frac{1}{2\pi} \bigg\{-1  -\frac{9\alpha_{1}}{20\pi}- \frac{25\alpha_{2}}{4\pi} -\frac{6\alpha_{3}}{\pi}  \nonumber\\
&&\qquad\qquad\qquad\quad + \frac{1}{8\pi^2}\,
\mbox{tr}\Big(3\, Y_{U} Y_{U}^+ + 3\, Y_{D} Y_{D}^+ + Y_{E} Y_{E}^+ \Big)\bigg\} + O\Big(\alpha^2,\alpha Y^2, Y^4\Big);\\
\label{Beta3_2Loops}
&& \frac{\beta_3(\alpha,Y)}{\alpha_{3}^2} = - \frac{1}{2\pi} \bigg\{3 -\frac{11\alpha_{1}}{20\pi} -\frac{9\alpha_{2}}{4\pi} -\frac{7\alpha_{3}}{2\pi}
\nonumber\\
&&\qquad\qquad\qquad\qquad\qquad\quad\, + \frac{1}{8\pi^2}\, \mbox{tr}\Big(2\, Y_{U} Y_{U}^+ + 2\, Y_{D} Y_{D}^+\Big)\bigg\} + O\Big(\alpha^2,\alpha Y^2, Y^4\Big).
\end{eqnarray}

\noindent
Similarly, the one-loop anomalous dimensions of the chiral MSSM superfields are given by the expressions

\begin{eqnarray}\label{Gamma_Q_1Loop}
&&\hspace*{-7mm} \gamma_Q(\alpha,Y)^T = - \frac{\alpha_1}{60\pi} - \frac{3\alpha_2}{4\pi} - \frac{4\alpha_3}{3\pi} + \frac{1}{8\pi^2}\Big(Y_{U} Y_{U}^+ + Y_{D} Y_{D}^+\Big)
+ O\Big(\alpha^2,\alpha Y^2, Y^4\Big);\qquad\\
\label{Gamma_U_1Loop}
&&\hspace*{-7mm} \gamma_U(\alpha,Y) = - \frac{4 \alpha_{1}}{15 \pi} - \frac{4\alpha_{3}}{3\pi} + \frac{1}{4\pi^2}\, Y_{U}^+ Y_{U}
+ O\Big(\alpha^2,\alpha Y^2, Y^4\Big);\\
\label{Gamma_D_1Loop}
&&\hspace*{-7mm} \gamma_D(\alpha,Y) = - \frac{\alpha_{1}}{15\pi} - \frac{4\alpha_{3}}{3\pi} + \frac{1}{4\pi^2}\, Y_{D}^+ Y_{D}
+ O\Big(\alpha^2,\alpha Y^2, Y^4\Big);\\
\label{Gamma_L_1Loop}
&&\hspace*{-7mm} \gamma_L(\alpha,Y)^T = - \frac{3\alpha_{1}}{20\pi} - \frac{3\alpha_{2}}{4\pi} + \frac{1}{8\pi^2}\, Y_{E} Y_{E}^+
+ O\Big(\alpha^2,\alpha Y^2, Y^4\Big);\\
\label{Gamma_E_1Loop}
&&\hspace*{-7mm} \gamma_E(\alpha,Y) = - \frac{3\alpha_{1}}{5\pi} + \frac{1}{4\pi^2}\, Y_{E}^+ Y_{E} + O\Big(\alpha^2,\alpha Y^2, Y^4\Big);\\
\label{Gamma_Hu_1Loop}
&&\hspace*{-7mm} \gamma_{H_u}(\alpha,Y) = - \frac{3\alpha_{1}}{20\pi} - \frac{3\alpha_{2}}{4\pi} + \frac{3}{8\pi^2}\, \mbox{tr}\Big(Y_{U}^+ Y_{U}\Big) + O\Big(\alpha^2,\alpha Y^2, Y^4\Big);\\
\label{Gamma_Hd_1Loop}
&&\hspace*{-7mm} \gamma_{H_d}(\alpha,Y) = - \frac{3\alpha_{1}}{20\pi} - \frac{3\alpha_{2}}{4\pi} + \frac{1}{8\pi^2}\, \mbox{tr}\Big(3\, Y_{D}^+ Y_{D} + Y_{E}^+ Y_{E}\Big) + O\Big(\alpha^2,\alpha Y^2, Y^4\Big).
\end{eqnarray}

It is also expedient to consider a modification of the MSSM containing some superfields which can be obtained from complete $SU(5)$ multiplets with the help of the branching rules with respect to the subgroup $SU(3)\times SU(2)\times U(1)\subset SU(5)$. Let us assume that the theory contains $n_5$ superfields coming from the representations $5$ and $\,\xbar{5}\,$, $n_{10}$ superfields coming from the representations $10$ and $\,\xbar{10}\,$,
$n_{15}$ superfields coming from $15$ and $\,\xbar{15}\,$, and $n_{24}$ superfields coming from the representation $24$. Certainly, we also assume that the anomaly cancellation condition remains satisfied. The larger representations of $SU(5)$ have large values of $T(R)$ and as a rule lead to the negative value of the inverse coupling constant at the unification scale (provided that their masses are essentially smaller than this scale). If we neglect the Yukawa interaction of the new superfields, then the two-loop $\beta$-functions are modified as

\begin{equation}
\frac{\beta_i(\alpha,Y)}{\alpha_i^2} \to \frac{\beta_i(\alpha,Y)}{\alpha_i^2} + \Delta\Big(\frac{\beta_i}{\alpha_i^2}\Big),
\end{equation}

\noindent
where $i=1,2,3$ and the additional contributions are given by the expressions

\begin{eqnarray}\label{Delta_Beta1}
&&\hspace*{-8mm} \Delta\Big(\frac{\beta_1}{\alpha_1^2}\Big) = \frac{1}{2\pi} \bigg\{\frac{1}{2} n_5 + \frac{3}{2} n_{10} + \frac{7}{2} n_{15} + 5 n_{24}
+ n_5\Big(\frac{7\alpha_1}{120\pi} + \frac{9\alpha_2}{40\pi} + \frac{4\alpha_3}{15\pi}\Big) + n_{10}\Big(\frac{23\alpha_1}{40\pi}
\nonumber\\
&&\hspace*{-8mm} + \frac{3\alpha_2}{40\pi} + \frac{6\alpha_3}{5\pi}\Big)
+ n_{15}\Big(\frac{181\alpha_1}{120\pi} + \frac{147\alpha_2}{40\pi} + \frac{82\alpha_3}{15\pi}\Big)
+ n_{24}\Big(\frac{25\alpha_1}{12\pi} + \frac{15\alpha_2}{4\pi} + \frac{20\alpha_3}{3\pi}\Big)\bigg\};\\
\label{Delta_Beta2}
&&\hspace*{-8mm} \Delta\Big(\frac{\beta_2}{\alpha_2^2}\Big) = \frac{1}{2\pi} \bigg\{\frac{1}{2} n_5 + \frac{3}{2} n_{10} + \frac{7}{2} n_{15} + 5 n_{24}
+ n_5\Big(\frac{3\alpha_1}{40\pi} + \frac{7\alpha_2}{8\pi}\Big)
+ n_{10}\Big(\frac{\alpha_1}{40\pi} + \frac{21\alpha_2}{8\pi}
\nonumber\\
&&\hspace*{-8mm} + \frac{2\alpha_3}{\pi}\Big) + n_{15}\Big(\frac{49\alpha_1}{40\pi} + \frac{69\alpha_2}{8\pi} + \frac{2\alpha_3}{\pi}\Big)
+ n_{24}\Big(\frac{5\alpha_1}{4\pi} + \frac{45\alpha_2}{4\pi} + \frac{4\alpha_3}{\pi}\Big)\bigg\};\\
\label{Delta_Beta3}
&&\hspace*{-8mm} \Delta\Big(\frac{\beta_3}{\alpha_3^2}\Big) = \frac{1}{2\pi} \bigg\{\frac{1}{2} n_5 + \frac{3}{2} n_{10} + \frac{7}{2} n_{15} + 5 n_{24}
+ n_5\Big(\frac{\alpha_1}{30\pi} + \frac{17\alpha_3}{12\pi}\Big)
+ n_{10}\Big(\frac{3\alpha_1}{20\pi} + \frac{3\alpha_2}{4\pi}
\nonumber\\
&&\hspace*{-8mm} + \frac{17\alpha_3}{4\pi}\Big) + n_{15}\Big(\frac{41\alpha_1}{60\pi} + \frac{3\alpha_2}{4\pi} + \frac{179\alpha_3}{12\pi}\Big)
+ n_{24}\Big(\frac{5\alpha_1}{6\pi} + \frac{3\alpha_2}{2\pi} + \frac{115\alpha_3}{6\pi}\Big)\bigg\}.
\end{eqnarray}

\noindent
They were calculated using the NSVZ equations and the one-loop expressions for the anomalous dimensions in theories with multiple gauge couplings presented in \cite{Korneev:2021zdz}. The terms containing $n_5$ and $n_{10}$ in Eqs. (\ref{Delta_Beta1}) --- (\ref{Delta_Beta3}) exactly agree with the result presented in \cite{Ghilencea:1997mu,Jack:2004ch}.

\medskip

For completeness, we also present the two-loop $\beta$-functions and the one-loop anomalous dimensions of the Yukawa couplings for the Standard Model. They were taken from Ref. \cite{Arason:1991ic} and are used for investigating the renormalization group running of various parameters below the supersymmetric threshold. In our notation the Standard Model $\beta$-functions are written as

\begin{eqnarray}\label{SM_Beta1_2Loops}
&& \frac{\beta_1(\alpha,Y)}{\alpha_{1}^2} = - \frac{1}{2\pi}\cdot \frac{3}{5} \bigg\{ -\frac{41}{6} -\frac{199\alpha_{1}}{120\pi} -\frac{9\alpha_{2}}{8\pi} -\frac{11\alpha_{3}}{3\pi}
\nonumber\\
&&\qquad\qquad\ + \frac{1}{8\pi^2} \mbox{tr}\Big(\frac{17}{12}\, Y_{U} Y_{U}^+ + \frac{5}{12}\, Y_{D} Y_{D}^+ + \frac{5}{4}\, Y_{E} Y_{E}^+ \Big)\bigg\} + O\Big(\alpha^2,\alpha Y^2, Y^4\Big);\qquad\vphantom{\frac{1}{2}}\\
\label{SM_Beta2_2Loops}
&& \frac{\beta_2(\alpha,Y)}{\alpha_{2}^2} = - \frac{1}{2\pi} \bigg\{\frac{19}{6}  -\frac{9\alpha_{1}}{40\pi}- \frac{35\alpha_{2}}{24\pi} -\frac{3\alpha_{3}}{\pi}  \nonumber\\
&&\qquad\qquad\quad + \frac{1}{8\pi^2}\,
\mbox{tr}\Big(\frac{3}{4}\, Y_{U} Y_{U}^+ + \frac{3}{4}\, Y_{D} Y_{D}^+ + \frac{1}{4}\,Y_{E} Y_{E}^+ \Big)\bigg\} + O\Big(\alpha^2,\alpha Y^2, Y^4\Big);\\
\label{SM_Beta3_2Loops}
&& \frac{\beta_3(\alpha,Y)}{\alpha_{3}^2} = - \frac{1}{2\pi} \bigg\{7 -\frac{11\alpha_{1}}{40\pi} -\frac{9\alpha_{2}}{8\pi} +\frac{13\alpha_{3}}{2\pi}
\nonumber\\
&&\qquad\qquad\qquad\qquad\qquad\quad\ \ + \frac{1}{8\pi^2}\, \mbox{tr}\Big( Y_{U} Y_{U}^+ +  Y_{D} Y_{D}^+\Big)\bigg\} + O\Big(\alpha^2,\alpha Y^2, Y^4\Big).
\end{eqnarray}

Similarly, the one-loop anomalous dimensions of the Yukawa couplings for the Standard Model in our notation take the form

\begin{eqnarray}\label{SM_Gamma_Yu}
&& \gamma_{Y_U}(\alpha,Y) \equiv (Y_U)^{-1} \frac{d Y_U}{d\ln\mu} = -\frac{17\alpha_1}{80\pi} - \frac{9\alpha_2}{16\pi} - \frac{2\alpha_3}{\pi} +\frac{1}{16\pi^2}\bigg[ \mbox{tr}\Big(3Y_U^+ Y_U
\nonumber\\
&&\qquad\qquad\quad + 3 Y_D^+ Y_D + Y_E^+ Y_E\Big) +\frac{3}{2}\Big(Y_U^+ Y_U - Y_D^+ Y_D\Big)\bigg] +  O\Big(\alpha^2,\alpha Y^2, Y^4\Big);\qquad\\
\label{SM_Gamma_Yd}
&& \gamma_{Y_D}(\alpha,Y) \equiv (Y_D)^{-1} \frac{d Y_D}{d\ln\mu} = -\frac{\alpha_1}{16\pi} - \frac{9\alpha_2}{16\pi} - \frac{2\alpha_3}{\pi} +\frac{1}{16\pi^2}\bigg[ \mbox{tr}\Big(3Y_U^+ Y_U
\nonumber\\
&&\qquad\qquad + 3 Y_D^+ Y_D + Y_E^+ Y_E\Big) +\frac{3}{2}\Big(-Y_U^+ Y_U + Y_D^+ Y_D\Big) \bigg] +  O\Big(\alpha^2,\alpha Y^2, Y^4\Big);\qquad\\
\label{SM_Gamma_Ye}
&& \gamma_{Y_E}(\alpha,Y) \equiv (Y_E)^{-1} \frac{d Y_E}{d\ln\mu} = -\frac{9\alpha_1}{16\pi} - \frac{9\alpha_2}{16\pi} +\frac{1}{16\pi^2}\bigg[ \mbox{tr}\Big(3Y_U^+ Y_U
\nonumber\\
&&\qquad\qquad\qquad\qquad\qquad\ + 3 Y_D^+ Y_D + Y_E^+ Y_E\Big) + \frac{3}{2} Y_E^+ Y_E \bigg] +  O\Big(\alpha^2,\alpha Y^2, Y^4\Big).\qquad
\end{eqnarray}

\section{Derivation of the RGIs (\ref{Yukawa_RGI1}), (\ref{Yukawa_RGI2}), and (\ref{I3_Tilde_RGI})}
\hspace{\parindent}\label{Appendix_RGIs_Derivation}

First, let us try to construct the RGI of the form

\begin{equation}
I_1 = \bigg|\Big(\frac{(Y_U)_{33}}{(Y_D)_{33}}\Big)^{z_1}\Big(\frac{(Y_U)_{22}}{(Y_D)_{22}}\Big)^{z_2}\bigg|,
\end{equation}

\noindent where $z_1$ and $z_2$ are certain numbers. They must be determined from the requirement that the expression $I_1$ depends on scale as little as possible. The derivative of $\ln I_1$ with respect to $\ln\mu$ calculated with the help of Eqs. (\ref{Yukawa_Beta_Functions}) and (\ref{Gamma_Q_1Loop}) --- (\ref{Gamma_Hd_1Loop}) reads as

\begin{eqnarray}
&& \frac{d\ln I_1}{d\ln\mu} = z_1 \Big[(\gamma_{Y_U})_{33} - (\gamma_{Y_D})_{33}\Big] + z_2 \Big[(\gamma_{Y_U})_{22} - (\gamma_{Y_D})_{22}\Big] \nonumber\\
&& = z_1\Big(-\frac{\alpha_1}{10\pi} + \frac{5}{16\pi^2} \left[(Y_U)_{33}\right]^2 - \frac{5}{16\pi^2} \left[(Y_D)_{33}\right]^2 - \frac{1}{16\pi^2} \left[(Y_E)_{33}\right]^2  \Big) \nonumber\\
&& + z_2\Big(-\frac{\alpha_1}{10\pi} + \frac{3}{16\pi^2} \left[(Y_U)_{33}\right]^2 - \frac{3}{16\pi^2} \left[(Y_D)_{33}\right]^2 - \frac{1}{16\pi^2} \left[(Y_E)_{33}\right]^2 \Big) + \ldots,\quad
\end{eqnarray}

\noindent
where dots denote the terms containing Yukawa couplings for the first and second generations and the higher order contributions. We see that the (potentially large) contributions proportional to $\alpha_3$ cancelled each other.
Therefore, the largest contributions to the right hand side come from the terms containing $\left[(Y_U)_{33}\right]^2$ and $\left[(Y_D)_{33}\right]^2$. These terms can be removed simultaneously by choosing $z_1/z_2 = -3/5$. In particular, it is possible (and convenient) to set $z_1=3$ and $z_2=-5$. In this case we obtain the approximate RGI (\ref{Yukawa_RGI1}), and its derivative with respect to $\ln\mu$ is given by Eq. (\ref{I1_Derivative}).

\medskip

Similarly, we may consider the product

\begin{equation}
I_2 \equiv \bigg|\frac{(Y_U)_{33}}{(Y_D)_{33}}\cdot \frac{(Y_D)_{22}}{(Y_U)_{22}}\cdot \bigg(\frac{(Y_E)_{33}}{(Y_D)_{33}}\bigg)^{z_1} \bigg(\frac{(Y_E)_{22}}{3(Y_D)_{22}}\bigg)^{z_2}\bigg|,
\end{equation}

\noindent
where $z_1$ and $z_2$ are real numbers to be determined. At low energies ($M_{\mbox{\scriptsize SUSY}}$) this expression can be expressed in terms of the Standard Model Yukawa couplings without involving the unknown (at present) value of $\mbox{tg}\,\beta$. Again, differentiating the expression $\ln I_2$ with respect to $\ln\mu$ and using Eqs. (\ref{Yukawa_Beta_Functions}) and (\ref{Gamma_Q_1Loop}) --- (\ref{Gamma_Hd_1Loop}) we obtain

\begin{eqnarray}\label{I2_Derivative_Original}
&&\hspace*{-5mm} \frac{d\ln I_2}{d\ln\mu} = (\gamma_{Y_U})_{33} - (1+z_1) (\gamma_{Y_D})_{33} - (\gamma_{Y_U})_{22} + (1-z_2) (\gamma_{Y_D})_{22} + z_1 (\gamma_{Y_E})_{33} + z_2 (\gamma_{Y_E})_{22} \nonumber\\
&&\hspace*{-5mm} = (z_1+z_2)\Big(-\frac{\alpha_1}{3\pi} + \frac{4\alpha_3}{3\pi}\Big) + \frac{1}{16\pi^2} \left[(Y_U)_{33}\right]^2 (2-z_1) - \frac{1}{16\pi^2} \left[(Y_D)_{33}\right]^2 (2+3z_1)
+ \frac{3}{16\pi^2}\nonumber\\
&&\hspace*{-5mm}\times \left[(Y_E)_{33}\right]^2 z_1 + \ldots\vphantom{\frac{1}{2}}
\end{eqnarray}

\noindent
(Note that the contributions proportional to $\alpha_2$ cancelled each other.) The largest contributions here are proportional to $\alpha_3$, $\left[(Y_U)_{33}\right]^2$, and $\left[(Y_D)_{33}\right]^2$. However, it is impossible to remove them all by choosing two free parameters $z_1$ and $z_2$. That is why we need to involve the additional assumptions that $|(Y_U)_{33}| = x |(Y_D)_{33}|$ at the unification scale and that the expression $||(Y_U)_{33}| - x |(Y_D)_{33}||$ remains small throughout the whole renormalization group evolution down to the low energies. In this case it is possible to set $|(Y_U)_{33}| \approx x |(Y_D)_{33}|$ in Eq. (\ref{I2_Derivative_Original}), so that the derivative of $\ln I_2$ with respect to $\ln\mu$ takes the form

\begin{equation}
\frac{d\ln I_2}{d\ln\mu} \approx \frac{1}{3\pi} (z_1+z_2)(- \alpha_1 + 4\alpha_3) + \frac{1}{16\pi^2}\left[(Y_D)_{33}\right]^2 \Big[2x^2-2-z_1(x^2+3)\Big].
\end{equation}

\noindent
This expression can be set to 0 if we choose

\begin{equation}
z_1 = - z_2 = \frac{2(x^2-1)}{x^2+3}.
\end{equation}

\noindent
The resulting approximate RGI is then given by Eq. (\ref{Yukawa_RGI2}).

\medskip

Next, for the MSSM let us calculate the derivative of the expression $\ln\widetilde I_3$, where

\begin{equation}
\widetilde I_3 \equiv \bigg|\frac{(Y_E)_{22}}{(Y_D)_{22}}\bigg|\cdot \Big(\frac{\alpha_1}{\alpha(M_X)}\Big)^{z_1} \Big(\frac{\alpha_2}{\alpha(M_X)}\Big)^{z_2} \Big(\frac{\alpha_3}{\alpha(M_X)}\Big)^{z_3},
\end{equation}

\noindent
with respect to $\ln\mu$ taking into account Eqs. (\ref{Beta1_2Loops}) --- (\ref{Beta3_2Loops}),

\begin{equation}
\frac{d\ln \widetilde I_3}{d\ln\mu} = -\frac{\alpha_1}{3\pi} + \frac{4\alpha_3}{3\pi} + \frac{33 z_1\alpha_1}{10\pi}\alpha_1 + \frac{z_2\alpha_2}{2\pi} - \frac{3z_3\alpha_3}{2\pi} + \ldots
\end{equation}

\noindent
Therefore, the largest contributions to the right hand side cancel each other for

\begin{equation}
z_1 = \frac{10}{99};\qquad z_2 = 0;\qquad z_3 = \frac{8}{9},
\end{equation}

\noindent
and we obtain the expression (\ref{I3_Tilde_RGI}).

\section{Derivation of the Yukawa relations (\ref{Yukawa_Relation_351_3}) and (\ref{Yukawa_Relation_351_2}) from the $E_6$ invariant $\,\xbar{27}\,\times\,\xbar{27}\,\times \,\xbar{351}\,{}'$}
\hspace{\parindent}\label{Appendix_Invariants}

In this appendix we describe the derivation of Eqs. (\ref{Yukawa_Relation_351_3}) and (\ref{Yukawa_Relation_351_2}) with the help of the group theory methods starting from the $E_6$ invariant $\,\xbar{27}\,\times\,\xbar{27}\,\times \,\xbar{351}\,{}'$. For this purpose we will need some intermediate invariants which are also constructed and discussed in what follows.

\subsection{The $E_6$ invariant $\,\xbar{27}\,\times \,\xbar{27}\,\times \,\xbar{351}\,'$}
\hspace*{\parindent}\label{Subsection_Invariant_E_6}

According to Eq. (\ref{Branchings_E_6_27}), the representation $\,\xbar{27}\,$ of the group $E_6$ can be presented as a set of the superfields $P$, $P_i$, and $P^a$ corresponding to $1(-4)$, $10(2)$, and $\,\xbar{16}\,(-1)$ of the subgroup $SO(10)\times U(1)$. Here the first number denotes the representation of $SO(10)$, and the number written in brackets is the $U(1)$ charge. Everywhere in this paper for the $U(1)$ charges we adopt the same normalization conditions as in Ref. \cite{Slansky:1981yr}. The vector indices (e.g., $i$ in the case under consideration) range from 1 to 10, while the spinor indices (e.g., $a$) range from 1 to 16. The lower spinor indices correspond to the representation 16, while the upper spinor indices correspond to the representation $\,\xbar{16}\,$. The normalization of the superfields forming the representations $27$ and $\,\xbar{27}\,$ of $E_6$ is fixed by choosing the $E_6$ invariant norm in the form

\begin{equation}\label{Normalization_27}
N_{27} \equiv \,\xbar{27}\,\times 27 \equiv P \bar P + P_i \bar P_i + P^a \bar P_a,
\end{equation}

\noindent
where the set $\{\bar P,\,\bar P_i\,\bar P_a\}$ forms the representation $27$.

Similarly, according to Eq. (\ref{Branchings_E_6_351'}), the representation $\,\xbar{351}\,'$ is given by the set

\begin{equation}
\,\xbar{351}\,' \sim \Big\{A,\, A_i,\,A_a,\,A_{ij},\,A_{ijklm},\,A_{ia}\Big\}.
\end{equation}

\noindent
The superfields of this set correspond to the representations $1(8)$, $10(2)$, $16(5)$, $54(-4)$, $126(2)$, and $\,\xbar{144}\,(-1)$ of the subgroup $SO(10)\times U(1)$, respectively. In particular, $A_{ij}$ is the symmetric traceless tensor (so that $A_{ij}=A_{ji}$ and $A_{ii}=0$) corresponding to the representation 54, and the representation 126 corresponds to the completely antisymmetric tensor $A_{ijklm}$ that satisfies the condition

\begin{equation}\label{126_Selfduality}
A_{i_1 i_2 i_3 i_4 i_5} = -\frac{i}{5!} \varepsilon_{i_1 i_2 i_3 i_4 i_5 j_1 j_2 j_3 j_4 j_5} A_{j_1 j_2 j_3 j_4 j_5}.
\end{equation}

\noindent
The superfield $A_{ia}$ satisfies the constraint $(B\Gamma_i)^{ab} A_{ib} = 0$ and describes the representation $\,\xbar{144}\,$. We fix the normalization of the representation $\,\xbar{351}\,{}'$ by choosing the $E_6$ invariant norm

\begin{equation}\label{Normalization_351'}
N_{351'} \equiv \,\xbar{351}\,'\times 351' \equiv  A \bar A + A_i \bar A_i + A_a \bar A^a + \frac{1}{2} A_{ij} \bar A_{ij} + \frac{1}{5!} A_{ijklm} \bar A_{ijklm} + A_{ia} \bar A_i^a.
\end{equation}

The expressions (\ref{Normalization_27}) and (\ref{Normalization_351'}) are manifestly $SO(10)\times U(1)$ invariant. However, this is not sufficient for invariance under the transformations of the entire group $E_6$. It is also necessary to take into account the hidden symmetry that can be observed by analysing the branching rule of the adjoint representation 78 \cite{Slansky:1981yr},

\begin{equation}
78 \Big|_{E_6} = 1(0) + 16(-3) + \,\xbar{16}\,(3) + 45(0)\Big|_{SO(10)\times U(1)}.
\end{equation}

\noindent
The terms $1(0)$ and $45(0)$ in the right hand side correspond to the subgroups $U(1)$ and $SO(10)$, respectively, for which the invariance of the expressions (\ref{Normalization_27}) and (\ref{Normalization_351'}) is manifest. However, it is also necessary to verify the invariance generated by the terms $16(-3)$ and $\,\xbar{16}\,(3)$ producing the hidden symmetry with the respective parameters $\varepsilon_a$ and $\varepsilon^a$. Under this symmetry various parts of the representation $\,\xbar{27}\,$ change as

\begin{eqnarray}\label{Transformations_27}
&& \delta P = \varepsilon_a P^a;\qquad \delta P_i = \frac{1}{\sqrt{2}} \varepsilon^a (\Gamma_i B)_{ab} P^b;\qquad \delta P^a = \varepsilon^a P + \frac{1}{\sqrt{2}} \varepsilon_b (B\Gamma_i)^{ab} P_i.
\end{eqnarray}

\noindent
Similarly, the transformations of various parts of the representation $\,\xbar{351}\,'$ take the form

\begin{eqnarray}\label{Transformations_351'}
&& \delta A = -\sqrt{2}\, \varepsilon^a A_a;\qquad\qquad\quad\  \delta A_i = -\frac{\sqrt{10}}{4} \varepsilon_a (B\Gamma_i)^{ab} A_b - \frac{\sqrt{5}}{2} \varepsilon^a A_{ia};\nonumber\\
&& \delta A_a = -\sqrt{2}\, \varepsilon_a A - \frac{\sqrt{10}}{4} \varepsilon^b (\Gamma_i B)_{ab} A_i - \frac{1}{2\sqrt{2}}\cdot \frac{1}{5!} \varepsilon^b (\Gamma_{ijklm} B)_{ab} A_{ijklm};\nonumber\\
&& \delta A_{ij} = -\frac{1}{\sqrt{2}} \Big[\varepsilon_a (B\Gamma_i)^{ab} A_{jb} + \varepsilon_a (B\Gamma_j)^{ab} A_{ib}\Big];\nonumber\\
&& \delta A_{ijklm} = - \frac{1}{2\sqrt{2}} \varepsilon_a (B\Gamma_{ijklm})^{ab} A_b - \frac{1}{2}\Big[\varepsilon^a (\Gamma_{ijkl})_a{}^b A_{mb} + \varepsilon^a (\Gamma_{mijk})_a{}^b A_{lb}
\nonumber\\
&&\qquad\qquad\qquad\qquad\qquad
+ \varepsilon^a (\Gamma_{lmij})_a{}^b A_{kb} + \varepsilon^a (\Gamma_{klmi})_a{}^b A_{jb} + \varepsilon^a (\Gamma_{jklm})_a{}^b A_{ib} \Big];\nonumber\\
&& \delta A_{ia} = - \frac{9}{4\sqrt{5}} \varepsilon_a A_i - \frac{1}{4}\cdot \frac{1}{4!} \varepsilon_b (\Gamma_{jklm})_a{}^b A_{ijklm} - \frac{1}{\sqrt{2}} \varepsilon^b (\Gamma_j B)_{ab} A_{ij}
\nonumber\\
&&\qquad\qquad\qquad\qquad\qquad\qquad\ \
+ \frac{1}{4\sqrt{5}} \varepsilon_b (\Gamma_{ij})_a{}^b A_j + \frac{1}{4}\cdot \frac{1}{5!} \varepsilon_b (\Gamma_{ijklmn})_a{}^b A_{jklmn}.\qquad
\end{eqnarray}

To construct the cubic invariant $\,\xbar{27}\,\times \,\xbar{27}\,\times \,\xbar{351}\,'$, we write down all manifestly $SO(10)\times U(1)$ invariant terms with unknown coefficients and determine their values by requiring the invariance under the transformations (\ref{Transformations_27}) and (\ref{Transformations_351'}).\footnote{More precisely, it is necessary to write the transformations with unknown coefficients and require the invariance of $N_{27}$, $N_{351'}$, and $\,\xbar{27}\,\times \,\xbar{27}\,\times \,\xbar{351}\,'$. The solution of the resulting system of equations produces Eqs. (\ref{Transformations_27}), (\ref{Transformations_351'}), and (\ref{Invariant_351'}).} The result for the invariant under consideration takes the form

\begin{eqnarray}\label{Invariant_351'}
&& \,\xbar{27}\,\times \,\xbar{27}\,\times \,\xbar{351}\,' = P^2 A + \frac{4}{\sqrt{10}}\, P P_i A_i + \sqrt{2}\, P \,P^a A_a + \frac{1}{\sqrt{2}}\, P_i P_j A_{ij}
\nonumber\\
&& + \sqrt{2}\, P_i P^a A_{ia} + \frac{1}{4\cdot 5!} P^a (\Gamma_{ijklm} B)_{ab} P^b A_{ijklm} + \frac{1}{4\sqrt{5}} P^a (\Gamma_i B)_{ab} P^b A_i.
\end{eqnarray}

\noindent
(Evidently, this expression is defined up to an arbitrary multiplicative constant.)

\subsection{The $SO(10)$ invariants $\,\xbar{16}\,\times \,\xbar{16}\,\times 10$, $\,\xbar{16}\,\times \,\xbar{16}\,\times 126$, and $10\times \,\xbar{16}\, \times \,\xbar{144}\,$}
\hspace*{\parindent}\label{Subsection_Invariant_SO(10)}

From Tables \ref{Table_Branching3} and \ref{Table_Branching2} it becomes clear that for deriving the Yukawa relation we need the $SO(10)$ invariants $\,\xbar{16}\,\times \,\xbar{16}\,\times 10$, $\,\xbar{16}\,\times \,\xbar{16}\,\times 126$, and $10\times \,\xbar{16}\, \times \,\xbar{144}\,$. Certainly, it is useful to construct explicit expressions for all of them. Although for the first two of them these expressions are known \cite{Nath:2001uw}, it is nevertheless expedient to describe briefly their derivation because the invariant $10\times \,\xbar{16}\, \times \,\xbar{144}\,$ will be obtained by the same method.

The $SO(10)$ representation $\,\xbar{16}\,$ can be considered as a set of the superfields $P^a\ \to\ \{\phi,\,\phi_\alpha,\,\phi^{\alpha\beta}\}$ (where $\phi^{\alpha\beta} = -\phi^{\beta\alpha}$) corresponding to $1(5)$, $5(-3)$, and $\,\xbar{10}\,(1)$ of $SU(5)\times U(1)$, respectively. Here the $SU(5)$ indices $\alpha$ and $\beta$ range from 1 to 5. The lower and upper indices correspond to the fundamental representation 5 and antifundamental representation $\,\xbar{5}\,$, respectively. In our conventions, the normalization of the representation $\,\xbar{16}\,$ is fixed by choosing the $SO(10)$ invariant norm

\begin{equation}\label{Normalization_16}
N_{16} \equiv \,\xbar{16}\, \times 16 \equiv P^a \bar P_a \equiv \phi\,\bar{\phi} + \phi_\alpha \bar\phi^\alpha + \frac{1}{2} \phi^{\alpha\beta} \bar\phi_{\alpha\beta},
\end{equation}

\noindent
where the superfields $\bar P_a \to \{\bar\phi,\,\bar\phi^\alpha,\,\bar\phi_{\alpha\beta}\}$ form the $SO(10)$ representation 16.

Similarly, the representation 10 of the group $SO(10)$ is formed by the superfields $\{R_\alpha,\,R^\alpha\}$ corresponding to $5(2)$ and $\,\xbar{5}\,(-2)$ in Eq. (\ref{Branchings_SO(10)_10}). For this representation, the normalization is fixed by the invariant

\begin{equation}\label{Normalization_10}
N_{10} \equiv 10_1\times 10_2\equiv (A_1)_i (A_2)_i \equiv (R_1)_\alpha (R_2)^\alpha + (R_1)^\alpha (R_2)_\alpha.
\end{equation}

According to Eq. (\ref{Branchings_SO(10)_126}), the representations 126 can be presented as the set

\begin{eqnarray}\label{Content_Of_126}
126 \sim A_{ijklm} \to \Big\{K,\,K^\alpha,\,K_{\alpha\beta},\,L^{\alpha\beta},\, K^\alpha_{\beta\gamma},\, K^{\mu\nu\sigma}_{\alpha\beta}\Big\}
\end{eqnarray}

\noindent
corresponding to the terms $1(-10)$, $\,\xbar{5}\,(-2)$, $10(-6)$, $\,\xbar{15}\,(6)$, $45(2)$, and $\,\xbar{50}\,(-2)$, respectively. They should satisfy the constraints

\begin{eqnarray}
&& K_{\alpha\beta} = - K_{\beta\alpha};\qquad L^{\alpha\beta} = L^{\beta\alpha};\qquad K^\alpha_{\beta\gamma} = - K^\alpha_{\gamma\beta};\qquad K^{\alpha}_{\alpha\beta} = 0;\qquad\vphantom{\Big(}\nonumber\\
&& K^{\mu\nu\sigma}_{\alpha\beta} = -K^{\mu\nu\sigma}_{\beta\alpha};\qquad\, K^{\mu\nu\sigma}_{\alpha\beta} = -K^{\nu\mu\sigma}_{\alpha\beta} = -K^{\mu\sigma\nu}_{\alpha\beta};\qquad\, K^{\alpha\mu\nu}_{\alpha\beta} = 0.
\vphantom{\Big(}
\end{eqnarray}

\noindent
(This in particular implies that the superfield $K^{\mu\nu\sigma}_{\alpha\beta}$ is antisymmetric in both upper and lower indices.) The normalization invariant for the representation 126 is chosen in the form

\begin{eqnarray}\label{Normalization_126}
&&\hspace*{-3mm} N_{126}\equiv \xbar{126}\times 126  \equiv \frac{1}{5!} \bar A_{ijklm} A_{ijklm}
\nonumber\\
&&\hspace*{-3mm} \qquad \equiv \,\xbar{K}\ K + \,\xbar{K}\,{}_\alpha K^\alpha + \frac{1}{2} \xbar{K}\,{}^{\alpha\beta} K_{\alpha\beta} + \frac{1}{2} \,\xbar{L}\,{}_{\alpha\beta} L^{\alpha\beta}
+ \frac{1}{2} \,\xbar{K}\,{}_\gamma^{\alpha\beta} K^\gamma_{\alpha\beta} + \frac{1}{12} \,\xbar{K}\,{}_{\alpha\beta\gamma}^{\mu\nu} K^{\alpha\beta\gamma}_{\mu\nu}.\qquad
\end{eqnarray}

Finally, according to Eq. (\ref{Branchings_SO(10)_144}) we present the representation $\,\xbar{144}\,$ as the set of $SU(5)$ tensors

\begin{eqnarray}\label{Content_Of_144}
&& \,\xbar{144}\, \sim A_{ia} \to \Big\{B_\alpha,\,B^\alpha,\,B^{\alpha\beta},\,C^{\alpha\beta},\,B_\alpha^\beta,\,B_{\alpha[\beta\gamma]},\,B^\gamma_{\alpha\beta} \Big\},
\end{eqnarray}

\noindent
which corresponds to the terms $5(-3)$, $\,\xbar{5}\,(-7)$, $\,\xbar{10}\,(1)$, $\,\xbar{15}\,(1)$, $24(5)$, $\,\xbar{40}\,(1)$, $45(-3)$, respectively. These superfields satisfy the constraints

\begin{eqnarray}
&& B^{\alpha\beta} = - B^{\beta\alpha};\qquad C^{\alpha\beta} = C^{\beta\alpha};\qquad\, B^\alpha_\alpha = 0;\qquad\, B_{\alpha[\beta\gamma]} = - B_{\alpha[\gamma\beta]};\qquad\vphantom{\Big(}\nonumber\\
&& B_{\alpha[\beta\gamma]} + B_{\beta[\gamma\alpha]} + B_{\gamma[\alpha\beta]} = 0;\qquad\quad B^\alpha_{\alpha\beta} = 0;\qquad\quad\ \ B^\gamma_{\alpha\beta} = - B^\gamma_{\beta\alpha},\vphantom{\Big(}
\end{eqnarray}

\noindent
while the normalization invariant is taken in the form

\begin{eqnarray}\label{Normalization_144}
&&\hspace*{-5mm} N_{144} \equiv \,\xbar{144}\,\times 144 \equiv A_{ia} \bar A^a_i \nonumber\\
&&\hspace*{-5mm}\qquad \equiv B_\alpha \,\xbar{B}\,{}^\alpha + B^\alpha \,\xbar{B}\,{}_\alpha + \frac{1}{2} B^{\alpha\beta} \,\xbar{B}\,{}_{\alpha\beta}
+ \frac{1}{2} C^{\alpha\beta} \,\xbar{C}\,{}_{\alpha\beta} + B_\alpha^\beta \,\xbar{B}\,{}_\beta^\alpha + \frac{1}{2} B_{\alpha[\beta\gamma]} \,\xbar{B}\,{}^{\alpha[\beta\gamma]}
+ \frac{1}{2} B^\gamma_{\alpha\beta} \,\xbar{B}\,{}^{\alpha\beta}_\gamma.\nonumber\\
\end{eqnarray}

Evidently, any $SO(10)$ invariant can be written in terms of $SU(5)$ representations. For this purpose it is necessary to write down all possible manifestly $SU(5)\times U(1)$ invariant terms with unknown coefficients and require the invariance of their sum under the transformations of the hidden symmetry. To construct this hidden symmetry, we again consider the branching rule for the adjoint representation, which in the case under consideration is given by the equation

\begin{equation}\label{Branchings_SO(10)_45}
45\Big|_{SO(10)} = 1(0) +10(4) + \,\xbar{10}\,(-4) + 24(0)\Big|_{SU(5)\times U(1)}.
\end{equation}

\noindent
The terms $24(0)$ and $1(0)$ correspond to the generators of the manifest $SU(5)\times U(1)$ symmetry. Two remaining terms $10(4)$ and $\,\xbar{10}\,(-4)$ generate the hidden symmetry, which is therefore parameterized by the antisymmetric tensors $\xi_{\alpha\beta}$ and $\xi^{\alpha\beta}$. The transformations of this symmetry are constructed by wring all possible manifestly $SU(5)\times U(1)$ invariant terms with such coefficients that lead to the invariance of the expressions under consideration (Eqs. (\ref{Normalization_16}), (\ref{Normalization_10}), (\ref{Normalization_126}), (\ref{Normalization_144}), (\ref{Invariant_10}), (\ref{Invariant_126}), and (\ref{Invariant_144})). For the representations $10$, $\,\xbar{16}\,$, $126$, and $\,\xbar{144}\,$ they are written as follows.

\begin{eqnarray}\label{Transformations_10}
&& \mbox{The representation 10:}\vphantom{\frac{1}{2}}\nonumber\\
&& \delta R_\alpha = \xi_{\alpha\beta} R^\beta;\qquad \delta R^\alpha = \xi^{\alpha\beta} R_\beta.\vphantom{\frac{1}{2}}\\
&&\vphantom{1}\nonumber\\
&& \mbox{The representation $\,\xbar{16}\,$:}\vphantom{\frac{1}{2}}\nonumber\\
\label{Transformations_16}
&& \delta\phi = \frac{1}{2} \xi_{\alpha\beta} \phi^{\alpha\beta};\qquad \delta\phi_\alpha = \frac{1}{4} \varepsilon_{\alpha\beta\gamma\mu\nu} \xi^{\beta\gamma} \phi^{\mu\nu};\qquad
\delta\phi^{\alpha\beta} = -\xi^{\alpha\beta}\phi - \frac{1}{2} \varepsilon^{\alpha\beta\gamma\mu\nu} \xi_{\mu\nu}\phi_\gamma.\qquad\\
&&\vphantom{1}\nonumber\\
&& \mbox{The representation 126:}\vphantom{\frac{1}{2}}\nonumber\\
\label{Transformations_126}
&& \delta K = \frac{1}{\sqrt{2}} \xi^{\alpha\beta} K_{\alpha\beta};\qquad \delta K^\alpha = \frac{1}{\sqrt{3}} \xi^{\beta\gamma} K^\alpha_{\beta\gamma} + \frac{1}{2\sqrt{3}} \varepsilon^{\alpha\beta\gamma\mu\nu} \xi_{\beta\gamma} K_{\mu\nu};\nonumber\\
&& \delta K_{\alpha\beta} = -\sqrt{2} \xi_{\alpha\beta} K - \frac{1}{\sqrt{3}} \varepsilon_{\alpha\beta\mu\nu\gamma} \xi^{\mu\nu} K^\gamma + \frac{1}{6\sqrt{2}} \varepsilon_{\alpha\beta\mu\nu\sigma} \xi^{\rho\tau} K^{\mu\nu\sigma}_{\rho\tau};\nonumber\\
&& \delta L^{\alpha\beta} = \frac{1}{4}\Big(\varepsilon^{\alpha\mu\nu\sigma\rho} \xi_{\mu\nu} K^\beta_{\sigma\rho} + \varepsilon^{\beta\mu\nu\sigma\rho} \xi_{\mu\nu} K^\alpha_{\sigma\rho}\Big);\nonumber\\
&& \delta K^\gamma_{\alpha\beta} = - \frac{2}{\sqrt{3}} \xi_{\alpha\beta} K^\gamma + \frac{1}{\sqrt{2}} \xi_{\mu\nu} K^{\gamma\mu\nu}_{\alpha\beta} - \frac{1}{2}\varepsilon_{\alpha\beta\mu\nu\sigma} \xi^{\mu\nu} L^{\sigma\gamma}
- \frac{1}{2\sqrt{3}}\Big(\delta_\alpha^\gamma \xi_{\beta\mu} K^\mu - \delta_\beta^\gamma \xi_{\alpha\mu} K^\mu\Big);\nonumber\\
&& \delta K^{\mu\nu\sigma}_{\alpha\beta} = -\sqrt{2} \xi_{\alpha\beta} K^{\mu\nu\sigma} - \frac{1}{\sqrt{2}} \Big(\delta^\mu_\alpha \xi_{\beta\gamma} K^{\gamma\nu\sigma} - \delta^\mu_\beta \xi_{\alpha\gamma} K^{\gamma\nu\sigma}
+ \delta^\nu_\alpha \xi_{\beta\gamma} K^{\gamma\sigma\mu} - \delta^\nu_\beta \xi_{\alpha\gamma} K^{\gamma\sigma\mu}\nonumber\\
&& + \delta^\sigma_\alpha \xi_{\beta\gamma} K^{\gamma\mu\nu} - \delta^\sigma_\beta \xi_{\alpha\gamma} K^{\gamma\mu\nu} \Big)
- \frac{1}{3\sqrt{2}}\xi_{\gamma\delta} \Big(
\delta^\mu_\alpha \delta^\nu_\beta K^{\gamma\delta\sigma} - \delta^\mu_\beta \delta^\nu_\alpha K^{\gamma\delta\sigma}
+ \delta^\sigma_\alpha \delta^\mu_\beta K^{\gamma\delta\nu} - \delta^\sigma_\beta \delta^\mu_\alpha K^{\gamma\delta\nu}
\nonumber\\
&& + \delta^\nu_\alpha \delta^\sigma_\beta K^{\gamma\delta\mu} - \delta^\nu_\beta \delta^\sigma_\alpha K^{\gamma\delta\mu} \Big)
-\sqrt{2}\Big(\xi^{\mu\nu} K^\sigma_{\alpha\beta} + \xi^{\sigma\mu} K^\nu_{\alpha\beta} + \xi^{\nu\sigma} K^\mu_{\alpha\beta}\Big)
- \frac{1}{\sqrt{2}}\Big(
\delta^\mu_\alpha \xi^{\nu\rho} K^\sigma_{\rho\beta}\nonumber\\
&& - \delta^\mu_\beta \xi^{\nu\rho} K^\sigma_{\rho\alpha}
-\delta^\nu_\alpha \xi^{\mu\rho} K^\sigma_{\rho\beta} + \delta^\nu_\beta \xi^{\mu\rho} K^\sigma_{\rho\alpha}
+ \delta^\sigma_\alpha \xi^{\mu\rho} K^\nu_{\rho\beta} - \delta^\sigma_\beta \xi^{\mu\rho} K^\nu_{\rho\alpha}
-\delta^\mu_\alpha \xi^{\sigma\rho} K^\nu_{\rho\beta} + \delta^\mu_\beta \xi^{\sigma\rho} K^\nu_{\rho\alpha}
\vphantom{\frac{1}{2}}\nonumber\\
&& + \delta^\nu_\alpha \xi^{\sigma\rho} K^\mu_{\rho\beta} - \delta^\nu_\beta \xi^{\sigma\rho} K^\mu_{\rho\alpha}
- \delta^\sigma_\alpha \xi^{\nu\rho} K^\mu_{\rho\beta} + \delta^\sigma_\beta \xi^{\nu\rho} K^\mu_{\rho\alpha}\Big)
- \frac{1}{3\sqrt{2}} \xi^{\rho\tau} \Big(\delta^\mu_\alpha \delta^\nu_\beta K^\sigma_{\rho\tau} - \delta^\nu_\alpha \delta^\mu_\beta K^\sigma_{\rho\tau}
\nonumber\\
&& + \delta^\sigma_\alpha \delta^\mu_\beta K^\nu_{\rho\tau} - \delta^\mu_\alpha \delta^\sigma_\beta K^\nu_{\rho\tau}
+ \delta^\nu_\alpha \delta^\sigma_\beta K^\mu_{\rho\tau} - \delta^\sigma_\alpha \delta^\nu_\beta K^\mu_{\rho\tau}\Big),\\
&&\mbox{where we use the notation}\quad K^{\mu\nu\sigma} \equiv \frac{1}{2} \varepsilon^{\mu\nu\sigma\alpha\beta} K_{\alpha\beta}.\nonumber\\
&&\vphantom{1}\nonumber\\
&& \mbox{The representation $\,\xbar{144}\,$:}\vphantom{\frac{1}{2}}\nonumber\\
\label{Transformations_144}
&& \delta B_\alpha = \frac{\sqrt{5}}{2} \xi_{\alpha\beta} B^\beta + \frac{1}{4\sqrt{6}} \varepsilon_{\alpha\beta\gamma\mu\nu} \xi^{\beta\gamma} B^{\mu\nu} + \frac{\sqrt{5}}{4} \xi^{\beta\gamma} B_{\alpha[\beta\gamma]};\nonumber\\
&& \delta B^\alpha = \frac{\sqrt{5}}{2} \xi^{\alpha\beta} B_\beta + \frac{1}{2} \xi^{\beta\gamma} B^\alpha_{\beta\gamma};\nonumber\\
&& \delta B^{\alpha\beta} = \sqrt{\frac{5}{6}}\Big(\xi^{\alpha\gamma} B^\beta_\gamma - \xi^{\beta\gamma} B^\alpha_\gamma\Big) - \frac{1}{2\sqrt{6}} \varepsilon^{\alpha\beta\gamma\mu\nu} \xi_{\mu\nu} B_\gamma
-\frac{1}{4} \sqrt{\frac{5}{6}} \xi_{\gamma\delta} \Big(\varepsilon^{\alpha\gamma\delta\mu\nu}  B^\beta_{\mu\nu} - \varepsilon^{\beta\gamma\delta\mu\nu} B^\alpha_{\mu\nu}\Big);\nonumber\\
&& \delta C^{\alpha\beta} = \frac{1}{4\sqrt{2}}\xi_{\gamma\delta}\Big(\varepsilon^{\alpha\gamma\delta\mu\nu} B^\beta_{\mu\nu} + \varepsilon^{\beta\gamma\delta\mu\nu} B^\alpha_{\mu\nu} \Big) + \frac{1}{\sqrt{2}}\Big(\xi^{\alpha\gamma} B^\beta_\gamma + \xi^{\beta\gamma} B^\alpha_\gamma\Big);\nonumber\\
&& \delta B_\alpha^\beta = -\sqrt{\frac{5}{6}} \Big(\xi_{\alpha\gamma} B^{\beta\gamma} - \frac{1}{5} \delta_\alpha^\beta \xi_{\mu\nu} B^{\mu\nu}\Big) + \frac{1}{\sqrt{2}} \xi_{\alpha\gamma} C^{\beta\gamma}
+\frac{1}{4} \varepsilon^{\beta\gamma\delta\mu\nu} \xi_{\gamma\delta} B_{\alpha[\mu\nu]};\nonumber\\
&& \delta B_{\alpha[\beta\gamma]} = -\frac{1}{2} \varepsilon_{\beta\gamma\delta\mu\nu} \xi^{\mu\nu} B^\delta_\alpha -\frac{1}{3} \varepsilon_{\alpha\beta\gamma\mu\nu} \xi^{\mu\sigma} B_\sigma^\nu
-\frac{\sqrt{5}}{3} \xi_{\beta\gamma} B_\alpha - \frac{\sqrt{5}}{6}\Big(\xi_{\alpha\gamma} B_\beta - \xi_{\alpha\beta} B_\gamma\Big)\nonumber\\
&& + \frac{2}{3} \xi_{\alpha\sigma} B^\sigma_{\beta\gamma} + \frac{1}{3}\Big(\xi_{\beta\sigma} B^\sigma_{\alpha\gamma} - \xi_{\gamma\sigma} B^\sigma_{\alpha\beta}\Big);\nonumber\\
&& \delta B^\gamma_{\alpha\beta} = -\xi_{\alpha\beta} B^\gamma + \frac{1}{4} B^\sigma \Big(\delta^\gamma_\beta \xi_{\alpha\sigma} - \delta^\gamma_\alpha \xi_{\beta\sigma} \Big)
- \frac{1}{4}\sqrt{\frac{6}{5}} \varepsilon_{\alpha\beta\mu\nu\sigma} \xi^{\mu\nu} B^{\gamma\sigma} + \frac{1}{\sqrt{30}} \varepsilon_{\alpha\beta\mu\nu\sigma} \xi^{\gamma\sigma} B^{\mu\nu}
\nonumber\\
&& - \frac{1}{8\sqrt{30}}\xi^{\mu\nu} B^{\sigma\rho} \Big(\delta^\gamma_\alpha \varepsilon_{\beta\mu\nu\sigma\rho}  - \delta^\gamma_\beta\varepsilon_{\alpha\mu\nu\sigma\rho}\Big) + B_{\sigma[\alpha\beta]} \xi^{\gamma\sigma}
- \frac{1}{8}\xi^{\mu\nu}\Big(\delta^\gamma_\alpha B_{\beta[\mu\nu]} - \delta^\gamma_\beta B_{\alpha[\mu\nu]}\Big)\nonumber\\
&& - \frac{1}{2\sqrt{2}} \varepsilon_{\alpha\beta\mu\nu\sigma} \xi^{\mu\nu} C^{\gamma\sigma}.
\end{eqnarray}

The invariants under consideration are constructed by writing all possible $SU(5)$ and $U(1)$ manifestly symmetric terms and imposing the condition of the invariance under the transformations of the hidden symmetry corresponding to $10(4)$ and $\,\xbar{10}\,(-4)$ in Eq. (\ref{Branchings_SO(10)_45}). The results are given by the expressions

\begin{eqnarray}\label{Invariant_10}
&& I_{10}\equiv \xbar{16}\times \xbar{16} \times 10 = c_{10}\Big(\phi\,\phi_{\alpha} R^\alpha + \phi_\alpha \phi^{\alpha\beta} R_\beta + \frac{1}{8} \varepsilon_{\alpha\beta\mu\nu\sigma} \phi^{\alpha\beta} \phi^{\mu\nu} R^\sigma\Big);\\
&&\vphantom{1}\nonumber\\
\label{Invariant_126}
&& I_{126}\equiv \xbar{16}\times \xbar{16} \times 126 = c_{126}\Big(\phi^2 K + \sqrt{\frac{3}{2}}\phi \phi_\alpha K^\alpha + \frac{1}{\sqrt{2}} \phi \phi^{\alpha\beta} K_{\alpha\beta} + \frac{1}{\sqrt{2}} \phi_\alpha\phi_\beta L^{\alpha\beta} \qquad\nonumber\\
&& + \frac{1}{\sqrt{2}} \phi_\gamma \phi^{\alpha\beta} K^\gamma_{\alpha\beta} - \frac{1}{8\sqrt{6}} \varepsilon_{\mu\nu\alpha\beta\gamma} \phi^{\mu\nu} \phi^{\alpha\beta} K^\gamma
+ \frac{1}{24} \varepsilon_{\mu\nu\alpha\beta\gamma} \phi^{\mu\nu} \phi^{\sigma\rho} K_{\sigma\rho}^{\alpha\beta\gamma}\Big);\\
&&\vphantom{1}\nonumber\\
\label{Invariant_144}
&& I_{144}\equiv 10\times \,\xbar{16}\,\times \,\xbar{144}\, = c_{144}\Big(R^\alpha \phi B_\alpha + \frac{\sqrt{5}}{2} R^\alpha \phi_\beta B_\alpha^\beta
+ \frac{1}{4\sqrt{6}} \varepsilon_{\alpha\beta\gamma\mu\nu} R^\alpha \phi^{\beta\gamma} B^{\mu\nu} \nonumber\\
&& + \frac{\sqrt{5}}{4} R^\alpha \phi^{\beta\gamma} B_{\alpha[\beta\gamma]} + \frac{\sqrt{5}}{2} R_\alpha \phi B^\alpha + \sqrt{\frac{5}{8}} R_\alpha \phi_\beta C^{\alpha\beta} + \frac{\sqrt{5}}{4} R_\alpha \phi^{\beta\gamma} B^\alpha_{\beta\gamma} + \frac{1}{4} R_\alpha \phi^{\alpha\beta} B_\beta\nonumber\\
&& + \frac{\sqrt{6}}{4} R_\alpha\phi_\beta B^{\alpha\beta}\Big),
\end{eqnarray}

\noindent
where $c_{10}$, $c_{126}$, and $c_{144}$ are the normalization constants. In fact, the normalization of all these expression is fixed by Eqs. (\ref{SO(10)_Normalization}), (\ref{Normalization_16}), (\ref{Normalization_10}), (\ref{Normalization_126}), and (\ref{Normalization_144}). Therefore, these constants have specific values that need to be determined. Evidently, this cannot be done with the help of method based on the hidden symmetry. In the next subsection we find the values of $c_{10}$, $c_{126}$, and $c_{144}$ using a different technique. The result is given by Eq. (\ref{Normalization_Constants}) below.

\subsection{Normalization of the invariants}
\hspace{\parindent}\label{Subsection_Normalizaion}

The method based on the hidden symmetry allows for constructing the invariants only up to an indefinite numerical factor. However, these factors are very essential for obtaining the equations which relate the Yukawa couplings. That is why here we will calculate some of the above invariants by a different method. Using this method the invariants $\,\xbar{16}\,\times \,\xbar{16}\,\times 10$ and $\,\xbar{16}\,\times\,\xbar{16}\,\times 126$ have already been constructed in \cite{Nath:2001uw}. However, for completeness, we briefly recall how this is made. After that, using this method we construct the invariant $10\times\,\xbar{16}\,\times \,\xbar{144}\,$. The results should coincide with the ones obtained in the previous section up to a numerical factor and (possible) redefinition of fields. Of course, this can be considered as a test of the calculation correctness.

Let us start with the explicit construction of the branching (\ref{Branchings_SO(10)_10}). The element of the representation $10$ can be presented as a ten-component real column. This column can equivalently be rewritten in the form

\begin{equation}\label{10_Branching}
A_i\ \to\ \frac{1}{\sqrt{2}}\left(
\begin{array}{c}
R_\alpha + R^\alpha\vphantom{\Big(}\\
-i(R_\alpha-R^\alpha)\vphantom{\Big(}
\end{array}
\right),
\end{equation}

\noindent
where the fields $R_\alpha$ and $R^\alpha = (R_\alpha)^*$ lie in the representations $5$ and $\,\xbar{5}\,$ with respect to the subgroup $SU(5) \subset SO(10)$. Equivalently, Eq. (\ref{10_Branching}) can be presented in the form

\begin{equation}\label{Tensor_Projection}
\left(
\begin{array}{c}
R_\alpha\\ R^\alpha
\end{array}
\right) = \frac{1}{\sqrt{2}}\left(
\begin{array}{cc}
1 & i\\
1 & -i
\end{array}
\right)
\left(
\begin{array}{c}
A_i\\ A_{i+5}
\end{array}
\right),
\end{equation}

\noindent
where $i=1,\ldots,5$. This equation admits evident generalization to the case of a tensor with an arbitrary number of indices. In particular, as a simple consequence, we obtain

\begin{equation}
\varepsilon_{\alpha_1\alpha_2\alpha_3\alpha_4\alpha_5}{}^{\beta_1\beta_2\beta_3\beta_4\beta_5} = - \varepsilon^{\beta_1\beta_2\beta_3\beta_4\beta_5}{}_{\alpha_1\alpha_2\alpha_3\alpha_4\alpha_5} =\ldots
= -i \varepsilon_{\alpha_1\alpha_2\alpha_3\alpha_4\alpha_5}\cdot\varepsilon^{\beta_1\beta_2\beta_3\beta_4\beta_5}.
\end{equation}

The normalization invariant $10\times 10$ written in terms of the $SU(5)$ (super)fields takes the form

\begin{equation}
(A_1)_i (A_2)_i = (R_1)_\alpha (R_2)^\alpha + (R_1)^\alpha (R_2)_\alpha
\end{equation}

\noindent
and coincides with the expression (\ref{Normalization_10}).

A similar construction for the representations $16$ and $\,\xbar{16}\,$ (see, e.g., \cite{Mohapatra:1979nn}) is more complicated. First, it is necessary to introduce the ten-dimensional Euclidean gamma-matrices. They will be denoted by $\bm{\Gamma}_i$ and have the size $32\times 32$. Constructing them we follow the notations of Ref. \cite{Stepanyantz:2023vat}. The resulting matrixes are Hermitian. They are symmetric and real for odd values of $i$, and antisymmetric and purely imaginary for even values of $i$. Next, we introduce the matrixes

\begin{equation}
\chi_\alpha = \frac{1}{2}\left(\bm{\Gamma}_\alpha+i\bm{\Gamma}_{5+\alpha}\right);\qquad \chi^{+\alpha} = \frac{1}{2}\left(\bm{\Gamma}_\alpha-i\bm{\Gamma}_{5+\alpha}\right),
\end{equation}

\noindent
where the index $\alpha$ ranges from 1 to 5. Using the equation $\{\bm{\Gamma}_i,\,\bm{\Gamma}_j\} = 2\delta_{ij}\cdot 1_{32}$, where $1_{32}$ is the $32\times 32$ identity matrix, it is easy to see that the matrices $\chi_\alpha$ and $\chi^{+\alpha}$ satisfy the anticommutation relations  of the Clifford algebra,

\begin{equation}
\{\chi_\alpha,\,\chi^{+\beta}\} = \delta_\alpha^\beta;\qquad \{\chi_\alpha,\,\chi_\beta\} = 0;\qquad \{\chi^{+\alpha},\,\chi^{+\beta}\} = 0
\end{equation}

\noindent
and the condition $(\chi_\alpha)^T = -(-1)^\alpha \chi^{+\alpha}$.

After that, we define the 32-component column $|0\rangle$ (which of course is not equal to 0) by imposing the conditions

\begin{equation}
\chi_\alpha|0\rangle \equiv 0\quad\mbox{for all}\quad \alpha=1,\ldots,5;\qquad  \langle 0 | 0 \rangle = 1,\quad\mbox{where}\quad \langle0| \equiv \left(|0\rangle\right)^T.
\end{equation}

\noindent
(Note that from these definitions one can easily obtain that $\langle 0|\chi^{+\alpha} = 0$ for all $\alpha=1,\ldots,5$.) Taking into account that $\chi^{*\alpha} = \pm \chi_\alpha$ it is easy to see
that $|0\rangle$ can be chosen to be real. Moreover, in our notation, $|0\rangle$ appears to be the right spinor,  $\bm{\Gamma}_{11}|0\rangle = |0\rangle$. The $SO(10)$ spinors forming the representations $16$ and $\,\xbar{16}\,$, can be presented in the form

\begin{eqnarray}\label{16_Branching}
&&\hspace*{-7mm} |16\rangle = \bar\phi |0\rangle - \frac{1}{2}\bar\phi_{\alpha\beta}\chi^{+\alpha}\chi^{+\beta}|0\rangle
- \frac{1}{4!} \varepsilon_{\alpha\beta\gamma\mu\nu} \bar\phi^\alpha \chi^{+\beta}\chi^{+\gamma}\chi^{+\mu}\chi^{+\nu}|0\rangle;\nonumber\\
&&\hspace*{-7mm} |\,\xbar{16}\,\rangle = \bm{B}\Big(-\phi_\alpha \chi^{+\alpha} |0\rangle + \frac{1}{2!\,3!} \varepsilon_{\alpha\beta\mu\nu\sigma} \phi^{\alpha\beta} \chi^{+\mu} \chi^{+\nu} \chi^{+\sigma} |0\rangle
\nonumber\\
&&\hspace*{-7mm}\qquad\qquad\qquad\qquad\qquad\qquad\qquad\qquad\quad + \frac{1}{5!} \phi\, \varepsilon_{\alpha\beta\mu\nu\sigma} \chi^{+\alpha} \chi^{+\beta} \chi^{+\mu} \chi^{+\nu} \chi^{+\sigma} |0\rangle\Big),\qquad
\end{eqnarray}

\noindent
respectively. Note that here we use the 32-component spinors, but 16 lower components vanish. It is also easy to see that $|16\rangle = |\xbar{16}\rangle^*$ if $\bar\phi = \phi^*$,
$\bar\phi^\alpha = \phi^{*\alpha}$, and $\bar\phi_{\alpha\beta} = \phi^*_{\alpha\beta}$.

The normalization invariant for the representations $16$ and $\,\xbar{16}\,$ can be written in the form

\begin{equation}
N_{16} = 16\times \,\xbar{16}\, \equiv \langle 16 | \,\xbar{16}\,\rangle \equiv (| 16 \rangle)^T | \,\xbar{16}\,\rangle
= \bar\phi \phi + \bar\phi^\alpha \phi_\alpha + \frac{1}{2} \bar\phi_{\alpha\beta} \phi^{\alpha\beta}
\end{equation}

\noindent
and coincides with the expression (\ref{Normalization_16}).

Now, it is possible to find properly normalized invariant $I_{10}$,

\begin{eqnarray}\label{I10}
&& I_{10} \equiv \,\xbar{16}\,\times\,\xbar{16}\,\times 10 = P^a (\Gamma_i B)_{ab} P^b A_i
= \sqrt{2}\,\langle \,\xbar{16}\, | \chi_\alpha \bm{B} |\,\xbar{16}\,\rangle R^\alpha + \sqrt{2}\,\langle \,\xbar{16}\, | \chi^{+\alpha} \bm{B} |\,\xbar{16}\,\rangle R_\alpha\quad\nonumber\\
&& = -2\sqrt{2} \Big(\phi\,\phi_{\alpha} R^\alpha + \phi_\alpha \phi^{\alpha\beta} R_\beta + \frac{1}{8} \varepsilon_{\alpha\beta\mu\nu\sigma} \phi^{\alpha\beta} \phi^{\mu\nu} R^\sigma\Big).
\end{eqnarray}

\noindent
Comparing this expression with Eq. (\ref{Invariant_10}) we see that the normalization constant appears to be $c_{10} = -2\sqrt{2}$. Note that the expression (\ref{I10}) up to notations agrees with the analogous
result of Ref. \cite{Nath:2001uw}.

\medskip

The invariant $I_{126}$ can be considered in a similar way \cite{Nath:2001uw}. The antisymmetric $SO(10)$ tensor with five indices that satisfies the condition (\ref{126_Selfduality})
corresponding to the representation $126$ produces the $SU(5)$ tensors

\begin{eqnarray}\label{126_Branching}
&&\hspace*{-4mm} A_{\alpha\beta\gamma\mu\nu} = 0;\qquad\ \ A_{\alpha\beta\gamma\mu}{}^\nu = \frac{1}{\sqrt{2}}\,\varepsilon_{\alpha\beta\gamma\mu\sigma} L^{\sigma\nu};\qquad\ \
A^{\alpha\beta\gamma\mu\nu} = \varepsilon^{\alpha\beta\gamma\mu\nu} K;\nonumber\\
&&\hspace*{-4mm} A_{\alpha\beta\gamma}{}^{\mu\nu} = -\frac{1}{\sqrt{2}}\Big(\delta_\alpha^\mu K_{\beta\gamma}^\nu - \delta_\alpha^\nu K_{\beta\gamma}^\mu
+ \delta_\gamma^\mu K_{\alpha\beta}^\nu - \delta_\gamma^\nu K_{\alpha\beta}^\mu
+ \delta_\beta^\mu K_{\gamma\alpha}^\nu - \delta_\beta^\nu K_{\gamma\alpha}^\mu \Big);\nonumber\\
&&\hspace*{-4mm} A_{\alpha\beta}{}^{\mu\nu\sigma} = K_{\alpha\beta}^{\mu\nu\sigma} +\frac{1}{\sqrt{6}}\Big(\delta_\alpha^\mu \delta_\beta^\nu K^\sigma - \delta_\alpha^\nu \delta_\beta^\mu K^\sigma
+ \delta_\alpha^\sigma \delta_\beta^\mu K^\nu - \delta_\alpha^\mu \delta_\beta^\sigma K^\nu
+ \delta_\alpha^\nu \delta_\beta^\sigma K^\mu - \delta_\alpha^\sigma \delta_\beta^\nu K^\mu \Big);\nonumber\\
&&\hspace*{-4mm} A_\alpha{}^{\beta\gamma\mu\nu} = \frac{1}{2\sqrt{2}}\Big(\delta_\alpha^\beta \varepsilon^{\gamma\mu\nu\sigma\rho} K_{\sigma\rho}
- \delta_\alpha^\gamma \varepsilon^{\mu\nu\beta\sigma\rho} K_{\sigma\rho}
+ \delta_\alpha^\mu \varepsilon^{\nu\beta\gamma\sigma\rho} K_{\sigma\rho}
- \delta_\alpha^\nu \varepsilon^{\beta\gamma\mu\sigma\rho} K_{\sigma\rho}\Big).
\end{eqnarray}

\noindent
Similar structures for the representation $\,\xbar{126}\,$ are written as

\begin{eqnarray}\label{126Bar_Branching}
&&\hspace*{-5mm} \bar A^{\alpha\beta\gamma\mu\nu} = 0;\qquad\ \ \bar A^{\alpha\beta\gamma\mu}{}_\nu = \frac{1}{\sqrt{2}}\,\varepsilon^{\alpha\beta\gamma\mu\sigma} \,\xbar{L}\,{}_{\sigma\nu};\qquad\ \
\bar A_{\alpha\beta\gamma\mu\nu} = \varepsilon_{\alpha\beta\gamma\mu\nu} \,\xbar{K}\,;\nonumber\\
&&\hspace*{-5mm} \bar A^{\alpha\beta\gamma}{}_{\mu\nu} = -\frac{1}{\sqrt{2}}\Big(\delta^\alpha_\mu \,\xbar{K}\,{}^{\beta\gamma}_\nu - \delta^\alpha_\nu \,\xbar{K}\,{}^{\beta\gamma}_\mu
+ \delta^\gamma_\mu \,\xbar{K}\,{}^{\alpha\beta}_\nu - \delta^\gamma_\nu \,\xbar{K}\,{}^{\alpha\beta}_\mu
+ \delta^\beta_\mu \,\xbar{K}\,{}^{\gamma\alpha}_\nu - \delta^\beta_\nu \,\xbar{K}\,{}^{\gamma\alpha}_\mu \Big);\nonumber\\
&&\hspace*{-5mm} \bar A^{\alpha\beta}{}_{\mu\nu\sigma} = \,\xbar{K}\,{}^{\alpha\beta}_{\mu\nu\sigma} +\frac{1}{\sqrt{6}}\Big(\delta^\alpha_\mu \delta^\beta_\nu \,\xbar{K}\,{}_\sigma - \delta^\alpha_\nu \delta^\beta_\mu \,\xbar{K}\,{}_\sigma
+ \delta^\alpha_\sigma \delta^\beta_\mu \,\xbar{K}\,{}_\nu - \delta^\alpha_\mu \delta^\beta_\sigma \,\xbar{K}\,{}_\nu
+ \delta^\alpha_\nu \delta^\beta_\sigma \,\xbar{K}\,{}_\mu - \delta^\alpha_\sigma \delta^\beta_\nu \,\xbar{K}\,{}_\mu \Big);\nonumber\\
&&\hspace*{-5mm} \bar A^\alpha{}_{\beta\gamma\mu\nu} = \frac{1}{2\sqrt{2}}\Big(\delta^\alpha_\beta \varepsilon_{\gamma\mu\nu\sigma\rho} \,\xbar{K}\,{}^{\sigma\rho}
- \delta^\alpha_\gamma \varepsilon_{\mu\nu\beta\sigma\rho} \,\xbar{K}\,{}^{\sigma\rho}
+ \delta^\alpha_\mu \varepsilon_{\nu\beta\gamma\sigma\rho} \,\xbar{K}\,{}^{\sigma\rho}
- \delta^\alpha_\nu \varepsilon_{\beta\gamma\mu\sigma\rho} \,\xbar{K}\,{}^{\sigma\rho}\Big).
\end{eqnarray}

\noindent
The coefficients in these equations are chosen so that the normalization invariant for the representation $126$ is reproduced correctly,

\begin{eqnarray}
&&\hspace*{-5mm} N_{126} = \frac{1}{5!} A_{ijklm} \bar A_{ijklm}\nonumber\\
&&\hspace*{-5mm}\qquad\ = \,\xbar{K}\ K + \,\xbar{K}\,{}_\alpha K^\alpha + \frac{1}{2} \xbar{K}\,{}^{\alpha\beta} K_{\alpha\beta} + \frac{1}{2} \,\xbar{L}\,{}_{\alpha\beta} L^{\alpha\beta}
+ \frac{1}{2} \,\xbar{K}\,{}_\gamma^{\alpha\beta} K^\gamma_{\alpha\beta} + \frac{1}{12} \,\xbar{K}\,{}_{\alpha\beta\gamma}^{\mu\nu} K^{\alpha\beta\gamma}_{\mu\nu}.\qquad
\end{eqnarray}

\noindent
After that, we calculate the invariant $I_{126}$ using the technique similar to that applied for deriving Eq. (\ref{I10}),

\begin{eqnarray}\label{I126}
&&\hspace*{-5mm} \,\xbar{16}\,\times\,\xbar{16}\,\times 126 =  \frac{1}{5!} P^a (\Gamma_{ijklm} B)_{ab} P^b A_{ijklm}
= 4\sqrt{2}\Big(\phi^2 K + \sqrt{\frac{3}{2}}\phi \phi_\alpha K^\alpha + \frac{1}{\sqrt{2}} \phi \phi^{\alpha\beta} K_{\alpha\beta}\nonumber\\
&&\hspace*{-5mm} + \frac{1}{\sqrt{2}} \phi_\alpha\phi_\beta L^{\alpha\beta} + \frac{1}{\sqrt{2}} \phi_\gamma \phi^{\alpha\beta} K^\gamma_{\alpha\beta}
- \frac{1}{8\sqrt{6}} \varepsilon_{\mu\nu\alpha\beta\gamma} \phi^{\mu\nu} \phi^{\alpha\beta} K^\gamma
+ \frac{1}{24} \varepsilon_{\mu\nu\alpha\beta\gamma} \phi^{\mu\nu} \phi^{\sigma\rho} K_{\sigma\rho}^{\alpha\beta\gamma}\Big).\nonumber\\
\end{eqnarray}

\noindent
Up to notations, this result agrees with the one obtained in \cite{Nath:2001uw}. Comparing it with Eq. (\ref{Invariant_126}) we conclude that $c_{126} = 4\sqrt{2}$.

\medskip

Now, let us apply this technique for constructing the invariant $I_{144}$ defined by Eq. (\ref{SO(10)_Normalization}). In this case the superfield $P_i$ in the $SO(10)$ representation $10$ produces two fields $R_\alpha$ and
$R^\alpha$ in the $SU(5)$ representations $5$ and $\,\xbar{5}\,$, respectively. They are defined exactly as in Eq. (\ref{10_Branching}), but $A_i$ of course should replaced with $P_i$. The $SO(10)$ representation
$\,\xbar{144}\,$ in terms of the $SU(5)$ (super)fields can be presented as two sets

\begin{eqnarray}\label{144_Branching}
&& |A_\alpha\rangle = \bigg(-\frac{2}{\sqrt{5}} B_\alpha + \frac{1}{2\sqrt{30}}\varepsilon_{\alpha\beta\gamma\mu\nu} B^{\beta\gamma} \chi^{+\mu} \chi^{+\nu}\nonumber\\
&&\qquad\qquad\qquad\qquad\qquad
+ \frac{1}{2} B_{\alpha[\mu\nu]} \chi^{+\mu} \chi^{+\nu} + \frac{1}{4!}\varepsilon_{\beta\mu\nu\sigma\rho} B_\alpha^\beta \chi^{+\mu} \chi^{+\nu} \chi^{+\sigma} \chi^{+\rho} \bigg) |0\rangle;\nonumber\\
&& |A^\alpha\rangle = \bigg(-B^\alpha +\frac{1}{2\sqrt{5}} B_\beta \chi^{+\alpha} \chi^{+\beta} + \frac{1}{2} B^\alpha_{\mu\nu} \chi^{+\mu} \chi^{+\nu}\nonumber\\
&&\qquad\qquad\qquad\qquad\qquad
+ \frac{1}{4!}\varepsilon_{\beta\mu\nu\sigma\rho} \Big(\sqrt{\frac{3}{10}} B^{\alpha\beta} + \frac{1}{\sqrt{2}} C^{\alpha\beta}\Big)\chi^{+\mu} \chi^{+\nu} \chi^{+\sigma} \chi^{+\rho}\bigg) |0\rangle,\qquad
\end{eqnarray}

\noindent
which (as they should) satisfy the condition

\begin{equation}
\chi^{+\alpha} |A_\alpha\rangle + \chi_{\alpha} |A^\alpha\rangle = 0
\end{equation}

\noindent
following from the constraint $(B\Gamma_i)^{ab} A_{ib} = 0$. The coefficients in Eq. (\ref{144_Branching}) are chosen in such a way that the normalization invariant takes the form (\ref{Normalization_144}),

\begin{eqnarray}
&&\hspace*{-10mm} N_{144} = A_{ia}\,\bar A^a_i = \langle A_\alpha | \bar A^\alpha \rangle + \langle A^\alpha | \bar A_\alpha\rangle \vphantom{\frac{1}{2}}\nonumber\\
&&\hspace*{-10mm} = B_\alpha \,\xbar{B}\,{}^\alpha + B^\alpha \,\xbar{B}\,{}_\alpha + \frac{1}{2} B^{\alpha\beta} \,\xbar{B}\,{}_{\alpha\beta}
+ \frac{1}{2} C^{\alpha\beta} \,\xbar{C}\,{}_{\alpha\beta} + B_\alpha^\beta \,\xbar{B}\,{}_\beta^\alpha + \frac{1}{2} B_{\alpha[\beta\gamma]} \,\xbar{B}\,{}^{\alpha[\beta\gamma]}
+ \frac{1}{2} B^\gamma_{\alpha\beta} \,\xbar{B}\,{}^{\alpha\beta}_\gamma.\nonumber\\
\end{eqnarray}

\noindent
The invariant under consideration can be written in the form

\begin{equation}\label{I144_Original}
I_{144} \equiv 10 \times\,\xbar{16}\,\times \,\xbar{144}\, = P_i P^a A_{ia}
= R_\alpha \langle \,\xbar{16}\, | A^\alpha \rangle + R^\alpha \langle \,\xbar{16}\, | A_\alpha\rangle.
\end{equation}

\noindent
Substituting the decompositions (\ref{16_Branching}) and (\ref{144_Branching}) into the expression (\ref{I144_Original}) after some (rather tedious) calculations we obtain

\begin{eqnarray}\label{I144}
&&\hspace*{-8mm} 10\times \,\xbar{16}\,\times \,\xbar{144}\, = P_i P^a A_{ia}\vphantom{\frac{1}{2}}\nonumber\\
&&\hspace*{-8mm} = - \frac{2}{\sqrt{5}} \Big(R^\alpha \phi B_\alpha + \frac{\sqrt{5}}{2} R^\alpha \phi_\beta B_\alpha^\beta
+ \frac{1}{4\sqrt{6}} \varepsilon_{\alpha\beta\gamma\mu\nu} R^\alpha \phi^{\beta\gamma} B^{\mu\nu} + \frac{\sqrt{5}}{4} R^\alpha \phi^{\beta\gamma} B_{\alpha[\beta\gamma]}
\nonumber\\
&&\hspace*{-8mm}
+ \frac{\sqrt{5}}{2} R_\alpha \phi B^\alpha + \sqrt{\frac{5}{8}} R_\alpha \phi_\beta C^{\alpha\beta}
+ \frac{\sqrt{5}}{4} R_\alpha \phi^{\beta\gamma} B^\alpha_{\beta\gamma} + \frac{1}{4} R_\alpha \phi^{\alpha\beta} B_\beta
+ \frac{\sqrt{6}}{4} R_\alpha\phi_\beta B^{\alpha\beta}\Big).
\end{eqnarray}

\noindent
This result exactly coincides with Eq. (\ref{Invariant_144}) derived by a different method, but now the normalization constant is fixed. The coincidence confirms the correctness of the calculation, while
comparing Eqs. (\ref{Invariant_144}) and (\ref{I144}) we see that the constant $c_{144}$ in Eq. (\ref{Invariant_144}) is equal to $-2/\sqrt{5}$.

\medskip

Thus, the values of the normalization constants in Eqs. (\ref{Invariant_10}), (\ref{Invariant_126}), and (\ref{Invariant_144}) are given by

\begin{equation}\label{Normalization_Constants}
c_{10} = -2\sqrt{2};\qquad c_{126}= 4\sqrt{2};\qquad c_{144} = - \frac{2}{\sqrt{5}},
\end{equation}

\noindent
respectively.

\subsection{Relations between the Yukawa couplings}
\hspace{\parindent}\label{Subsection_Normalizaion}

Here we derive the relations (\ref{Yukawa_Relation_351_3}) and (\ref{Yukawa_Relation_351_2}) using the expressions for the invariants obtained in the previous subsections. As a starting point, we recall how the $SU(5)$ invariants $\,\xbar{10}\,\times\,\xbar{10}\,\times\,\xbar{5}\,$, $5\times\,\xbar{10}\,\times 5$, and $5\times\,\xbar{10}\,\times 45$ produce various parts present in the MSSM superpotential (\ref{Superpotential_For_MSSM}). In terms of the MSSM superfields various parts of the $SO(10)$ representation $\,\xbar{16}\,$ are written in form

\begin{equation}
\phi = N;\qquad \phi_\alpha = \left(
\begin{array}{c}
D_1 \\ D_2\\ D_3 \\ \widetilde E\\ -\widetilde N
\end{array}
\right);\qquad \phi^{\alpha\beta} = \left(
\begin{array}{ccccc}
0 & U_3 & -U_2 & \widetilde U^1 & \widetilde D^1\\
- U_3 & 0 & U_1 & \widetilde U^2 & \widetilde D^2\\
U_2 & -U_1 & 0 & \widetilde U^3 & \widetilde D^3\\
-\widetilde U^1 & -\widetilde U^2 & -\widetilde U^3 & 0 & E\\
-\widetilde D^1 & -\widetilde D^2 & -\widetilde D^3 & -E & 0
\end{array}
\right).
\end{equation}

\noindent
It is important that in this case the normalization invariant (\ref{Normalization_16}) gives the correct normalization of the MSSM superfields,

\begin{eqnarray}
&& \phi^* \phi + \phi^{*\alpha} \phi_\alpha + \frac{1}{2} \phi^*_{\alpha\beta} \phi^{\alpha\beta} \nonumber\\
&&\qquad \to \left(
\begin{array}{c}
\widetilde U\\
\widetilde D
\end{array}
\right)^+ \left(
\begin{array}{c}
\widetilde U\\
\widetilde D
\end{array}
\right) + U^+ U + D^+ D + \left(
\begin{array}{c}
\widetilde N\\
\widetilde E
\end{array}
\right)^+ \left(
\begin{array}{c}
\widetilde N\\
\widetilde E
\end{array}
\right) + E^+ E + N^+ N.\
\qquad
\end{eqnarray}

In our notation, the invariant $\,\xbar{10}\,\times\,\xbar{10}\,\times\,\xbar{5}\,$ is defined by Eq. (\ref{SU(5)_Normalization}), and the Higgs superfield $H_u$ comes from the doublet part of $\,\xbar{5}\,{}^\gamma$,

\begin{equation}
\,\xbar{5}\,{}^\gamma \to \left(
\begin{array}{c}
0 \\ 0 \\ 0 \\ H_{u1} \\ H_{u2}
\end{array}
\right).
\end{equation}

\noindent
After this replacement, the invariant under consideration gives one of the Yukawa terms present in Eq. (\ref{Superpotential_For_MSSM}),

\begin{equation}\label{10x10x5}
\,\xbar{10}\,\times \,\xbar{10}\,\times \,\xbar{5}\,\equiv \frac{1}{8}\varepsilon_{\mu\nu\alpha\beta\gamma} \phi^{\mu\nu} \phi^{\alpha\beta} \,\xbar{5}\,{}^\gamma\ \to\
\left(\widetilde U\ \widetilde D \right)^{a}
\left(
\begin{array}{cc}
0 & 1\\
-1 & 0
\end{array}
\right)
\left(
\begin{array}{c}
H_{u1}\\ H_{u2}
\end{array}
\right) U_{a}.
\end{equation}

The invariant $5\times\,\xbar{10}\,\times 5 = \phi_\alpha\,\phi^{\alpha\beta}\,5_\beta$ is considered similarly. In this case the Higgs superfield $H_d$ appears from the doublet part of $5_\beta$,

\begin{equation}
5_\beta \to \left(
\begin{array}{c}
0 \\ 0 \\ 0 \\ H_{d2} \\ -H_{d1}
\end{array}
\right),
\end{equation}

\noindent
so that

\begin{equation}\label{5x10x5}
5\times\,\xbar{10}\,\times 5\ \to\
\left(\widetilde N\ \widetilde E \right)
\left(
\begin{array}{cc}
0 & 1\\
-1 & 0
\end{array}
\right)
\left(
\begin{array}{c}
H_{d1}\\ H_{d2}
\end{array}
\right) E
+ \left(\widetilde U\ \widetilde D \right)^{a}
\left(
\begin{array}{cc}
0 & 1\\
-1 & 0
\end{array}
\right)
\left(
\begin{array}{c}
H_{d1}\\ H_{d2}
\end{array}
\right) D_{a}.
\end{equation}

\noindent
This in particular implies that $Y_D = Y_E$ (or, more precisely, $Y_E = (Y_D)^T$ in the case of several generations).

As a simple exercise, let us verify the (well-known) Yukawa relations coming from the $SO(10)$ invariant $\,\xbar{16}\,\times\,\xbar{16}\,\times 10$. Taking into account that

\begin{equation}
1\times 5\times\,\xbar{5}\,\equiv \phi\,\phi_\alpha \,\xbar{5}\,{}^\alpha\ \to\ - \left(\widetilde N\ \widetilde E \right)
\left(
\begin{array}{cc}
0 & 1\\
-1 & 0
\end{array}
\right)
\left(
\begin{array}{c}
H_{u1}\\ H_{u2}
\end{array}
\right) N,
\end{equation}

\noindent
we obtain that

\begin{eqnarray}
&&\hspace*{-5mm} Y\cdot\,\xbar{16}\,\times\,\xbar{16}\,\times 10\ \to\ -2\sqrt{2}\, Y\Big(\phi\,\phi_\alpha \,\xbar{5}\,{}^\alpha + \phi_\alpha\phi^{\alpha\beta} 5_\beta + \frac{1}{8}\varepsilon_{\alpha\beta\mu\nu\sigma}\phi^{\alpha\beta}\phi^{\mu\nu} \,\xbar{5}\,{}^\sigma \Big)\nonumber\\
&&\hspace*{-5mm} \ \to\  2\sqrt{2} Y\, \left(\widetilde N\ \widetilde E \right)
\left(
\begin{array}{cc}
0 & 1\\
-1 & 0
\end{array}
\right)
\left(
\begin{array}{c}
H_{u1}\\ H_{u2}
\end{array}
\right) N - 2\sqrt{2}\, Y \left(\widetilde N\ \widetilde E \right)
\left(
\begin{array}{cc}
0 & 1\\
-1 & 0
\end{array}
\right)
\left(
\begin{array}{c}
H_{d1}\\ H_{d2}
\end{array}
\right) E \qquad
\nonumber\\
&&\hspace*{-5mm} - 2\sqrt{2}\, Y \left(\widetilde U\ \widetilde D \right)^{a}
\left(
\begin{array}{cc}
0 & 1\\
-1 & 0
\end{array}
\right)
\left(
\begin{array}{c}
H_{d1}\\ H_{d2}
\end{array}
\right) D_{a}
- 2\sqrt{2}\, Y \left(\widetilde U\ \widetilde D \right)^{a}
\left(
\begin{array}{cc}
0 & 1\\
-1 & 0
\end{array}
\right)
\left(
\begin{array}{c}
H_{u1}\\ H_{u2}
\end{array}
\right) U_{a}.
\end{eqnarray}

\noindent
Therefore, (in the case of several generations) the Yukawa relations take the form (see, e.g., \cite{Mohapatra:1986uf})

\begin{equation}\label{16_Yukawa_Relations}
Y_U = Y_D = Y_E = -Y_\nu= -2\sqrt{2}\, Y,
\end{equation}

\noindent
where we took into account that the matrix $Y$ should be symmetric.

It remains to consider the invariant $5\times\,\xbar{10}\,\times 45$, where the $SU(5)$ representation $45$ produces the Higgs superfield $H_d$. It is convenient to split the range of index values into two parts, e.g., $\alpha = (a,i)$, where $a=1,\ldots,3$ and $i=4,5$. After setting to 0 all superfields inside $45$ except for the ones containing $H_{d1}$ and $H_{d2}$, the nontrivial components of this representation will be given by the expressions

\begin{equation}\label{45_Higgs}
45^i_{jk} = \frac{\sqrt{3}}{2}\Big[\delta^i_j (H_d)_k - \delta^i_k (H_d)_j \Big];\qquad 45^a_{bi} = - 45^a_{ib} = - \frac{1}{2\sqrt{3}}\delta^a_b (H_d)_i,
\end{equation}

\noindent
where $(H_d)_4 = H_{d2}$ and $(H_d)_5 = -H_{d1}$. Eq. (\ref{45_Higgs}) is obtained if one takes into account the constraints $45^\alpha_{\beta\gamma} = - 45^\alpha_{\gamma\beta}$ and $45^\alpha_{\alpha\gamma} = 0$ together with  the normalization condition. Namely, in this case the correct normalization of the Higgs superfield is obtained because

\begin{equation}
\frac{1}{2} (45^*)^{\beta\gamma}_\alpha\, 45^\alpha_{\beta\gamma}\ \to\ \frac{1}{2} (45^*)^{jk}_i\, 45^i_{jk} + (45^*)^{bi}_a\, 45^a_{bi} = \left(
\begin{array}{c}
H_{d1}\\ H_{d2}
\end{array}
\right)^+
\left(
\begin{array}{c}
H_{d1}\\ H_{d2}
\end{array}
\right).
\end{equation}

\noindent
Then the invariant under consideration takes the form

\begin{eqnarray}\label{5x10x45}
&&\hspace*{-7mm} 5\times\,\xbar{10}\,\times 45 = \frac{1}{2} \phi_\alpha\,\phi^{\beta\gamma}\,45^\alpha_{\beta\gamma}\vphantom{\frac{1}{2}}\nonumber\\
&&\hspace*{-7mm}
\to\  \frac{1}{2}\phi_i\,\phi^{jk}\,45^i_{jk} + \phi_a\,\phi^{bi}\,45^a_{bi} = \frac{\sqrt{3}}{2} \phi_i \phi^{ik} (H_d)_k - \frac{1}{2\sqrt{3}} \phi_a\,\phi^{ai}(H_d)_i\nonumber\\
&&\hspace*{-7mm} = \frac{\sqrt{3}}{2} \left(\widetilde N\ \widetilde E \right)
\left(
\begin{array}{cc}
0 & 1\\
-1 & 0
\end{array}
\right)
\left(
\begin{array}{c}
H_{d1}\\ H_{d2}
\end{array}
\right) E
- \frac{1}{2\sqrt{3}}
\left(\widetilde U\ \widetilde D \right)^{a}
\left(
\begin{array}{cc}
0 & 1\\
-1 & 0
\end{array}
\right)
\left(
\begin{array}{c}
H_{d1}\\ H_{d2}
\end{array}
\right) D_{a},
\end{eqnarray}

\noindent
so that $Y_E = -3 Y_D$ (again, $Y_E = - 3(Y_D)^T$ in the case of several generations).

\medskip

Now everything is ready for deriving the Yukawa relations (\ref{Yukawa_Relation_351_3}) and (\ref{Yukawa_Relation_351_2}). Let us start with the Yukawa relations for the third generation. According to Eqs. (\ref{Invariant_351'}), (\ref{I126}), and (\ref{10x10x5}), in this case the Yukawa coupling $(Y_U)_{33}$ is obtained as follows:

\begin{eqnarray}\label{Y_U_33_Derivation}
&& Y_3\cdot\,\xbar{27}\,\times\,\xbar{27}\,\times\,\xbar{351}\,{}'\Big|_{E_6}\ \nonumber\\
&& \to\  \frac{1}{4} Y_3\cdot \frac{1}{5!} P^a (\Gamma_{ijklm} B)_{ab} P^b A_{ijklm}
= \frac{1}{4} Y_3\cdot \,\xbar{16}\,\times\,\xbar{16}\,\times 126\Big|_{SO(10)}\ \nonumber\\
&& \to\ \frac{1}{4} Y_3\cdot 4\sqrt{2}\cdot \Big(- \frac{1}{8\sqrt{6}}\Big) \varepsilon_{\mu\nu\alpha\beta\gamma} \phi^{\mu\nu} \phi^{\alpha\beta} K^\gamma = -\frac{1}{\sqrt{3}}\,Y_3\cdot \,\xbar{10}\,\times\,\xbar{10}\,\times\,\xbar{5}\,\Big|_{SU(5)}\ \qquad
\nonumber\\
&& \to\ -\frac{1}{\sqrt{3}}\,Y_3\cdot \left(\widetilde U\ \widetilde D \right)^{a}
\left(
\begin{array}{cc}
0 & 1\\
-1 & 0
\end{array}
\right)
\left(
\begin{array}{c}
H_{u1}\\ H_{u2}
\end{array}
\right) U_{a}.
\end{eqnarray}

\noindent
A similar chain producing the Yukawa couplings $(Y_E)_{33}$ and $(Y_D)_{33}$ can be written using Eqs. (\ref{Invariant_351'}), (\ref{I144}), and (\ref{5x10x5}),

\begin{eqnarray}\label{Y_ED_33_Derivation}
&&\hspace*{-5mm} Y_3\cdot\,\xbar{27}\,\times\,\xbar{27}\,\times\,\xbar{351}\,{}'\Big|_{E_6}\ \nonumber\\
&&\hspace*{-5mm} \to\  \sqrt{2}\,Y_3\cdot P_i P^a A_{ia}
= \sqrt{2}\,Y_3\cdot 10\times \,\xbar{16}\,\times\,\xbar{144}\,\Big|_{SO(10)}\ \nonumber\\
&&\hspace*{-5mm} \to\ \sqrt{2}\,Y_3\cdot \Big(-\frac{2}{\sqrt{5}}\Big)\cdot \frac{1}{4} R_\alpha \phi^{\alpha\beta} B_\beta = -\frac{1}{\sqrt{10}}\,Y_3\cdot 5\times \,\xbar{10}\,\times\,5 \Big|_{SU(5)}\ \qquad
\nonumber\\
&&\hspace*{-5mm} \to\ -\frac{1}{\sqrt{10}}\,Y_3 \left(\widetilde N\ \widetilde E \right)
\left(
\begin{array}{cc}
0 & 1\\
-1 & 0
\end{array}
\right)
\left(
\begin{array}{c}
H_{d1}\\ H_{d2}
\end{array}
\right) E
-\frac{1}{\sqrt{10}}\, Y_3 \left(\widetilde U\ \widetilde D \right)^{a}
\left(
\begin{array}{cc}
0 & 1\\
-1 & 0
\end{array}
\right)
\left(
\begin{array}{c}
H_{d1}\\ H_{d2}
\end{array}
\right) D_{a}.\nonumber\\
\end{eqnarray}

\noindent
Therefore, the equations relating the Yukawa couplings of the third generation take the form

\begin{equation}
Y_3 = -\sqrt{3}\, (Y_U)_{33} = -\sqrt{10}\,(Y_E)_{33} = -\sqrt{10}\,(Y_D)_{33}
\end{equation}

\noindent
and are equivalent to Eq. (\ref{Yukawa_Relation_351_3}). The other relations present in Eqs. (\ref{Y_U_33_Derivation}) and (\ref{Y_ED_33_Derivation}) are collected in Table \ref{Table_Branching3}.

\medskip

For the second generation, the Yukawa coupling $(Y_U)_{22}$ is obtained with the help of Eqs. (\ref{Invariant_351'}), (\ref{I10}), and (\ref{10x10x5}) through the chain

\begin{eqnarray}\label{Y_U_22_Derivation}
&& Y_2\cdot\,\xbar{27}\,\times\,\xbar{27}\,\times\,\xbar{351}\,{}'\Big|_{E_6}\ \nonumber\\
&& \to\  Y_2\cdot \frac{1}{4\sqrt{5}}\, P^a (\Gamma_i B)_{ab} P^b A_i
= \frac{1}{4\sqrt{5}}\, Y_2\cdot \,\xbar{16}\,\times\,\xbar{16}\,\times 10\Big|_{SO(10)}\ \nonumber\\
&& \to\ \frac{1}{4\sqrt{5}}\, Y_2\cdot (-2\sqrt{2})\cdot \frac{1}{8} \varepsilon_{\mu\nu\alpha\beta\gamma} \phi^{\mu\nu} \phi^{\alpha\beta} R^\gamma = -\frac{1}{\sqrt{10}}\,Y_2\cdot \,\xbar{10}\,\times\,\xbar{10}\,\times\,\xbar{5}\,\Big|_{SU(5)}\ \qquad
\nonumber\\
&& \to\ -\frac{1}{\sqrt{10}}\,Y_2\cdot \left(\widetilde U\ \widetilde D \right)^{a}
\left(
\begin{array}{cc}
0 & 1\\
-1 & 0
\end{array}
\right)
\left(
\begin{array}{c}
H_{u1}\\ H_{u2}
\end{array}
\right) U_{a}.
\end{eqnarray}

\noindent
Similarly, using Eqs. (\ref{Invariant_351'}), (\ref{I126}), and (\ref{5x10x45}) we derive the equations relating $(Y_E)_{22}$ and $(Y_D)_{22}$ to the coupling $Y_2$,

\begin{eqnarray}\label{Y_ED_22_Derivation}
&&\hspace*{-5mm} Y_2\cdot\,\xbar{27}\,\times\,\xbar{27}\,\times\,\xbar{351}\,{}'\Big|_{E_6}\ \nonumber\\
&&\hspace*{-5mm} \to\  \frac{1}{4} Y_2\cdot \frac{1}{5!} P^a (\Gamma_{ijklm} B)_{ab} P^b A_{ijklm}
= \frac{1}{4} Y_2\cdot \,\xbar{16}\,\times\,\xbar{16}\,\times 126\Big|_{SO(10)}\ \nonumber\\
&&\hspace*{-5mm} \to\ \frac{1}{4}\, Y_2\cdot 4\sqrt{2}\cdot \frac{1}{\sqrt{2}} \phi_\gamma \phi^{\alpha\beta} K^\gamma_{\alpha\beta} = 2Y_2\cdot 5\times \,\xbar{10}\,\times 45 \Big|_{SU(5)}\ \qquad
\nonumber\\
&&\hspace*{-5mm} \to\ \sqrt{3}\,Y_2 \left(\widetilde N\ \widetilde E \right)
\left(
\begin{array}{cc}
0 & 1\\
-1 & 0
\end{array}
\right)
\left(
\begin{array}{c}
H_{d1}\\ H_{d2}
\end{array}
\right) E
- \frac{1}{\sqrt{3}}\,Y_2
\left(\widetilde U\ \widetilde D \right)^{a}
\left(
\begin{array}{cc}
0 & 1\\
-1 & 0
\end{array}
\right)
\left(
\begin{array}{c}
H_{d1}\\ H_{d2}
\end{array}
\right) D_{a}.\nonumber\\
\end{eqnarray}

\noindent
Therefore, the Yukawa relations for the second generation following from Eqs. (\ref{Y_U_22_Derivation}) and (\ref{Y_ED_22_Derivation}) can be written in the form

\begin{equation}
Y_2 = -\sqrt{10}\, (Y_U)_{22} = \frac{1}{\sqrt{3}}\,(Y_E)_{22} = -\sqrt{3}\,(Y_D)_{22}
\end{equation}

\noindent
and are equivalent to Eq. (\ref{Yukawa_Relation_351_2}). The other relations derived in Eqs. (\ref{Y_U_22_Derivation}) and (\ref{Y_ED_22_Derivation}) are presented in Table \ref{Table_Branching2}.

\end{document}